\documentclass{aa}  

\usepackage{graphicx}
\usepackage{txfonts}
\usepackage [normalem] {ulem}
\usepackage{booktabs} 
\usepackage{amsmath}
\usepackage{placeins} 

\newcommand{\teff}{$T_{\mathrm{eff}}$}
\newcommand{\av}{$A_{\mathrm{V}}$}
\newcommand{\veil}{$r_{\mathrm{veil}}$}
\newcommand{\logg}{$\log{g}$}
\newcommand{\veilmap}{$r^{\mathrm{MAP}}_{\mathrm{veil}}$}
\newcommand{\veiltf}{$r^{\mathrm{TF}}_{\mathrm{veil}}$}

\newcommand{\nai}{\mbox{Na\,\textsc{i}}}
\newcommand{\ki}{\mbox{K\,\textsc{i}}}

\newcommand{\msol}{M$_{\odot}$}

\newcommand{\Clean}{Clean}
\newcommand{\Unc}{Uncertain}

\defcitealias{Kang+23b}{Paper 1}
\defcitealias{Itrich+2024}{IT24}
\usepackage[colorlinks=true, citecolor=blue, linkcolor=blue]{hyperref}
\begin{document}

   \title{Spectral classification of young stars using conditional invertible neural networks}

   \subtitle{II. Application to Trumpler 14 in Carina}

   \author{Da Eun Kang\inst{1,2}
          \and
          Dominika Itrich\inst{3,4,5}
          \and
          Victor F. Ksoll\inst{2}          
          \and
          Leonardo Testi\inst{1,6}
          \and
          Ralf S. Klessen\inst{2,7, 8, 9}
          \and
          Sergio Molinari\inst{10}
          }

   \institute{
            Alma Mater Studiorum Università di Bologna, Dipartimento di Fisica e Astronomia (DIFA), Via Gobetti 93/2, I-40129, \\
            Bologna, Italy \\
            \email{daeun.kang@unibo.it}
            \and
            Universit\"{a}t Heidelberg, Zentrum f\"{u}r Astronomie, 
            Institut f\"{u}r Theoretische Astrophysik, 
            Albert-Ueberle-Stra{\ss}e 2,\\
            D-69120 Heidelberg, Germany
            \and
            European Southern Observatory, Karl-Schwarzschild-Str. 2, 85748 Garching bei M\"{u}nchen, Germany
            \and
            Universit\"{a}ts-Sternwarte, Ludwig-Maximilians-Universit\"{a}t, Scheinerstrasse 1, 81679 M\"{u}nchen, Germany
            \and
            Steward Observatory, The University of Arizona, Tucson, AZ 85721, USA
            \and
            INAF-Osservatorio Astrofisico di Arcetri, Largo E. Fermi 5, I-50125, Firenze, Italy 
            \and
            Universit\"{a}t Heidelberg, Interdisziplin\"{a}res Zentrum f\"{u}r Wissenschaftliches Rechnen, Im Neuenheimer Feld 205,\\
            D-69120 Heidelberg, Germany
            \and 
            Harvard-Smithsonian Center for Astrophysics, 60 Garden Street, Cambridge, MA 02138, USA
            \and
            Elizabeth S. and Richard M. Cashin Fellow at the Radcliffe Institute for Advanced Studies at Harvard University, 10 Garden Street, Cambridge, MA 02138, USA
            \and
            INAF - Istituto di Astrofisica e Planetologia Spaziali, Via Fosso del Cavaliere 100, I-00133 Roma, Italy
            }
             
 
  \abstract
  {}
   {We introduce an updated version of our deep learning tool that predicts effective temperature, surface gravity, extinction, and veiling from the optical spectra of young low-mass stars with intermediate spectral resolution. We determine the stellar parameters of 2051 stars in Trumpler 14 (Tr14) in the Carina Nebula Complex, observed with VLT/MUSE.}
   {We adopt a conditional invertible neural network (cINN) architecture to infer the posterior distribution of stellar parameters and train our cINN on two Phoenix stellar atmosphere model libraries (Settl and Dusty). Compared to the cINNs presented in our first study, the updated cINN considers the influence of the relative flux error on the parameter estimation and predicts an additional fourth parameter, veiling. We test the prediction performance of cINN on synthetic test models to quantify the intrinsic error of the cINN as a function of relative flux error and on 36 class III template stars to validate the performance on real spectra.
   Using our cINN, we estimate the stellar parameters of young stars in Tr14 and compare them with those derived using the classical template fitting method applied to the same data in a previous study.}
   {We provide \teff, \logg, \av, and \veil\ values of 2051 stars in Tr14 measured by our cINN as well as stellar ages and masses derived from the Hertzsprung–Russell diagram based on the measured parameters. Our parameter estimates generally agree well with those measured by template fitting. However, for K- and G-type stars, the \teff\ derived from template fitting is, on average, 2–3 subclasses hotter than the cINN estimates, while the corresponding veiling values from template fitting appear to be underestimated compared to the cINN predictions.
   We obtain an average age of 0.7$^{+3.2}_{-0.6}$~Myr for the Tr14 stars. By examining the impact of veiling on the equivalent width-based classification, we demonstrate that the main cause of temperature overestimation for K- and G-type stars in the previous study is that veiling and effective temperature are not considered simultaneously in their process.
    }
   {Our cINN performs comparably to the multi-dimensional template fitting method while being significantly faster and capable of consistently analysing stars across a wide temperature range (2600–7000~K).
   }

   \keywords{methods: statistical --
                stars: late-type --
                stars: pre-main sequence -- 
                ISM: HII regions --
                Galaxy: open clusters and associations: individual: Trumpler 14, Carina Nebula Complex
               }

   \maketitle

\section{Introduction}
Low-mass stars, whose masses are similar to or lower than the solar mass, account for the majority of the stars in star-forming regions~\citep{Bochanski+10} and about half of the total stellar mass~\citep{Kroupa01, Chabrier03}. As low-mass stars remain in the pre-main-sequence phase even when the massive stars are dead, studying low-mass stars is key to understanding the early phases of stellar evolution, protoplanetary disk, and planet formation. 
Therefore, it is essential to accurately determine the fundamental physical parameters of low-mass stars to gain insights into their structure and evolutionary processes.

Stellar parameters are estimated from photometric or spectroscopic data by using characteristic features (i.e. spectral indices) that appear differently depending on the type of stars~\citep[e.g.][]{Luhman+1997, Luhman1999, Luhman2003, Riddick2007, Herczeg2014, Rugel+2018} or by fitting the observed spectrum with already well-analysed stars, referred to as templates~\citep[e.g.][]{Fang+2021, Itrich+2024}. Thus, selecting the appropriate spectral indices tailored to the star under consideration is important in these methodologies. In addition, the range of applicable methods may be restricted by the wavelength and specification of the instrument. In this study, we apply a deep learning-based method to consistently analyse young low-mass stars over a wide temperature range.

Recently, numerous studies utilised artificial neural networks~\citep[NNs;][]{Goodfellow+16} in various ways, e.g. to predict physical parameters~\citep[e.g.][]{Fabbro+2018, Ksoll+20,Olney2020, Rhea+2020, Sharma2020, Kang+22, Shen+2022}, to efficiently analyse images such as identifying structures and exoplanets~\citep[e.g.][]{Abraham+18, DeBeurs+2022}, and to classify observations~\citep[e.g.][]{Wu+19, Wei+2020, Whitmore+2021, Walmsley+2021}.
In addition to the time-efficient nature of machine learning techniques, NNs are advantageous in solving complicated problems that are difficult to solve with classical methods.
In this study, we adopt the conditional invertible neural network (cINN) architecture~\citep{Ardizzone2019b}. Based on a supervised learning approach, the cINN is suitable for tasks such as regression or classification because it finds the complex links between the observable quantities and physical properties hidden in the training data. Moreover, another advantage of the cINN is that it always provides a posterior distribution for the target parameters without any additional calculations. For this reason, the cINN is especially well suited for solving degenerate inverse problems and has been used to analyse various complicated observations in astronomy~\citep[e.g.][]{Ksoll+20, Ksoll+2024,  Bister+2022, Kang+22, Kang+23a, Haldemann2022, Candebat+2024}.

We develop a cINN that estimates four stellar parameters (\teff, \logg, \av, and \veil) from the optical stellar spectrum with intermediate spectral resolution and apply it to analyse numerous low-mass stars in Trumpler 14 (Tr14) in the Carina Nebula observed with the Multi Unit Spectroscopic Explorer (MUSE) of the Very Large Telescope (VLT). As a first step, in \citet[][hereafter \citetalias{Kang+23b}]{Kang+23b}, we pretested whether a cINN can extract physical properties well from the stellar spectrum. In \citetalias{Kang+23b}, we introduced three cINNs trained on different Phoenix stellar atmosphere model libraries (Settl, NextGen, and Dusty) and examined the performance of these networks on 36 class III template stars that are well-analysed in the literature~\citep{Manara2013, Stelzer2013, Manara2017}. We confirmed that cINNs estimated basic stellar parameters (\teff, \logg, and \av) from the optical spectrum with a sufficiently high accuracy, achieving an average error of 3\%, with a maximum error range of  5-10\%.

In this paper, we introduce a new cINN to estimate the stellar parameters of 2051 low-mass stars in Tr14. The new cINN additionally estimates spectrum veiling (\veil) due to the excess emission by the accretion or disk emission. Moreover, it considers the influence of the flux error on the parameter estimation because the signal-to-noise ratios (S/N) of the spectra analysed in this study are on average lower than that of the template stars used in \citetalias{Kang+23b}, so that the influence of the flux error is not negligible anymore. We first test the overall performance of the new cINN using synthetic test models and 36 class III template stars and then determine the stellar parameters of Tr14 stars. We compare the parameters estimated by our cINN with those obtained in \citet[][hereafter \citetalias{Itrich+2024}]{Itrich+2024}, which were derived using classical template fitting methods on the same observational data.

The paper is structured as follows. We describe the observational data of Tr14 and the template fitting method used in \citetalias{Itrich+2024} in Sect.~\ref{sec:sample} and introduce the structure of the cINN, our network setup and training methodologies in Sect.~\ref{sec:neural_network}. Sect.~\ref{sec:training_data} presents the compilation of our training data. We test and validate the performance of our new cINN in Sect.~\ref{sec:validation}. In Sect.~\ref{sec:result}, we present our overall results on Tr14 stars. In Sect.~\ref{sec:discussion}, we further discuss the methodological limitations of both the cINN and template fitting method and the validity of the measured veiling. We summarise the results in Sect.~\ref{sec:summary}.


\section{Young stars in Tr14}
\label{sec:sample}

\subsection{Observation and samples}
\label{sec:observation_samples}
The Carina Nebula Complex (CNC) is one of the most massive star-forming regions in the Galaxy containing more than 70 O-type stars~\citep{Smith2006, Berlanas+2023}, emitting intense ultraviolet radiation~\citep{Smith2006}, and located at a distance of 2.35 kpc~\citep{Goppel&Preibisch2022} from the Sun in the plane of the Galactic disc. Interstellar extinction towards the CNC is known to be low but exhibits an abnormal reddening law, especially around Trumpler 14 and 16 ($R_{\rm{V}} = 4 - 5$, e.g. \citealt{Tapia+2003, Carraro+2004, Hur+2012, Hur+2023}). Among the three main star clusters in the CNC (Trumpler 14, 15, and 16), Tr14 is the youngest and most compact one with an age estimated around 1 Myr (\citealt{Penny+1993, Vazquez+1996, Carraro+2004, Smith&Brooks2008}, \citetalias{Itrich+2024}).

Tr14 has been observed with the MUSE \citep{Bacon2010} on the VLT under the programme ID 097.C-0137 (PI: A. McLeod). MUSE is an integral-field unit (IFU) instrument and offers spatially sampled medium-resolution spectroscopy ($R\sim4000$) in the optical regime (4650-9300~\AA) with the relative wavelength accuracy expected to be below 0.1\AA\ \citep{Weilbacher2020}. Observations were performed in Wide Field Mode with a field of view of $1\arcmin\times1\arcmin$ and a total integration time of 39~min per pointing. The whole mosaic covering the Tr14 cluster consists of 22 pointings with the seeing ranging from 0.5" to 1.6". 
\citetalias{Itrich+2024} reduced the observations with the dedicated ESO pipeline v.~2.8.3 \citep{Weilbacher2020}, part of the {\tt EsoReflex} environment \citep{Freudling2013}. The data reduction included wavelength and flux calibration, as well as astrometry correction. Stellar spectra were extracted from datacubes using {\tt Source-Extractor} \citep{Bertin1996} with the 50\% level of completeness at 15.5~mag based on the $J$-band HAWK-I magnitudes matched to the sources (\citetalias{Itrich+2024}). 
They estimated stellar parameters (\teff, \av, and \veil) from the stellar spectra using class III template stars observed with VLT/X-Shooter and analysed by \cite{Manara2013, Manara2017}. 
In this paper, we refer to measurements of \citetalias{Itrich+2024} as TF parameters (template-fitting parameters) or TF values and compare them to the stellar parameters obtained in this work.

The CNC is a bright H{\sc ii} region with spatially highly variable nebular emission. This emission hinders spectral analysis because nebular lines contaminate the stellar ones and make their measurement particularly challenging. \citetalias{Itrich+2024} adopted a conservative approach to this problem. 
After removing spurious sources (e.g. objects with $I$-band photometric uncertainty > 0.1~mag, when one or more pixels were saturated, or when the position of the object was too close to the edge of the image, etc.), \citetalias{Itrich+2024} excluded sources where nebular emission variation around the star was too large compared to the stellar flux.
They measured the fluctuation of $I$-band nebular emission within a radius of 20\arcsec\ and compared it to the stellar emission. 
\citetalias{Itrich+2024} then applied an S/N ratio cut (mean S/N >10) for the robustness and derived a \lq\lq\Clean\rq\rq\ sample consisting of all the sources, whose $I$-band flux is higher than three times the fluctuation of nebular emission. 

Among the $\sim$800 \Clean\ samples, \citetalias{Itrich+2024} excluded foreground and background stars based on parallaxes corrected for bias ($\overline{\omega}$), as described in \cite{Lindegren+2021}. First, \citetalias{Itrich+2024} selected stars from the Clean samples with good Gaia astrometry~\citep{Gaia2016, Gaia2023} to establish an exclusion condition using the following criteria: goodness of fit parameter, RUWE $\leq$ 1.4~\citep{Lindegren2018}, astrometric\_gof\_al $\leq$ 5~\citep{Lindegren+2021}, a parallax over error larger than 5, and uncertainty of the proper motion below 20\%.\footnote{We confirmed that the change of proper motion quality cut to proper motion error smaller than 0.2 mas/yr does not change the final good astrometry sample of Gaia counterparts.} The distribution of corrected parallaxes for the selected 175 stars satisfying the above criteria was then fitted with a Gaussian profile (see Fig. 3 of \citetalias{Itrich+2024}) which centre value fell on 0.43~mas, which corresponds well to the distance of 2.35~kpc found in \cite{Goppel&Preibisch2022}, and a 1-$\sigma$ width of 0.04~mas. Here, the exclusion criteria, $\overline{\omega}_{\rm{min}}$ and $\overline{\omega}_{\rm{max}}$, are set to values $\mp$1-$\sigma$ away from the centre, which correspond to 2.61~kpc and 2.13~kpc, respectively. Stars whose parallax values considering the 3-sigma uncertainty were lower than $\overline{\omega}_{\rm{min}}$ or larger than $\overline{\omega}_{\rm{max}}$ were defined as background stars or foreground stars, respectively. After excluding foreground and background stars, 780 stars remained in the final \Clean\ sample in \citetalias{Itrich+2024}.

Due to the strict limits on contamination by nebular emission described above, more than 1800 sources were excluded in the main analysis of \citetalias{Itrich+2024}, mostly stars with $M_*< 1\,M_\odot$. However, \citetalias{Itrich+2024} also identified probable members among these sources, characterised their stellar parameters, and reported a separate catalogue for these stars (see Table D.1 of \citetalias{Itrich+2024}), following the same member selection approach and the template fitting method used for the \Clean\ samples. \citetalias{Itrich+2024} did not apply an additional S/N ratio cut to these sources but confirmed that they have mean S/N > 2, with approximately 80\%\ having mean S/N > 10.
This \lq\lq\Unc\rq\rq\ sample of 1867 objects constitutes a significant fraction of the \lq\lq All\rq\rq\ sample (71\%) which could add to the statistical significance of the study of \citetalias{Itrich+2024}. Here, we aim to enhance these previous efforts, use the full potential of MUSE capabilities with our updated neural network architecture and classify all Tr14 sources observed with MUSE. 

Among the 780 \Clean\ samples and 1867 \Unc\ samples reported by \citetalias{Itrich+2024}, we only use stars that have $J$-band magnitudes from the HAWK-I~\citep{Preibisch+2011a, Preibisch+2011b} or VISTA~\citep{Preibisch+2014} catalogues and also have spectral type assessed by \citetalias{Itrich+2024} for a direct comparison. This retains 727 and 1324 sources in the \Clean\ and \Unc\ samples, respectively. Because of several quality cuts applied in \citetalias{Itrich+2024}, the $I$-band magnitudes of the final samples range from 12.7 to 22.3~mag. This corresponds to a lower mass limit of $\sim$0.06~\msol\ ($\sim$2850~K, M7) at 1~Myr~\citep{Baraffe2015} when applying a distance modulus of 11.86~mag, equivalent to 2.35~kpc~\citep{Goppel&Preibisch2022}, and a global visual extinction towards Tr14 of $\sim$2.6~mag (\citetalias{Itrich+2024}) with $R_{\mathrm{V}}$ of 4.4~\citep{Hur+2012}.

\begin{figure}
    \centering
    \includegraphics[width=\columnwidth]{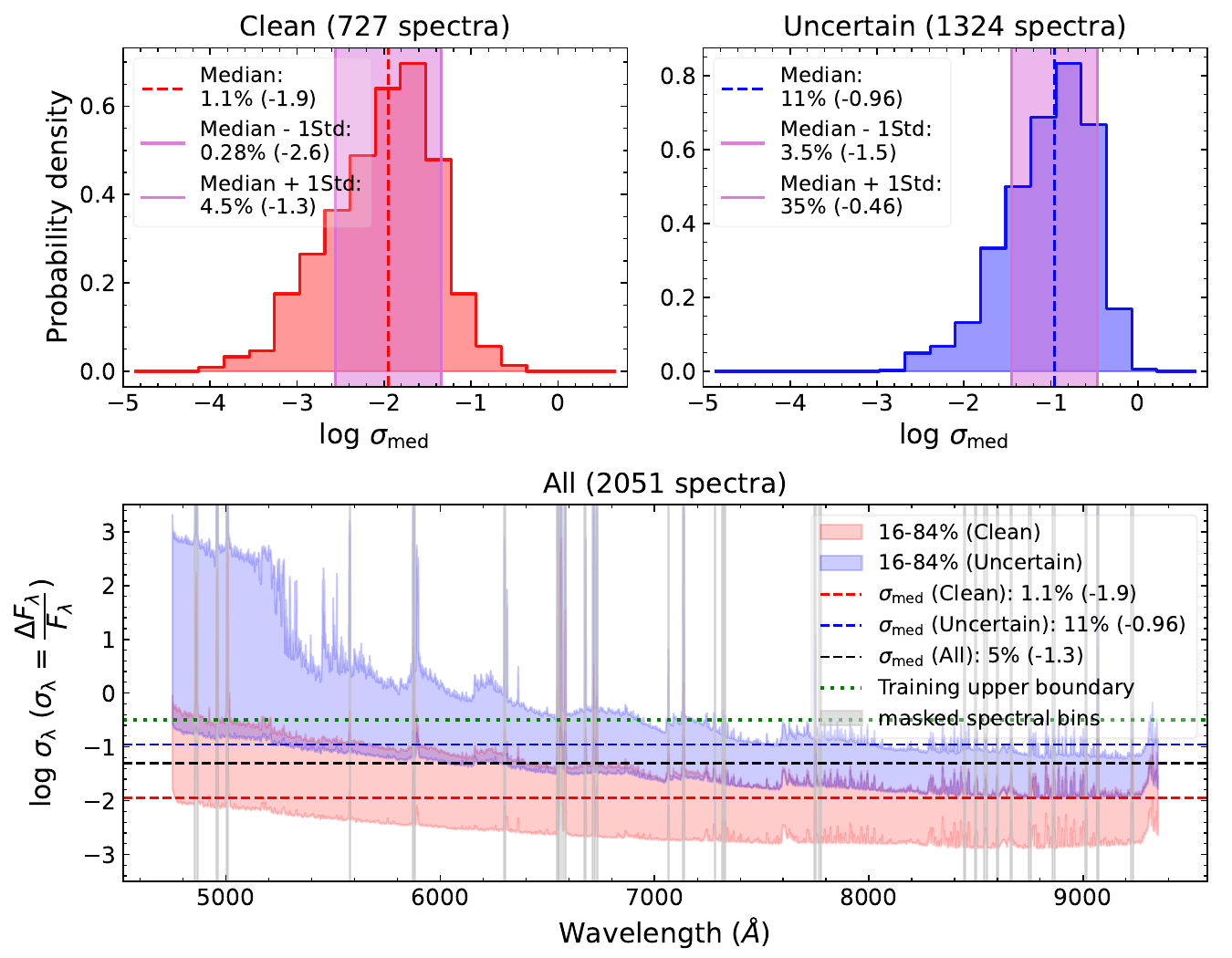}
    \caption{Upper panels: Histograms of median relative flux error ($\sigma_{\rm{med}}$) for the spectra in the \Clean\ samples (left) and the \Unc\ samples (right). Dashed lines and purple shaded areas indicate the median values and $\pm$1 standard deviations of the distributions.
    Lower panel: Distribution of relative flux error at each spectral bin ($\sigma_{\lambda} =  \Delta \rm{Flux}(\lambda)/\rm{Flux}(\lambda)$) for the \Clean\ (red) and \Unc\ (blue) groups. The shaded areas indicate the interquantile ranges from 16\%\ to 84\%. The red and blue dashed lines show the median of $\sigma_{\rm{med}}$ for each group as shown in the upper panels and the black dashed line shows the median $\sigma_{\rm{med}}$ for all 2051 spectra. 
    The green dotted line indicates the upper boundary of the training range (log $\sigma$=-0.5) of our cINN. The narrow grey-shaded vertical areas indicate the masked spectral bins not used in our network due to the presence of emission lines. 
     } \label{fig:flux_err_distr} 
\end{figure}

As the stellar spectra of several Tr14 sources exhibit low S/N ratios, we design a cINN that takes both the flux and flux error into account in this paper (see Sect.~\ref{sec:noise-net} for more details). This cINN adopts the dimensionless relative flux error ($\sigma_{\lambda}$), that is the flux error divided by the flux, $\sigma_{\lambda} = \Delta \rm{Flux}(\lambda)/\rm{Flux}(\lambda)$, the same as noise-to-signal ratio.
The bottom panel of Fig.~\ref{fig:flux_err_distr} shows the $\sigma_{\lambda}$ distributions of the \Clean\ and \Unc\ samples as a function of wavelength. Although $\sigma_{\lambda}$ is sensitive to wavelength, we decided to use a representative value, the median value across the wavelength range ($\sigma_{\rm{med}}$), in the following, because we encountered issues in training the cINN (see Sect.~\ref{sec:noise-net}) and difficulties in modelling $\sigma_{\lambda}$. The top panels of Fig.~\ref{fig:flux_err_distr} present the $\sigma_{\rm{med}}$ distributions for \Clean\ and \Unc\ samples, respectively. $\sigma_{\rm{med}}$ is widely distributed on a logarithmic scale from $10^{-4}$ to $10^{0}$ (0.01 -- 10~\%). The \Unc\ samples have on average about 1~dex higher $\sigma_{\rm{med}}$ than the \Clean\ samples. The median of $\sigma_{\rm{med}}$ for the \Clean\ samples is about 1\%. We provide in Table~\ref{table:catalog} the measured $\sigma_{\rm{med}}$ of all sources.

With this approach, $\sigma_{\lambda}$ can be underestimated at shorter wavelengths, because $\sigma_{\rm{med}}$ is mostly determined at 6500--7500~\AA\ for Tr14 data. 
To assess this, we calculate the median $\sigma_{\lambda}$ at 4750–5500~\AA\ ($\sigma_{5000}$) per star. Using spectral classifications from \citetalias{Itrich+2024}, we find that M-type stars have an average $\sigma_{5000}$/$\sigma_{\rm{med}}$ of 13, which decreases to 7.5 for the \Clean\ sample. In contrast, K- and G-type stars show lower ratios of 4.8 and 2.8, respectively, indicating a smaller S/N ratio variation across the spectrum.
K/G-type stars also exhibit higher overall S/N ratios, with median $\sigma_{\rm{med}}$ values of 0.88\%\ (K-type) and 0.27\%\ (G-type), compared to 9.84\%\ for M-type stars. The median $\sigma_{\lambda}$ at $\lambda$ < 5500~\AA\ is 3.8\%\ (K-type) and 0.66\%\ (G-type), both lower than the 3.4\%\ median $\sigma_{\lambda}$ at $\lambda$ > 8500~\AA\ for M-type stars. While underestimating $\sigma_{\lambda}$ at shorter wavelengths could impact parameter accuracy, this effect is likely minor for earlier-type stars, given their higher S/N ratios and smaller spectral S/N ratio variation. In future studies, we will improve our approach to account for wavelength-dependent flux errors.

\begin{table*}
    \centering
    \caption{Catalouge of low-mass stars in Trumpler 14 with stellar parameters measured by cINN.}
    \resizebox{\textwidth}{!}{
    \begin{tabular}{c cccc cccc cccc}
    \toprule
    ID  & Coordinates   & Sample group  & $\sigma_{\mathrm{med}}$  &  $T_{\mathrm{eff}}$ & $\log(g/\mathrm{cm\,s}^{-2})$ & $A_{\mathrm{V}}$ & $r_{\mathrm{veil}}$ & Library & SpT & log $(L_{\mathrm{bol}}/L_{\odot})$ & $M_{*}$ & Age \\
        & (h:m:s d:m:s) &  & (\%) & (K) &  & (mag) & & & &  & ($M_{\odot}$) & (Myr) \\
    \midrule
    F01N010 & 10:44:08.38 -59:29:12.02 & Clean & 0.6 & 4469$\pm$28 & 4.34$\pm$0.07 & 1.43$\pm$0.03 & 0.01$\pm$0.02 & Settl & K4.5$^{+0.1}_{-0.1}$ & -0.29 & 1.02 & 9.5 \\ 
    F01N100 & 10:44:06.81 -59:29:40.96 & Clean & 0.1 & 4216$\pm$26 & 3.34$\pm$0.07 & 2.73$\pm$0.03 & 0.10$\pm$0.02 & Settl & K5.9$^{+0.2}_{-0.2}$ & 1.01 & 0.77 & 0.1 \\
    \bottomrule
    \end{tabular}} 
    \label{table:catalog} 
    \tablefoot{The first column gives IDs of the sources, the second one lists coordinates, the third one lists the sample group and the fourth one lists the median relative flux error along the wavelength that is fed into the cINN. The following nine columns list measured parameters: effective temperature, surface gravity, visual extinction, veiling factor at 7500~\AA, Phoenix library origin flag, spectral type, bolometric luminosity, and stellar mass and age estimated from PARSEC~\citep{Bressan+2012} evolutionary tracks. A full version of this table is available at CDS. The ﬁrst few rows are shown as an example.}
\end{table*}

\subsection{Spectral classification used in \citetalias{Itrich+2024}}
\label{sec:template fitting}

In this section, we briefly summarise the spectral classification methods used in \citetalias{Itrich+2024}. Some approaches such as for veiling and reddening are applied identically to the training data of our cINN in this paper. \citetalias{Itrich+2024} found the best-matching template for the observed spectra using 37 class III template stars, which cover a range from M9.5 to G8 type and are all subject to a negligible amount of extinction of below 0.3~mag. These stars are observed with VLT/X-shooter and published by \cite{Manara2013, Manara2017}. Because of the difference in spectral coverage between MUSE and the optical arm of X-Shooter, \citetalias{Itrich+2024} convolved the template spectra with a Gaussian kernel to match the MUSE resolution and re-sampled them to the common spectral range between 5600 and 9350~\AA, which is a bit narrower than the entire spectral range of the MUSE spectra (4750--9350~\AA). Spectra of both Tr14 samples and template stars are normalised by the flux at 7500~\AA.

\citetalias{Itrich+2024} preselected M-type stars by using the spectral indices based on the TiO and VO absorption bands~\citep{Riddick2007, Jeffries2007, Herczeg2014} and then applied different classification methods to M-type stars and K/G-type stars. 
For M-type stars, \citetalias{Itrich+2024} found the best-fit template simultaneously considering the spectral type, extinction, and veiling from a grid of template spectra with extinction (\av) covering a range from 0 to 7 mag in steps of 0.1~mag and veiling (\veil) covering an interval from 0 to 1.9 in steps of 0.02. They adopted a simple constant veiling model, similar to the approach used in \cite{Fang+2021}, where the additional flux is determined by a veiling factor at 7500~\AA\ (\veil) and the flux at 7500~\AA\ ($F_{7500}$):
\begin{equation}
\label{eq:veiling}
    F_{\lambda, \rm{veiled}} = F_{\lambda} + F_{7500} \times r_{\rm{veil}}.
\end{equation}
They first veiled the spectrum and redden it with a given visual extinction value (\av) following the extinction law of \cite{Cardelli+1989}, adopting the $R_{\rm{V}} = 4.4$ for Tr14~\citep{Hur+2012}.

On the other hand, for K- or G-type stars, \citetalias{Itrich+2024} first determined the spectral type using the equivalent widths (EWs) of absorption lines and then determined \av\ and \veil\ using the template star with the closest spectral type. Assuming a linear relationship between EWs and spectral types, \citetalias{Itrich+2024} derived the correlation from the template stars and then applied it to the Tr14 data. 
Spectral types were translated into effective temperatures, \teff, using scales from \cite{Luhman2003} for M-type stars, and from \cite{Kenyon&Hartmann1995} for earlier type stars.

\section{Neural Network}
\label{sec:neural_network}
Following the approach introduced in \citetalias{Kang+23b}, we employ a conditional invertible neural network \citep[cINN;][]{Ardizzone2019a, Ardizzone2019b} in this study. cINNs belong to the family of deep learning architectures called Normalising Flows \citep[NFs,][]{Tabak2010, Tabak2013, Dinh2015, Rezende2015}, encompassing methods that employ sequences of invertible transformations of simple known probability distributions to model complex target distributions \citep[for a review see e.g.][]{Kobyzev2021}. 

\subsection{The conditional invertible neural network}
\label{sec:cinn}
The cINN marks a NF variant that is particularly well-tailored towards solving degenerate inverse problems, that is tasks where the underlying physical properties $\mathbf{x}$ of a system are to be recovered from a set of observable quantities $\mathbf{y}$. In nature, this inverse mapping $\mathbf{x} \leftarrow \mathbf{y}$ is often subject to degeneracy due to an inherent loss of information in the corresponding forward process $\mathbf{x} \rightarrow \mathbf{y}$, such that different sets of physical properties may appear similar or even entirely the same in observations, rendering attempts to solve the inverse problem highly challenging.

The cINN approach tackles these challenges by encoding the information loss of the forward mapping in a set of unobservable, latent variables $\mathbf{z}$ that follow a known prior distribution $P(\mathbf{z})$. To do so, the cINN learns a mapping $f$ from the physical parameters $\mathbf{x}$ to the latent variables $\mathbf{z}$ conditioned on the observations $\mathbf{y}$, that is
\begin{equation}
    f\left(\mathbf{x}; \mathbf{c} = \mathbf{y}\right) = \mathbf{z}, 
\end{equation}
such that $\mathbf{z}$ captures the variance of $\mathbf{x}$ that is not explained by $\mathbf{y}$, while maintaining the prescribed shape $P(\mathbf{z})$ for the prior of $\mathbf{z}$. To characterise a new observation $\mathbf{y}'$ at prediction time, the encoded variance can then be queried by sampling the latent space according to $P(\mathbf{z})$, allowing the cINN to estimate the full posterior distribution $p\left(\mathbf{x}|\mathbf{y}'\right)$ by running in reverse following
\begin{equation}
    p\left(\mathbf{x}|\mathbf{y}'\right) \sim g\left(\mathbf{z}; c = \mathbf{y}'\right), \,\, \mathrm{with}\,\, \mathbf{z} \propto P\left(\mathbf{z}\right),
\end{equation}
where $f^{-1}(\cdot, \mathbf{c}) = g(\cdot, \mathbf{c})$ denotes the inverse of the learned forward mapping for fixed condition $\mathbf{c}$. In practise a multivariate normal distribution with zero mean and unit covariance is the simplest choice for the prior $P(\mathbf{z})$, while the dimension of the latent space is by design of the cINN equal to that of the target parameter space, that is $\dim(\mathbf{z}) = \dim(\mathbf{x})$.

\subsection{Noise-Net}
\label{sec:noise-net}

In \citetalias{Kang+23b}, we applied our approach to class III template stars, for which the flux uncertainties are mostly low enough because of their high S/N ratios. On the other hand, the stellar spectra of most Tr14 sources in this paper have lower S/N ratios than the class III templates, so we cannot ignore the flux error. Therefore, we use a type of cINN called Noise-Net~\citep{Kang+23a} designed to consider the errors in the observable.

The Noise-Net, first introduced in \cite{Kang+23a}, is a version of the cINN that estimates posterior distributions $p(\mathbf{x}\,|\,\mathbf{y}, \boldsymbol{\sigma}$) that account for the error (i.e. random noise, $\boldsymbol{\sigma}$) in observations by adding them as an additional condition to the network and learning the influence of the noise during the training process.
\cite{Kang+23a} showed that the Noise-Net reflects the observational error well in the predicted posterior distribution while maintaining a good overall performance. By comparing the Noise-Net with the regular cINN that treats observational error via post-processing, \cite{Kang+23a} showed that Noise-Net achieves higher accuracy than the regular cINN for observations with non-negligible errors. On this account, we adopt the Noise-Net to build the cINN for our analysis in this paper.

There are two general differences between the Noise-Net and the regular cINN. Firstly, in the Noise-Net, both observations $\mathbf{y} = \{y_1, \ldots, y_M\}$ and corresponding errors $\boldsymbol{\sigma} = \{\sigma_1, \ldots, \sigma_M\}$ are used as the condition ($\mathbf{c}$) of the cINN. 
We define the error ($\sigma$) as the dimensionless relative error, that is the one-sigma Gaussian error divided by the observation. In this paper, as we use stellar spectra with 3433 spectral bins, $\sigma_i$ is obtained by dividing the flux error by the flux for each spectral bin. Accordingly, the dimension of the condition ($\mathbf{c}$) of the Noise-Net is twice that of the regular cINN.
Secondly, two new steps are added in the training process for the Noise-Net to account for the uncertainties. The first step is to randomly sample errors for the training models at each training epoch. In order for the Noise-Net to learn a wide range of error values, we sample the errors in a logarithmic scale from the uniform distribution,
\begin{equation} 
\label{eq:sample_sigma}
    p(\mathrm{log}\:\sigma_{i}) = U(a, b),
\end{equation}
where the upper and lower bounds $a$ and $b$ are fixed during the training. 
Considering the relative flux error of the Tr14 spectra, we choose a lower bound of $10^{-5}$ and an upper bound of $10^{-0.5}$.

The next step is to perturb the flux by adding random Gaussian noise based on the error sampled in the first step. The true flux $\mathbf{y^*}$ from the original training data is perturbed following
\begin{equation} 
\label{eq:perturb_lum}
  y_{i}' = y^{*}_{i} \: (1 + r_{i}), \mathrm{where} \; r_{i} \in N(0, \sigma^{2}_{i}).
\end{equation}
We clip the perturbed fluxes to the minimum flux value of the given spectrum to avoid generating negative fluxes. Finally, the Noise-Net is trained to learn the forward process, that is the mapping from the target parameter values ($\mathbf{x^*}$) to the latent variables ($\mathbf{z}$) using the combination of the perturbed spectrum and corresponding randomly sampled errors as the conditioning input for the cINN, $\mathbf{c} \:= [\mathbf{y'}, \boldsymbol{\sigma}]$.
The two random sampling processes in the training procedure ensure that the Noise-Net learns about different errors and flux values in every training epoch, which has a similar effect as extending the overall size of the training data.

In the first step, we can sample the errors of each spectral bin ($\sigma_i$) in various ways, such as sampling all bins independently or having the errors following a correlation function. In this paper, we decide to use one single value for all spectral bins per one model (i.e. $\sigma_i = \sigma$). The relative error in the real data depends on the wavelength as shown in Fig.~\ref{fig:flux_err_distr} but modelling the correlation function of $\sigma_{\lambda}$ is not easy. By sampling errors at all spectral bins independently, the network can in principle learn about the widest cases, but in our experiments, we found that training fails with this setup because the observable space becomes too complex. Therefore, we decided to sample one value and use the same relative error for all spectral bins per observation. As we perturb the observation (flux) at each spectral bin in the second step, the network learns about different perturbations at different wavelengths even though we use the same relative error. We plan to improve the error sampling and training methodology in the follow-up study to consider the correlations between errors.

When applying the trained Noise-Net to new observations, we always need to provide both the flux ($\mathbf{y}$) and corresponding relative error ($\boldsymbol{\sigma}$) as an input to the Noise-Net. The Noise-Net then returns the error-considered posterior distribution of the parameters $p(\mathbf{x}\,|\,\mathbf{y}, \boldsymbol{\sigma}$). We can either feed errors differently at each spectral bin or feed the representative error value. In this paper, when we apply our network to Tr14 stars, we use the median relative error along the wavelength ($\sigma_i = \sigma_{\mathrm{med}}$) per observation instead of using the error at each spectral bin. Using synthetic test models and Tr14 stars in the \Clean\ samples, we tested and confirmed that the change in the results is negligible whether we use the median relative error or different errors at each bin.

\subsection{Network setup}
\label{sec:network_setup}

\subsubsection{Network construction}

As in \citetalias{Kang+23b}, we employ so-called (conditional) affine coupling blocks \citep{Dinh2016} in order to build an invertible network architecture for the cINN. Given the halves $\mathbf{u}_1$ and $\mathbf{u}_2$ of the block input vector $\mathbf{u}$, each of these blocks performs two complementary affine transformations
\begin{equation}
    \label{eq:acb_forward}
    \begin{split}
        \mathbf{v}_1 &= \mathbf{u}_1 \odot \exp\left(s_2(\mathbf{u}_2, \mathbf{c})\right) \oplus t_2(\mathbf{u}_2, \mathbf{c}) \\
        \mathbf{v}_2 &= \mathbf{u}_2 \odot \exp\left(s_1(\mathbf{v}_1, \mathbf{c})\right) \oplus t_1(\mathbf{v}_1, \mathbf{c}),
    \end{split}
\end{equation}
which are easily inverted given the halves $\mathbf{v}_1$, $\mathbf{v}_2$ of the output vector $\mathbf{v}$ following
\begin{equation}
    \label{eq:acb_backward}
    \begin{split}
        \mathbf{u}_2 &= \left(\mathbf{v}_2 \ominus t_1(\mathbf{v}_1, \mathbf{c})\right) \odot \exp\left(-s_1(\mathbf{v}_1, \mathbf{c})\right) \\
        \mathbf{u}_1 &= \left(\mathbf{v}_1 \ominus t_2(\mathbf{u}_2, \mathbf{c})\right) \odot \exp\left(-s_2(\mathbf{u}_2, \mathbf{c})\right).
    \end{split}
\end{equation}
Here $s_i$ and $t_i$ ($i \in \{1, 2\}$) denote arbitrarily complex transformations that do not need to be invertible themselves (being only evaluated in the forward direction in both Eqs.~\ref{eq:acb_forward} and \ref{eq:acb_backward}) and can also be learned directly by the cINN itself when represented by small sub-networks \citep{Ardizzone2019a, Ardizzone2019b}. 

In this paper, we construct a cINN consisting of 8 conditional affine coupling blocks in the GLOW \citep[Generative Flow;][]{Kingma2018} configuration, where 
a single subnetwork is adopted as the internal transformation of each affine coupling block. For each subnetwork, we employ a simple fully connected architecture with 3 layers and a width of 512, using the rectified linear units (ReLU, $\mathrm{ReLU}(x) = \max(0, x)$) as the activation functions. 
After each affine coupling block, we add an invertible random permutation layer to mix the information streams ($\mathrm{u}_1$ and $\mathrm{u}_2$). The permutation layer is a random orthogonal matrix and is fixed during the training~\citep{Ardizzone2019a, Ardizzone2019b}.

With the flux and corresponding errors entering as a condition and simply being concatenated to the input of the subnetworks $s_i$ and $t_i$ in each affine coupling layer, the cINN architecture has the additional advantage that a) the dimension of the input $\mathbf{c}$ can become arbitrarily large and b) a conditioning network $h$ can be introduced (trained together with the cINN itself), which transforms the input condition into a learned, more informative representation $\tilde{\mathbf{c}} = h(\mathbf{c})$ for the cINN \citep{Ardizzone2019b}. 
For the conditioning network $h$, we also adopt a simple fully connected feed-forward network with four layers and a width of 256 that extracts $256$ features in the final layer.

After constructing the network based on this setup, we train the network by minimising the maximum log-likelihood loss as described in \cite{Ardizzone2019b}, using the Adam \citep{Kingma2014} optimiser for stochastic gradient descent with a step-wise learning rate adjustment.

\subsubsection{Data pre-processing}
Following the same approach introduced in \citetalias{Kang+23b}, we transform both parameters $\mathbf{x} = \{x_1, \ldots, x_N\}$ and observations $\mathbf{y} = \{y_1, \ldots, y_M\}$ by using linear transformations, prior to the training, to ensure the distributions of parameters and observables have zero mean and unit standard deviation. Each target property $x_i$ and input observable $y_i$ is rescaled following,
\begin{equation}
    \begin{split}
        \hat{x}_i &= \frac{x_i - \mu_{x_i}}{s_{x_i}}, \\
        \hat{y}_i &= \frac{y_i - \mu_{y_i}}{s_{y_i}},
    \end{split}
\label{eq:linear_transformation}
\end{equation}
where $\mu_{x_i}$, $\mu_{y_i}$ and $s_{x_i}$, $s_{y_i}$, denote the means and standard deviations of the respective parameter/observable across the training data. 

In the case of the relative flux errors $\boldsymbol{\sigma} = \{\sigma_1, \ldots, \sigma_M\}$, we first convert each component $\sigma_i$ into a logarithmic scale because they are sampled from a wide distribution in the log space during the training. Then we rescale them using a similar form of linear transformation as Eq.~\ref{eq:linear_transformation}.
Because the relative flux errors are randomly sampled from $p(\rm{log} \sigma)$ during the training (Eq.~\ref{eq:sample_sigma}), we use the mean and the standard deviation of Eq.~\ref{eq:sample_sigma} for $\mu_{\sigma_i}$ and $s_{\sigma_i}$.
The transformation coefficients ($\mu_{x_i}$, $\mu_{y_i}$, $\mu_{\sigma_i}$, and $s_{x_i}$, $s_{y_i}$, $s_{\sigma_i}$) determined from the training data are applied in the same way to new query data.


\section{Training data}
\label{sec:training_data}

As in \citetalias{Kang+23b}, we train the cINN based on the Phoenix stellar atmosphere model~\citep{Allard+2011, Allard2012} with solar abundances, which covers a wide range of temperature and gravity appropriate for pre-main-sequence stars and provides a sufficient spectral resolution. We generate training data by using the same Phoenix libraries and interpolation pipeline used in \citetalias{Kang+23b}. Each Phoenix library is spaced uniformly with a 100~K interval for \teff\ and 0.5 for $\log(g/\mathrm{cms}^{-2})$. For a given temperature and gravity values, we first interpolate linearly in \logg\ for the two nearest temperatures for the given \teff\ and then interpolate in the temperature linearly between the two resulting spectra. Next, we add veiling and redden the spectra, adopting the same approaches used in \citetalias{Itrich+2024}, which is described in Sect.~\ref{sec:template fitting}. In our simple veiling model, the flux excess due to the accretion or disk emission is constant along the wavelength and determined by the flux at 7500~\AA\ and a veiling factor, \veil\ (Eq.~\ref{eq:veiling}). Lastly, the spectrum is reddened with a given visual extinction value (\av) following the extinction law of \cite{Cardelli+1989}, adopting the $R_{\rm{V}} = 4.4$ for Tr14~\citep{Hur+2012}.

In \citetalias{Kang+23b}, we tested three spectral libraries, namely Settl (i.e. BT-Settl CIFIST), NextGen and Dusty, for training our cINN and found that a cINN trained on the Settl library provided the best overall performance and widest parameter coverage. However, we also found that a cINN based on the Dusty library performs very well within the more limited temperature range ($T_\mathrm{eff} < 4000$ K). Based on this, we adopt a hybrid approach for this work, combining Settl and Dusty libraries for our training data set, instead of using a single spectral library.
We first sample a large list of in total $131,072$ different combinations of $\log(g/\mathrm{cms}^{-2})$ and $\log(T_\mathrm{eff}/\mathrm{K})$, following \citetalias{Kang+23b}, with uniform random sampling in log space: $\log(g)$ from $2.5$ to $5$ and $\log(T_\mathrm{eff})$ from $3.415$ to $3.845$ ($2600$--$7000$ K). Afterwards, we synthesise the corresponding spectra using the interpolation scheme. For $\log(g)$–$\log(T_\mathrm{eff})$ combinations where Dusty models exist ($3 \leq \log(g) \leq 5$ and $2600 \leq T_\mathrm{eff} \leq 4000$ K), we randomly select between Dusty and Settl models, ensuring equal representation. Outside this range, only Settl models are used.
To keep track of the library origin we add a flag to each training instance indicating the model used to synthesise a given spectrum, where a flag of 0 indicates a Settl origin and 1 denotes Dusty-based spectra. This flag is also treated as a target parameter, allowing the cINN to infer the most suitable model for a given query spectrum. We split this database and use 80\%\ of it as a training set and the rest as a test set.

Unlike effective temperature and surface gravity, the extinction (\av) and the veiling (\veil) are sampled and applied to the synthetic spectra during the training. As \av\ and \veil\ values differ in every training epoch, the cINN can learn more diverse cases with a limited number of models. We sample \veil\ from 0 to 2 and sample \av\ from 0 to 10~mag. 
When we randomly sample the parameters within the given ranges, the cINN tends to perform poorly at the boundaries of the training range. To improve the cINN performance at the boundary values without giving unphysical values like negative extinction, we assign boundary values to a certain fraction of the training data while randomly sampling the rest.
For veiling, we expect some stars in Tr14 may not have veiling ($r_{\rm{veil}} = 0$) and most of their veiling values to remain below 2, based on \citetalias{Itrich+2024}. Thus, we set \veil\ of 0 for 30\%\ of the training data and sample the rest uniformly from 0 to 2. For extinction, \av\ is set to 0 for 10\%\ of the training data. These assignments are randomized at each training epoch.

For the synthesised spectra, we initially cover a wavelength range from $4750\,\AA$ to $9350\,\AA$, matching the spectral resolution of MUSE, that is we cover the wavelength interval with a total of $3681$ spectral bins with a width of $1.25\,\AA$. 
Afterwards, we mask a few wavelength intervals, which correspond to prominent emission line features that can occur in the real observed data from Tr14. We have to exclude these spectral bins because neither the Settl nor Dusty spectral libraries account for emission lines in their synthetic spectra. Consequently, the cINN cannot properly fit the data in these wavelength intervals when applied to the real observations as it has never seen examples with emission lines during training. Properly modelling the emission lines for the training data is a complex issue and at this stage out of the scope of this study. The omitted wavelength ranges are the following: 4853--4868, 4955--4963, 5002--5011.5, 5575--5582, 5872--5881, 6297--6303, 6544--6552, 6553--6587.5, 6673--6680, 6710--6722, 6729--6735, 7064--7067, 7130--7140, 7279--7282.5, 7316--7323, 7327--7331, 7745.5--7753, 7768--7780, 8441--8453, 8495--8504, 8536--8552, 8597--8605, 8659--8670, 8747--8758, 8858--8867, 9009--9016, 9064--9073 and 9222--9235~\AA\ (see grey areas in Fig.~\ref{fig:flux_err_distr}).
Before we feed the spectrum into the cINN, we normalise the spectrum by the sum of flux at all spectral bins, excluding the masked bins listed above.

Once trained, our network can sample posterior estimates very efficiently. When tested with an NVIDIA H100 graphics card, the network generates a posterior distribution with 4096 posterior samples per observation for 100 observations in 0.4 seconds (249 observations/s). When tested with an M2 pro CPU, the speed is about 13 observation/s.


\section{Validation}
\label{sec:validation}
\subsection{Performance test on synthetic models}
\label{sec:synthetic}

We first investigate the performance of our trained network using synthetic models. The accuracy of the Noise-Net varies according to the amount of the error of measured observables. \cite{Kang+23a} showed that the average accuracy and precision of the Noise-Net gradually deteriorated as a function of error. In this section, we examine how large the intrinsic estimation error is as a function of the relative flux error by using the test set of our database (i.e. models not used in the training).

We use the entire 26,214 models in the test set for this experiment. As the synthetic model spectra are by definition pure without any flux error, we produce mock flux errors under a simple assumption. Although the flux error in the real spectra varies with wavelength (see Fig.~\ref{fig:flux_err_distr}), we simply use a fixed $\sigma$ value at all wavelengths ($\sigma_{\rm{i}}=\sigma_{\rm{fix}}$). This means the ratio of flux error to flux is fixed, but not the flux error in the physical unit.
Considering the training range of $\sigma$ of our cINN ($10^{-5}$ -- $10^{-0.5}$) and the range of $\sigma_{\mathrm{med}}$ of Tr14 stars in Fig.~\ref{fig:flux_err_distr}, we select eleven $\sigma_{\rm{fix}}$ values from $10^{-3}$ to $10^{-0.5}$ in 0.25~dex intervals on a logarithmic scale. We sample 4096 posterior estimates to obtain a posterior distribution for each test model at each $\sigma_{\rm{fix}}$ value. The maximum a posteriori (MAP) point estimate of each parameter is determined by performing a Gaussian kernel density estimation on the corresponding marginalised 1D posterior distribution (see \citealt{Kang+22} for the detailed method). MAP estimates are used as representative estimates in this paper in many cases.

\begin{figure*}
    \includegraphics[width=2\columnwidth]{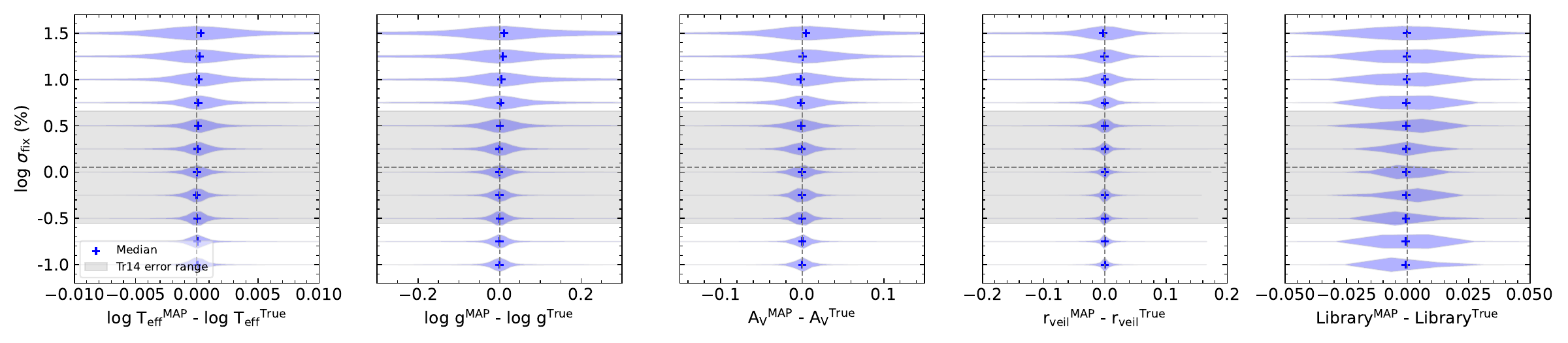}
    \caption{Histograms of the MAP accuracy for the five target parameters measured using 26241 synthetic test models at 11 different relative flux errors ($\sigma$). The blue plus symbol indicates the median value of each histogram. The grey horizontal dashed line and shaded area represent the median (1.12\%) and standard deviation (0.77~dex) of the median relative flux error of Tr14 samples in the \Clean\ group (see Fig~\ref{fig:flux_err_distr}).
     } \label{fig:violin} 
\end{figure*}

\begin{figure*}
	\includegraphics[width=2\columnwidth]{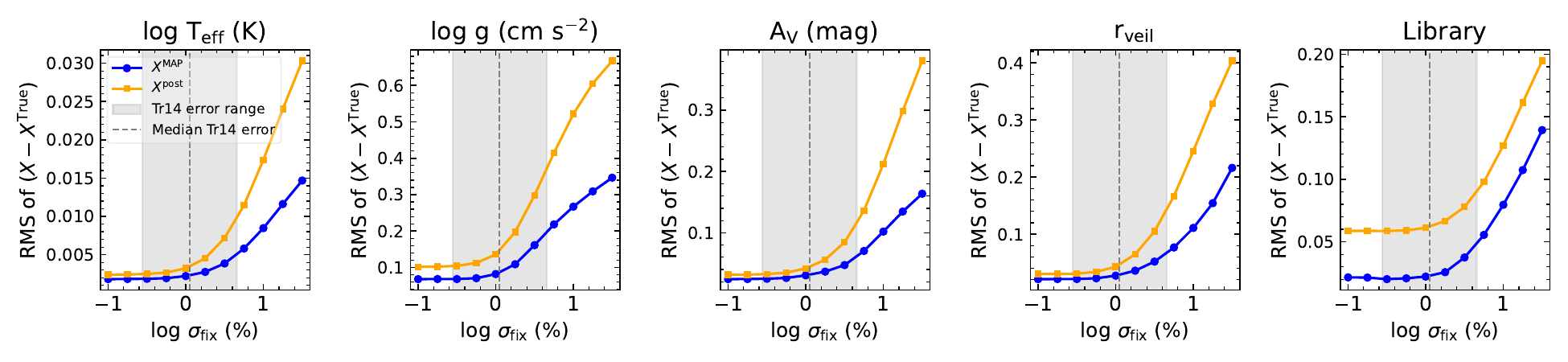}
    \caption{The RMSE of the MAP estimates (blue) and the entire posterior estimates (orange) for synthetic test models at 11 different median relative flux errors. The grey lines and shaded areas are the same as in Fig.~\ref{fig:violin}.
     } \label{fig:rmse_curve} 
\end{figure*}

We investigate how much the MAP estimates differ from the ground truth of the test models. Fig.~\ref{fig:violin} shows the distribution of the MAP accuracy, the error between the MAP estimate and the ground truth, of the cINN for our five target parameters for 11 different $\sigma_{\rm{fix}}$ values. The distribution broadens with increasing $\sigma_{\rm{fix}}$, which is a common trend for all five parameters. This is a typical characteristic of a Noise-Net as shown in the previous study~\citep{Kang+23a}, where the overall accuracy deteriorates with increasing observational error. However, although the distribution widens, the overall error value is very small even at the highest $\sigma_{\rm{fix}}$ value. Moreover, the median of the MAP accuracy, denoted by the plus symbol, reveals that the cINN is on average very accurate.

We calculate the root mean squared error (RMSE) of the posterior estimate with respect to the ground truth in order to quantitatively measure the expected accuracy of the cINN as a function of $\sigma_{\rm{fix}}$. The blue curve in Fig.~\ref{fig:rmse_curve} is the RMSE of the MAP estimates, while the orange curve shows the RMSE of all posterior samples in the posterior distributions, which is generally larger than the RMSE of the MAP estimates. In both curves, the RMSE increases slowly when $\sigma_{\rm{fix}}$ is smaller than about 1\%, but after 1\%, the RMSE increases more rapidly than before. This trend is common for all five target parameters.

Additionally, we split the test models into three groups, M type (2600 -- 4000~K), K type (4000 -- 5250~K), and G type (5250 -- 6000~K), and examine how the RMSE curves vary with \teff\ of the models. Figure~\ref{fig:rmse_curve_spt} shows the RMSE curves for the three groups along with the one for the entire sample (the same as the blue curve in Fig.~\ref{fig:rmse_curve}). For the four main parameters (\teff, \logg, \av, and  \veil), the RMSEs for the K/G-type models are significantly larger than that of the M-type models at the same $\sigma_{\rm{fix}}$. The RMSE for the entire test model is closer to that of the K/G-type models but is smaller than the RMSE of the G-type models when $\sigma_{\rm{fix}}$ is larger than 10\%. In particular, the RMSE of \logg\ for the G-type models is more than 0.2~dex larger than that of the later types even when $\sigma_{\rm{fix}}$ is small. On the other hand, the M-type models show a much smaller RMSE than the earlier types, especially when $\sigma_{\rm{fix}}$ is large.
These results imply that it is more difficult for cINN to measure parameters accurately for hotter stars, especially when the relative flux error is large. 
In the case of the library origin flag, RMSE is only observed for the M-type models, because only the models with \teff\ below 4000~K can have a choice of either Dusty or Settl.

When applying the cINN to real observations, it is important to interpret the obtained posterior estimates considering the intrinsic error of the cINN at the given relative flux error based on the result of this experiment. On this account, we indicate the median relative flux error of the Tr14 stars in the \Clean\ group in Figs~\ref{fig:violin} and \ref{fig:rmse_curve} (grey dashed lines). The grey shade denotes the interval of $\pm$1 standard deviation. As the $\sigma_{\rm{med}}$ of the Tr14 spectra is widely distributed (see Fig.~\ref{fig:flux_err_distr}), the range of the corresponding RMSE is also wide. We choose three $\sigma_{\rm{fix}}$ values (0.1, 1 and 10\%) close to the range of the Tr14 errors and list the RMSE of the MAP estimate at these errors in Table~\ref{table:rmse_map}.
Taking a 1\%\ value close to the median error of the Tr14 observations, the expected error of our cINN is sufficiently small, on the order of $10^{-3}$--$10^{-2}$.
However, at large errors above 10\%, the intrinsic error of the cINN is not negligible, especially in the case of surface gravity, extinction and veiling. 

The half-width of the posterior distribution is usually used as an uncertainty of the MAP estimates, but the predicted posterior distributions of our cINNs in this paper and in \citetalias{Kang+23b} are very narrow and mostly negligible. This is due to a large number of input data points (about 3000 data points) per observation containing sufficient information to predict precise stellar parameters. As we determine the average intrinsic error of the MAP estimate as a function of the relative flux error in this section, we will take this into account when applying our cINN to real observations. In the following sections, the uncertainty of the MAP estimate is determined by combining the half-width of the posterior distribution and the intrinsic parameter estimation error at the corresponding relative flux error. We determine the width of the 68\%\ confidence interval (i.e. $u_{68}$) and use half of it as a half-width of the 1D posterior distribution. Using the median relative flux error along the wavelength ($\sigma_{\mathrm{med}}$), we find the closest one among the 11 $\sigma_{\mathrm{fix}}$ values and use the intrinsic errors of stellar parameters at the corresponding $\sigma_{\mathrm{fix}}$. The final uncertainty of the MAP estimate is calculated as
\begin{equation}
\label{eq:param_error}
    \sigma_{\mathrm{param}} = \sqrt{(0.5\,u_{68} )^{2} + (\sigma_{\mathrm{param, intrs}})^{2}}.
\end{equation}

\renewcommand{\arraystretch}{1.25}
\begin{table}
    \center
    \caption{RMSE of the MAP estimates with respect to the ground truth for 26,214 synthetic test models for three different relative flux errors. 
    \label{table:rmse_map}}
    \resizebox{0.49\textwidth}{!}{
    \begin{tabular}{ c  c  c  c  c c}
        \toprule
        & \multicolumn{5}{c}{RMSE of MAP estimate} \\
        \cmidrule(rl){2-6}        
        $\sigma$ & $T_{\mathrm{eff}}$ (K) & $\log(g/\mathrm{cm\,s}^{-2})$ & $A_{\mathrm{V}}$ (mag) & $r_{\mathrm{veil}}$ & Library  \\
        \midrule
        0.1\% & 25.7 & 0.0682 & 0.0243 & 0.0212 & 0.0216  \\
        1\% & 31.1 & 0.0822 & 0.0305 & 0.0273 & 0.0222  \\
        10\% & 113 & 0.267 & 0.102 & 0.111 & 0.0796  \\
        \bottomrule
    \end{tabular} }
    \tablefoot{The relative flux errors are fixed along the wavelength range.}
\end{table}

\subsection{Test on class III template stars}
\label{sec:templates}

\begin{figure*}
    \centering
    \includegraphics[width=1.99\columnwidth]{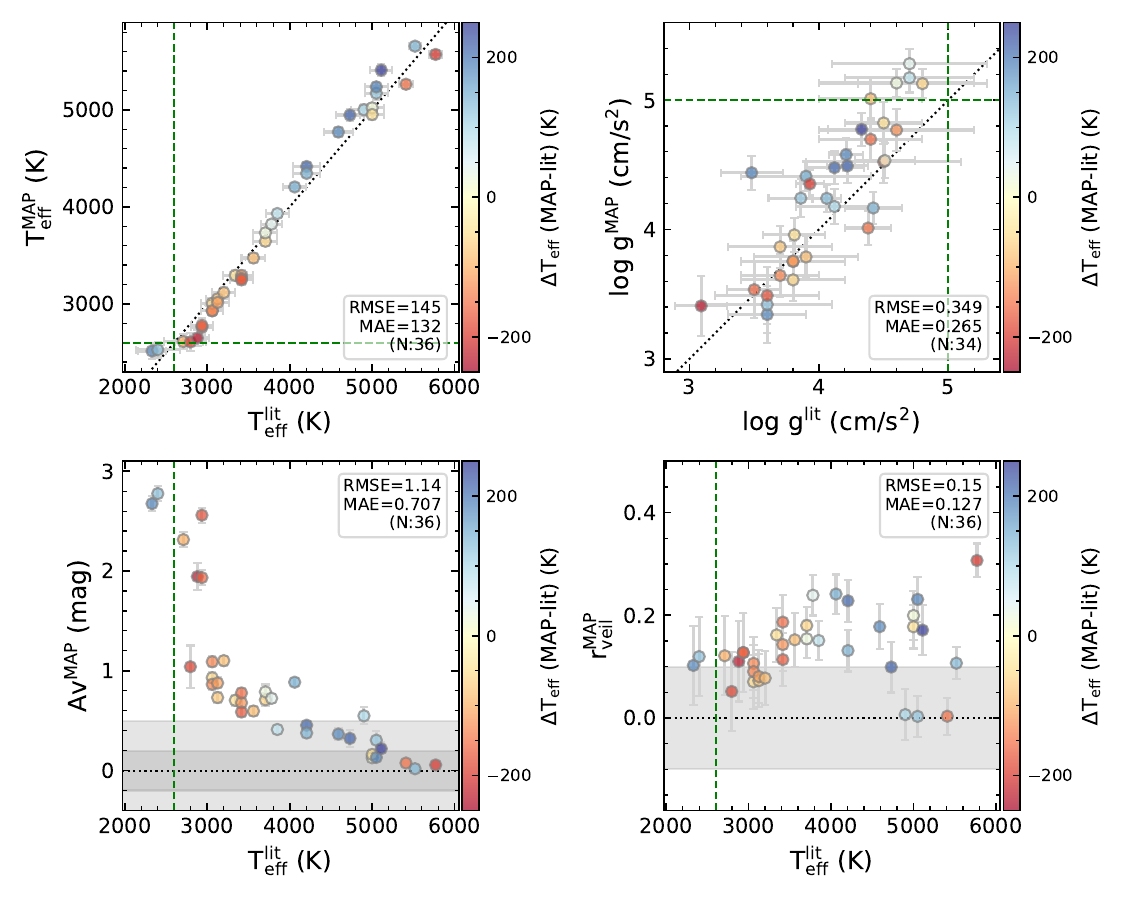}
    \caption{Comparison of MAP estimates with literature values for 36 class III template stars. The colour indicates the temperature difference between MAP estimates and literature values and the dotted green lines indicate the training range of the cINN. The texts in each panel show the root mean square errors and mean absolute errors between the MAP estimates and the literature values. Since these template stars are class III and have negligible extinction (\av\ < 0.5~mag) with an estimation uncertainty of 0.2~mag~\citep{Manara2017}, we regard $A^{\mathrm{lit}}_\mathrm{V}$=0 and $r^{\mathrm{lit}}_\mathrm{veil}$=0. The horizontal grey areas in the lower left panel represent $\pm$0.2~mag and $\pm$0.5~mag, while that in the lower right panel represents $\pm$0.1.
    }
    \label{fig:tpl_4param}
\end{figure*}

We test the performance of the cINN using the 36 class III template stars observed with VLT/X-shooter and analysed by \cite{Manara2013, Stelzer2013, Manara2017}. While we validated the prediction accuracy of cINNs for three parameters (\teff, \logg, and \av) in \citetalias{Kang+23b}, using the same 36 stars, we perform similar examinations again because we introduce a new cINN including veiling as a fourth parameter and taking relative flux errors into account. Unlike the previous networks that used the common spectral range between the MUSE and X-shooter VIS arm ($\sim$5600--9350~\AA), the cINN in this work employs the full MUSE wavelength range (4750--9350~\AA). Therefore, we combine the VIS and UV arm spectra of the template stars and degrade them to match MUSE spectral resolution and wavelength sampling. Since flux error information per spectral bin is not available for these data, we measure the standard deviation of flux at 20~\AA\ intervals within the 6500--7500~\AA\ range. The median of these standard deviations, normalised by the flux at 7000~\AA, is utilised as $\sigma_{\rm{med}}$ of the template spectrum. The average $\sigma_{\rm{med}}$ calculated in this way is 3.81\%\ for M-type stars, 1.7\%\ for K-type stars, and 1.11\%\ for G-type stars, which correspond well to the average S/N ratios reported in \cite{Manara2017}.

Fig.~\ref{fig:tpl_4param} shows comparisons between the MAP estimates measured by the cINN and the stellar parameters obtained in literature~\citep{Manara2013, Stelzer2013, Manara2017}. In Table~\ref{table:tpl_param}, we listed the literature stellar parameters and measured MAP estimates of these stars. We regard the literature extinction and veiling values as zero because previous studies reported that these stars are class III and have negligible extinction, mostly less than 0.2~mag and of up to 0.5~mag, with an extinction uncertainty of about 0.2~mag. Since the veiling of these stars was not measured in those studies, we arbitrarily assume the uncertainty of the literature veiling as 0.1. The estimation uncertainty of the MAP estimate is calculated following Eq.~\ref{eq:param_error}.
Similar to \citetalias{Kang+23b}, the new network accurately estimates effective temperature and surface gravity. For \teff, the mean absolute error (MAE) is about 130~K and the relative error is about 3.5\%, equivalent to a subclass difference of about 0.95. For \logg, the average error is about 0.3~dex, similar to the uncertainty of gravity measurements in the literature.

In the case of extinction, the network performs accurately for stars above 5000~K but overestimates extinction for cooler stars, particularly for stars below 3300~K, where the mean measurement is approximately 1.5~mag higher than the literature value. This trend was observed identically in \citetalias{Kang+23b}, even when we used other Phoenix models to train the cINN (Settl, NextGen, and Dusty). We interpret this as a limitation of the Phoenix models in accurately representing lower-temperature stars, i.e. a large simulation gap, rather than a problem in the predictive ability of the cINN architecture because we confirmed in \citetalias{Kang+23b} that the Phoenix spectra, synthesised based on the predicted parameters, agree well with the observed input spectra. However, the cINN in this study overestimates \av\ by approximately 0.6~mag for stars within the range of 3300--4200~K (M3.5 to K6), whereas the networks in \citetalias{Kang+23b} accurately measured \av\ to within 0.2~mag for these stars. This difference is related to the additional parameter, veiling, in the new network.
For the 36 stars, the mean absolute error of the veiling measurement is about 0.13. Stars below 3300~K exhibit a lower average error, under 0.1, whereas those in the 3300--4200~K range show a higher average error of 0.17. 
For stars between 3300--4200~K, the network appears to overestimate veiling relative to the other stars, which results in larger extinction measurements as a compensatory effect.

We further investigate how the parameter estimates change when the relative flux error ($\sigma_{\rm{med}}$) is increased by a factor of 5 and 10. For \teff, \logg, and \av, the MAP values do not change significantly except for stars above 5000~K. For these stars, the MAP estimates of \teff\ decrease slightly and become closer to the literature values. As a result, the mean error decreases from 131.6~K to 119.5~K and then to 105.3~K as $\sigma_{\rm{med}}$ increases. On the other hand, the most noticeable change is observed in the veiling measurements (Fig.~\ref{fig:tpl_veil}). For stars over 4200~K, as $\sigma_{\rm{med}}$ increases, the veiling estimates approach zero, while the veiling of the other stars does not change. The decrease in veiling for these stars results in a decrease in \teff\ but no noticeable change in \av. 

Through the tests, we validated that the new network performs well for real spectra similar to those in \citetalias{Kang+23b}. For the newly added parameter, \veil, we confirmed that the average error is 0.13, but for stars within the temperature range of 3300--4200~K, the \veil\ error is higher around 0.17, thus overestimating extinction.


\section{Application to Tr14}
\label{sec:result}

\subsection{Stellar parameter estimation}
\label{sec:parameter_estimation}

This section outlines the parameter estimation process for the Tr14 observational data.
We feed the stellar spectrum ($f_{\mathrm{\lambda}}$) normalised by the sum of fluxes at all spectral bins, excluding the masked bins (see Sect.~\ref{sec:training_data}), and a median relative flux error ($\sigma_{\rm{med}}$) as an input to the network and generate 4096 posterior samples per observation to get a posterior distribution. We measure the MAP estimates from the marginalised, one-dimensional posterior distribution for each stellar parameter and use them as a representative estimate. However, if degeneracy remains in the prediction, resulting in a multimodal posterior distribution, measuring the representative values from 1D posterior distributions may be problematic. Because the posterior estimates of each stellar parameter correlate to each other, MAP estimates independently measured from the marginalised 1D posterior distribution do not guarantee that they are from the same mode of the multimodal multivariate distribution. 
By examining the number of peaks (i.e. local maxima) in each 1D posterior distribution, we confirm that we can use the MAP estimate as a reasonable representative value in this work. The detailed analysis of the remaining degeneracy in the posterior distributions is described in Appendix~\ref{sec:posterior_peak}, providing some example posterior distributions.

Next, we investigate whether the MAP estimates fall well within its training range or if the predictions fall in unlearned areas. As mentioned in Sect.~\ref{sec:training_data}, our network is trained on 2600--7000~K for \teff\ and 2.5--5.0 for $\log(g/\mathrm{cms}^{-2})$ in the case of Settl models and 2600--4000~K for \teff\ and 3.0--5.0 for $\log(g/\mathrm{cms}^{-2})$ in the case of Dusty models, and all models are trained on \av\ of 0--10~mag and \veil\ of 0--2. To distinguish the library, we labelled Settl models as 0 and Dusty models as 1. However, the cINN returns continuous values, not a perfect integer, so we consider MAP estimates of the library origin within -0.5 to 1.5 to be valid measurements. In the case of \av\ and \veil, the cINN can return negative values very close to 0 if the ground truth is either 0 or very close to 0. Therefore, we accept predictions that fall within a $\pm$0.1~mag margin at the boundary condition for \av\ and within $\pm$0.05 at the boundary for \veil. 

In the entire sample, 90\%\ of sources have MAP estimates within the training ranges for all five parameters. The other 10\%\ of the sample have at least one extrapolated MAP estimate with most extrapolations happening in \logg. We checked the extrapolated parameters and confirmed that there are no extreme outliers. Most of the extrapolated values are close to the boundary of the training range. For example, MAP estimates of \logg\ larger than 5.0 are mostly distributed near 5.0--5.2 with a maximum of 5.5. For the effective temperature, there are 19 cases of extrapolation in total, with a maximum MAP estimate of 7790~K and a minimum value of 2495~K. However, most of the high \teff\ values are lower than 7500~K and most of the low \teff\ values are higher than 2580~K. 
We found 12 stars having either negative \av\ or negative \veil\ values. With a minimum \av\ of -0.9~mag and a minimum \veil\ of -0.06, however, these do not appear to be extreme outliers. We confirmed that this was caused by the low quality of the spectra rather than cINN performance, so that \citetalias{Itrich+2024} measured zero values for these stars as well.
Accordingly, we conclude not to exclude these extrapolated MAP estimates in our analysis. 

Next, we measure the uncertainty of the stellar parameter estimates (i.e.~estimation error) by combining the half-width of the posterior distribution and the intrinsic estimation error of the cINN at the corresponding observation error (Eq.~\ref{eq:param_error}). By including the intrinsic errors, the uncertainty increased by a factor of 2$\sim$4 compared to the half-width of the posterior distribution on average. The uncertainty of stellar parameters is small overall. 
For the entire sample, the median uncertainties of effective temperature and extinction are 80.6~K and 0.0738~mag, respectively. The amount of uncertainty for the \Clean\ group is even smaller than the other samples. The median uncertainty of each parameter for different sample groups is listed in Table~\ref{table:error_map}. However, it is important to note that while cINN can distinguish very small variations between synthetic models, resulting in small estimation uncertainty listed in Table~\ref{table:error_map}, this uncertainty does not quantitatively reflect the gap between the Phoneix models and real spectra.
The measured stellar parameters and their uncertainties are listed in Table~\ref{table:catalog}.

\renewcommand{\arraystretch}{1.25}
\begin{table}
    \center
    \caption{Median estimation uncertainty of MAP estimates for Tr14 samples in three different sample groups.
    \label{table:error_map}}
    \resizebox{0.49\textwidth}{!}{
    \begin{tabular}{ c  c  c  c  c c}
        \toprule
        & \multicolumn{5}{c}{Median uncertainty of MAP estimates} \\
        \cmidrule(rl){2-6}        
        Sample group & \teff\ (K) & $\log(g/\mathrm{cms}^{-2})$ & \av\ (mag) & \veil\ & Library \\
        \midrule
        All & 80.6 & 0.22 & 0.0738 & 0.0769 & 0.095 \\ 
        Clean & 35.5 & 0.0899 & 0.0368 & 0.03 & 0.0556 \\ 
        Uncertain & 116 & 0.298 & 0.12 & 0.114 & 0.122 \\
        \bottomrule
    \end{tabular} }
    \tablefoot{The estimation error is calculated based on the half-width of the 1D posterior distribution and the intrinsic estimation error of the cINN at the given relative flux error ($\sigma_{\rm{med}}$), following the Eq.~\ref{eq:param_error}. However, this error does not reflect the uncertainty caused by the discrepancy between the Phoenix models and real spectra.
    }
\end{table}

\subsection{Stellar parameter comparison}
\label{sec:param_comparison}

\begin{figure*} 
    \centering
    \includegraphics[width=2\columnwidth]{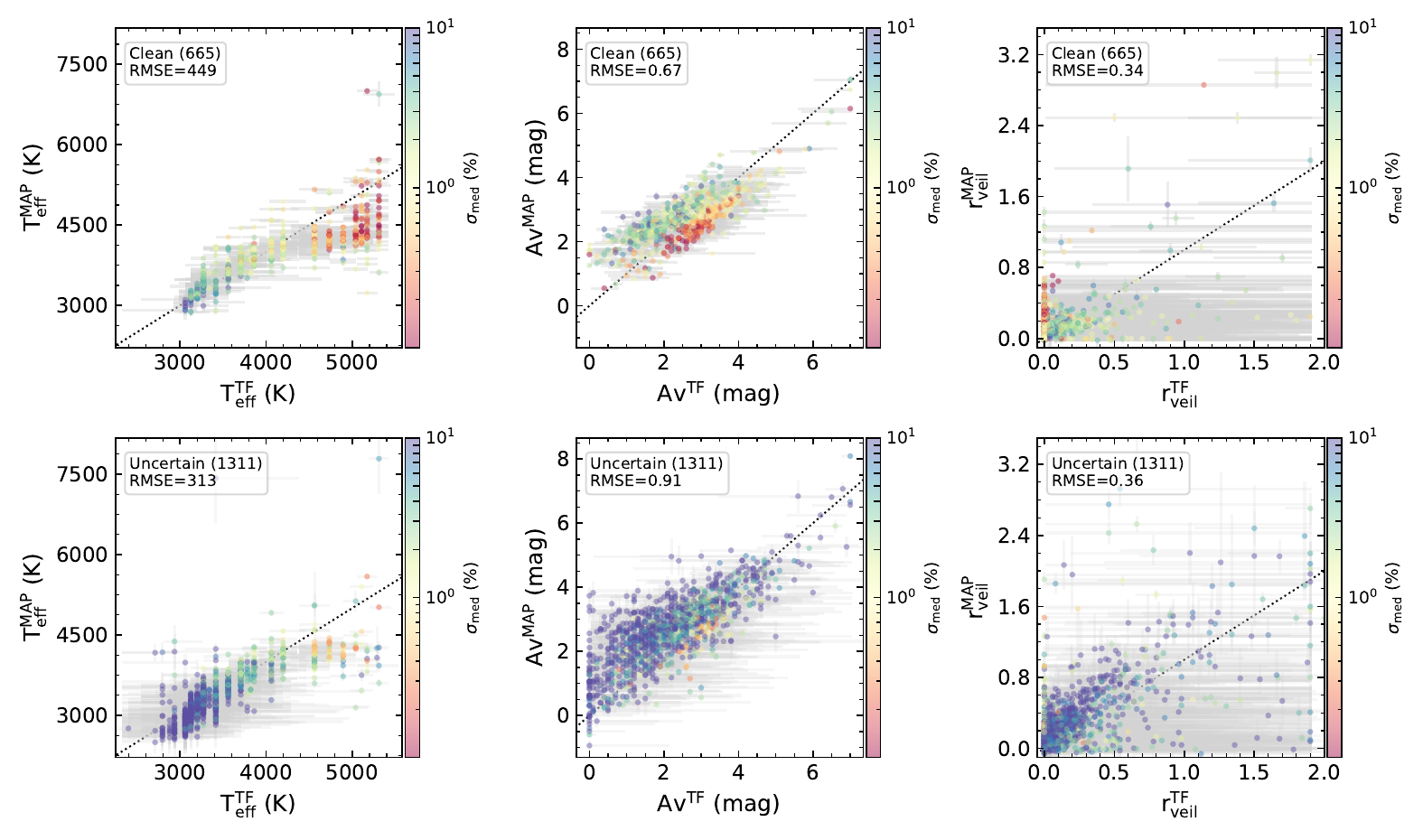}
    \caption{Comparison of the effective temperature (\teff), extinction (\av), and veiling factor (\veil) of stars in Tr14 measured by template fitting (TF parameters; \citetalias{Itrich+2024}) with our corresponding cINN-MAP estimates for the samples in the \Clean\ (top) and \Unc\ (bottom) groups. The colour code denotes the median relative flux error along the wavelength of each star which is used as an additional input to our cINN. 
     } \label{fig:group_1to1} 
\end{figure*}

\begin{figure*} 
    \centering
    \includegraphics[width=2\columnwidth]{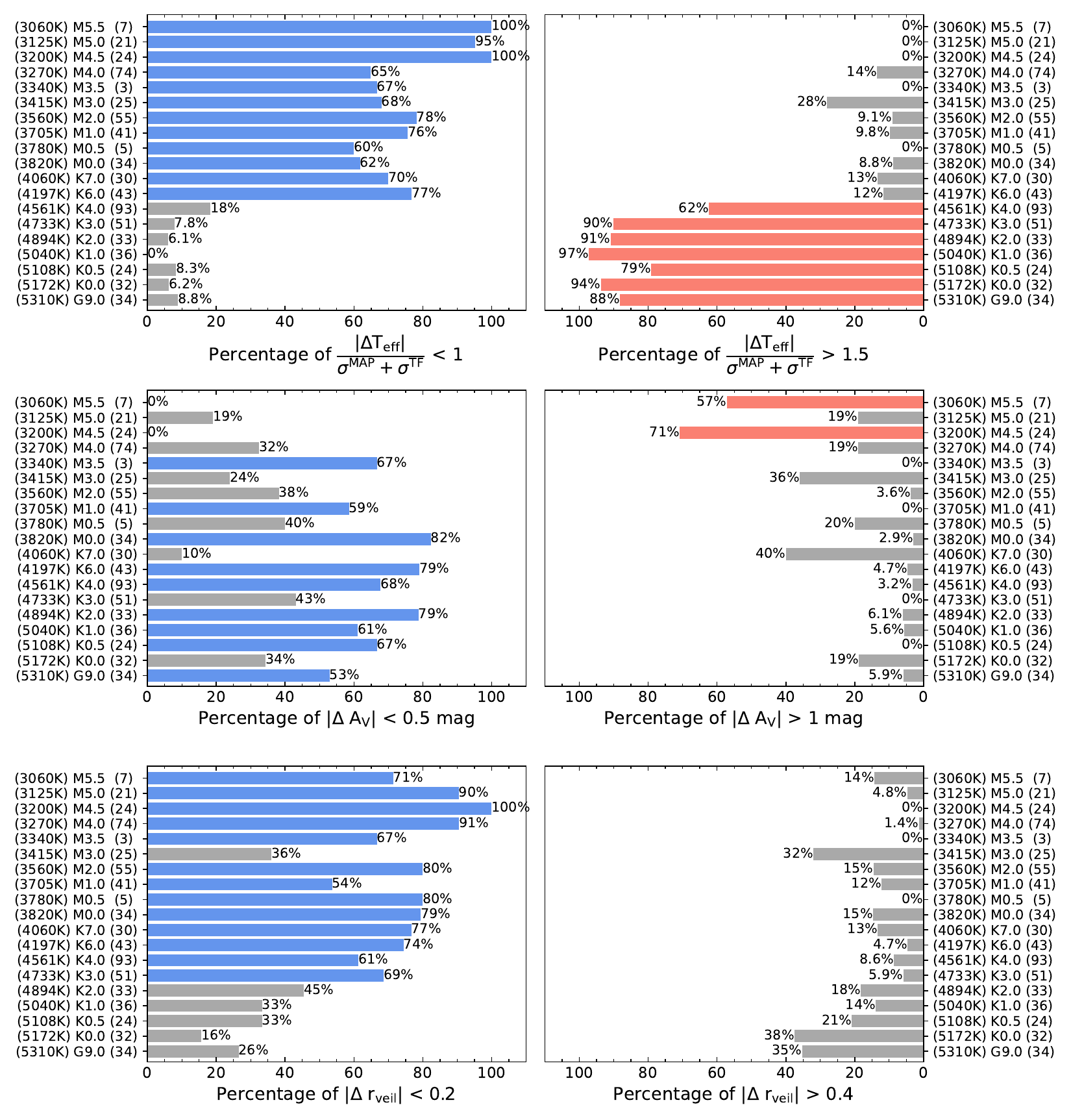}
    \caption{We calculate the fraction of samples where the difference between the template fitting parameter (TF parameter; \citetalias{Itrich+2024}) and the MAP value is either smaller than the given criterion (left panels) or larger than the given criterion (right panels) within each spectral type group based on the \teff\ from TF parameters. We only use the \Clean\ samples in this figure. The y-axis labels present the corresponding spectral type and the number of samples belonging to the spectral type group. We highlight the bar in blue (left panels) or red (right panels) when the fraction is higher than 50\%. 
     } \label{fig:perform_bar_clean} 
\end{figure*}

\begin{figure*} 
    \centering
    \includegraphics[width=2\columnwidth]{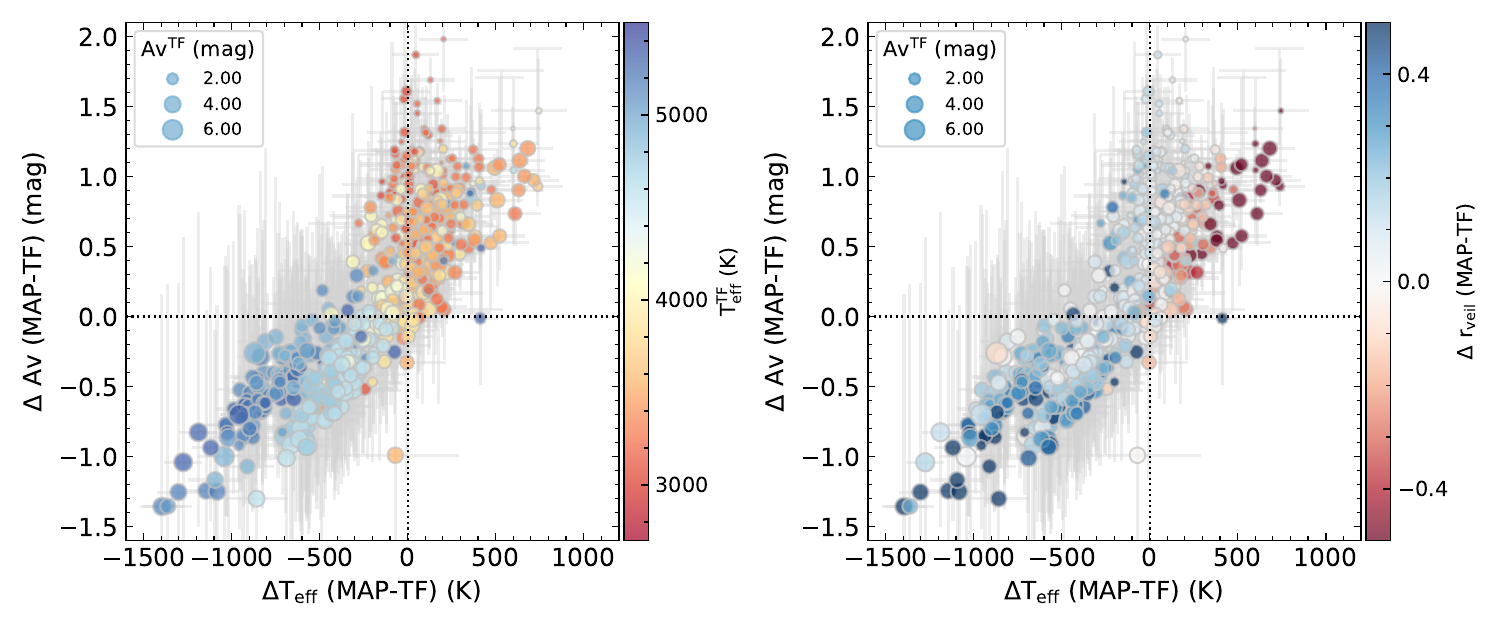}
    \caption{Extinction and temperature differences between cINN-MAP estimates and TF parameters for stars in the \Clean\ group using the different colour codes: \teff\ measured by template fitting (left) and the difference in \veil\ between MAP estimates and TF parameters (right). The size of the symbol indicates the extinction from TF parameters.
     } \label{fig:clean_3d} 
\end{figure*}

In this section, we compare the stellar parameters estimated by our cINN with those measured by template fitting in \citetalias{Itrich+2024} (i.e. TF parameters) from the same MUSE spectra. \citetalias{Itrich+2024} could not measure surface gravity because it usually requires higher spectral resolution than that of MUSE to measure surface gravity. Therefore, we only use \teff, \av, and \veil\ when comparing our MAP estimates with the TF parameters obtained in \citetalias{Itrich+2024}. We discuss the surface gravity of Tr14 stars estimated by our cINN separately in Appendix~\ref{sec:gravity}. Additionally, in this section, we exclude 75 stars classified as G8 type in \citetalias{Itrich+2024} because their \teff\ measured in \citetalias{Itrich+2024} are just a lower limit.

Fig.~\ref{fig:group_1to1} shows the 1-to-1 comparison of the three stellar parameters (\teff, \av, and \veil) for the \Clean\ and \Unc\ groups, where the colour coding denotes the median relative flux error of each star ($\sigma_{\rm{med}}$) used as an input to our cINN. The stellar parameters measured by the two different methods are similar overall, but, for Tr14 stars, the difference between the TF parameters and MAP values is on average larger than that observed for the class III template stars presented in Sect.~\ref{sec:templates} and \citetalias{Kang+23b}.
The RMSE of \teff\ across all stars is 365~K, which is $\sim$2.5 times larger than that we measured for the template stars. However, it is important to note that the TF parameters are not a perfect ground truth and their uncertainty is not negligible, unlike the class III template stars. The average uncertainty of the TF parameters for the \Clean\ samples is about 155~K, 0.48~mag, and 0.42 for \teff, \av, and \veil, respectively. The scatter between MAP estimates and TF parameters for \teff\ is smaller for cooler stars than for hotter stars. Across the \Clean\ and \Unc\ sets combined, the RMSE of \teff\ for cooler stars ($T^{\rm{TF}}_{\rm{eff}} < 4200$~K) is 237~K whereas the RMSE for hotter stars ($T^{\rm{TF}}_{\rm{eff}} \geq 4200$~K) is 659~K.

Fig.~\ref{fig:group_1to1} shows that \teff\ from template fitting ($T^{\rm{TF}}_{\rm{eff}}$) are discretised due to the limited number of template stars, whereas the MAP estimates of \teff\ lie on the interval between different subclasses. Interestingly, \citetalias{Itrich+2024} lacked templates between K6 and K4 types (4197 -- 4561~K) so they could not measure the temperature within this range, whereas the cINN measured temperatures between K6 and K4 types for several stars whose $T^{\rm{TF}}_{\rm{eff}}$ are mostly higher than their $T^{\rm{MAP}}_{\rm{eff}}$. 
\citetalias{Itrich+2024} also lacked templates hotter than G8 type (5430~K). 
There are 33 stars whose MAP-based \teff\ are higher than 5430~K in the \Clean\ samples. 27 of them were classified as G8 type, which are excluded in Fig.~\ref{fig:group_1to1}, 4 of them were classified as G9 type, and the remaining 2 were classified as K0 type by \citetalias{Itrich+2024}. It is highly likely that \citetalias{Itrich+2024} underestimated the temperature of these stars due to the absence of hotter template stars.

In the case of extinction, the RMSE between the cINN estimates and TF parameters is 0.67--0.9~mag depending on the sample groups, but for most samples, the cINN estimates agree well with the TF parameters within 1-$\sigma$ estimation uncertainty.
For the veiling factor, we find that TF parameters and MAP values for several stars fall nicely on the one-to-one line, while some stars lie along the y-axis. We found that \citetalias{Itrich+2024} measured zero veiling ($r_{\rm{veil}}^{\rm{TF}} = 0$) for 84\%\ of stars classified as K- or G-types (i.e. $T^{\rm{TF}}_{\rm{eff}}$ > 4000~K), whereas the cINN measured a non-negligible amount of veiling ($r_{\rm{veil}}^{\rm{MAP}} > 0.2$) for half of them. The RMSE of \veil\ between TF parameters and MAP values is 0.3 for stars cooler than 4000~K and 0.46 for stars hotter than 4000~K. This large \veil\ difference for stars hotter than 4000~K is related to large \teff\ difference in corresponding stars, which will be discussed in Sect.~\ref{sec:discussion_tf}.
It is important to note that both \citetalias{Itrich+2024} and this study utilise a wavelength-independent veiling model (Eq.~\ref{eq:veiling}), which can be problematic, particularly for stars with high veiling. Therefore we consider that veiling values are relatively less reliable for cases where \veil\ > 1 despite the low measurement uncertainty and accurate performance of cINN shown in Sect.~\ref{sec:validation}. This consideration also applies to the veiling values from \citetalias{Itrich+2024}.
The relative flux error represented by $\sigma_{\rm{med}}$ is usually smaller for hotter stars than cold stars. We could not find any relation between $\sigma_{\rm{med}}$ and the difference between TF parameters and MAP values within the similar temperature groups.

We further investigate the difference between MAP estimates and TF parameters for each spectral type group as our samples cover a wide range of spectral types.
In this analysis, we only use samples in the \Clean\ group which have relatively small parameter estimation errors. In Fig.~\ref{fig:perform_bar_clean}, we present the fraction of samples where the difference between the TF parameter and the MAP value is either smaller or larger than the indicated criteria in each spectral type group. 
For \teff, we use the parameter difference normalised by the sum of their 1-$\sigma$ estimation uncertainties, i.e. $|T_{\mathrm{eff}}^{\mathrm{MAP}} - T_{\mathrm{eff}}^{\mathrm{TF}}|/(\sigma_{T}^{\mathrm{MAP}} + \sigma_{T}^{\mathrm{TF}})$. We check if the \teff\ difference normalised by their uncertainty is smaller than 1 or larger than 1.5 to evaluate whether the MAP values are in good or bad agreement with the TF parameters. On the other hand, in the case of \av\ and \veil, we used fixed values for the evaluation, considering the average 1-$\sigma$ error of \av\ and \veil.

In Fig.~\ref{fig:perform_bar_clean}, the spectral types with small temperature differences and those with large temperature differences are clearly separated. Later-type stars up to K6 type (4197~K) mostly have normalised differences below 1, while earlier types from K4 type (4561~K) onwards show more samples with differences above 1.5. 
The difference in extinction is less dependent on the spectral type compared to the effective temperature, except for M5.5 (3060~K) and M4.5 (3200~K) types, where \av\ differences above 1 mag are more frequent. This is a systematic trend already shown in Sect.~\ref{sec:templates} and \citetalias{Kang+23b} that cINNs overestimate \av\ by about 1~mag for cold stars (\teff\ < 3200~K) even though estimates of the other parameters are accurate. 
In the case of \veil, cool stars up to K3 type (4733~K) have smaller differences than earlier type stars, similar to the case of \teff. The fractions of samples with a \veil\ difference larger than 0.4 are higher for earlier type stars above K0 (5172~K), but they are mostly lower than 40\%.
Based on Fig.~\ref{fig:group_1to1} and Fig.~\ref{fig:perform_bar_clean}, the MAP estimates and TF parameters show good agreement in all three stellar parameters for stars within M4 -- K6 types which corresponds to 3270--4197~K. Although it depends on the parameters, cool stars (\teff\ < 4500~K) show generally a good agreement.

To investigate if the degeneracy between stellar parameters affects the difference between MAP values and TF parameters, we examine the correlation between parameter deviations in Fig.~\ref{fig:clean_3d}, only using the \Clean\ samples. The deviation in temperature and the deviation in extinction show a clear positive correlation: for M-types ($T^{\rm{TF}}_{\rm{eff}} < 4000$~K), the cINN predicts higher \teff\ and \av\ than template fitting, while for K/G-type stars ($T^{\rm{TF}}_{\rm{eff}} > 4000$~K), the trend reverses. 
Interestingly, K/G-type stars exhibit a tighter correlation, whereas M-type stars show more scatter with two distinct trends, as seen in the right panel of Fig.~\ref{fig:clean_3d}, where colour indicates \veil\ differences.
For K/G-type stars, cINN tends to predict higher \veil\ than template fitting, forming a clear correlation: $\Delta$\teff\ < 0, $\Delta$\av\ < 0, and $\Delta$\veil\ > 0. Among M-type stars, one group (red circles) follows this trend with $\Delta$\veil\ < 0,  while another group, where $\Delta$\veil\ > 0 but small, clusters near the y-axis, meaning \av\ deviations remain large despite small \veil\ and \teff\ deviations. As noted earlier, this aligns with a systematic trend in cINN seen in this study and \citetalias{Kang+23b}, where extinction is overestimated for \teff\ < 3200~K, even when other parameters are accurately predicted.

\subsection{Hertzsprung–Russell diagram}

\begin{figure*} 
    \centering
    \includegraphics[width=1.7\columnwidth]{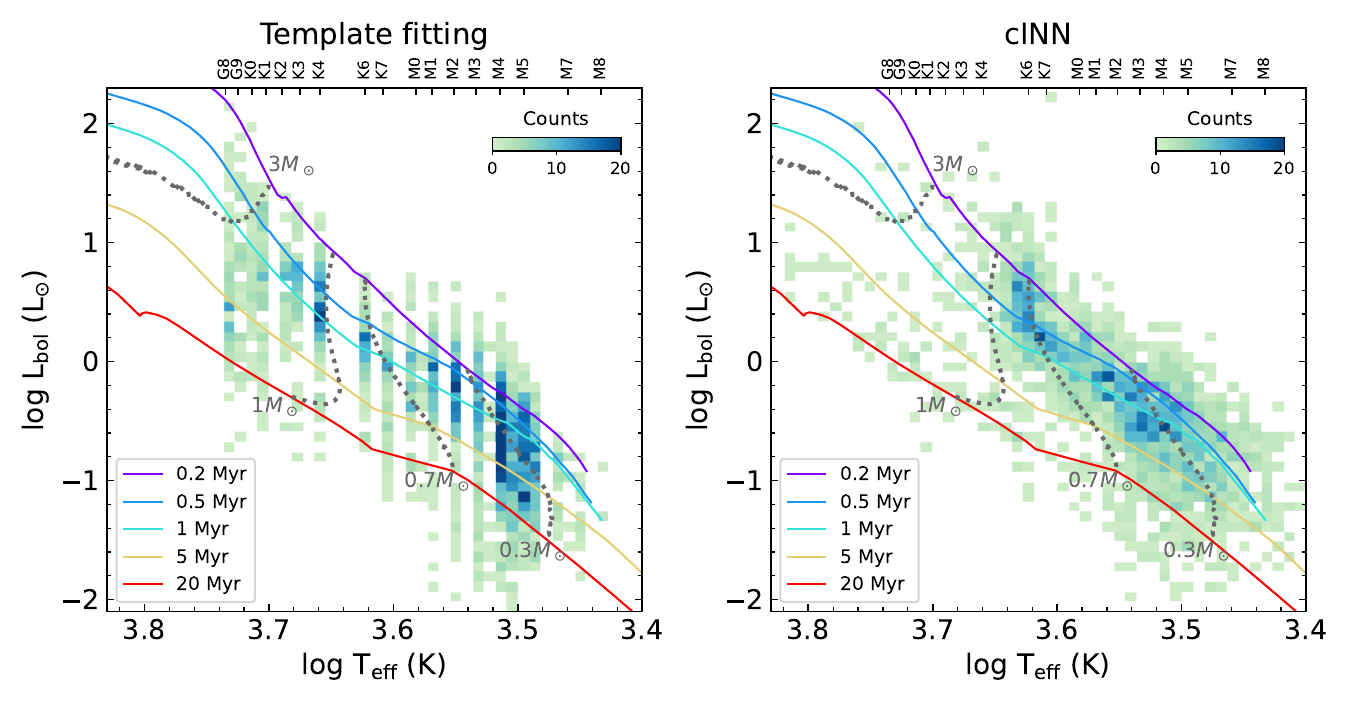}
    \includegraphics[width=1.7\columnwidth]{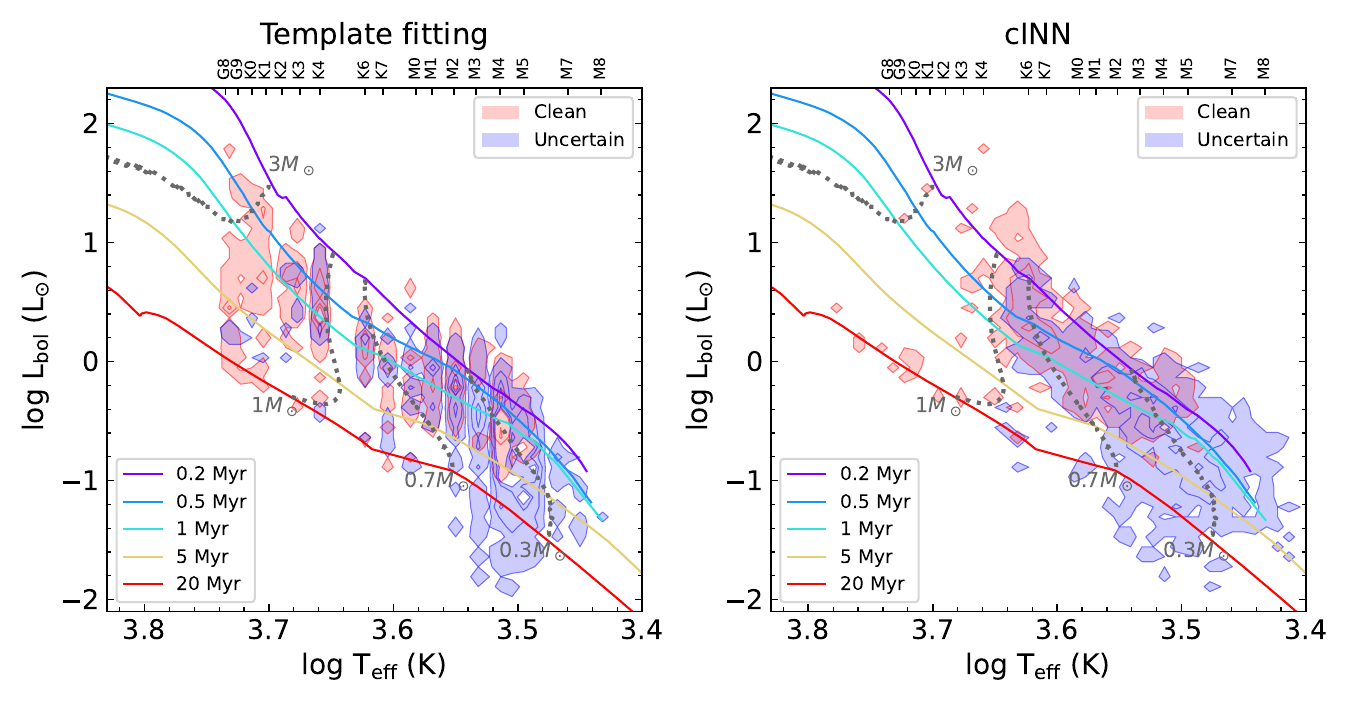}
    \caption{HR diagrams of Tr14 stars based on the stellar parameters from \citetalias{Itrich+2024} (left panels) and the stellar parameters from this work (right panels). In the upper panels, we plot 2-dimensional histograms using all samples, while in the lower panels, we plot contours dividing the samples into the \Clean\ (red) and the \Unc\ (blue) groups. We overplot PARSEC theoretical evolutionary tracks~\citep{Bressan+2012} where solid lines indicate isochrones from 0.2 to 20~Myr and dashed lines indicate isomasses of 0.3, 0.7, 1, and 3~\msol.
     } \label{fig:hr} 
\end{figure*}

We further estimate the bolometric luminosities of individual stars following the same procedure as \citetalias{Itrich+2024} did. 
To measure the bolometric luminosity, we first draw the $J$-band photometry from either the VISTA catalogue~\citep{Preibisch+2014} or the HAWK-I catalogue~\citep{Preibisch+2011a, Preibisch+2011b}. We choose the one with smaller uncertainty if the $J$-band magnitude is available in both catalogues. The $J$-band photometry is the most suitable for deriving bolometric luminosities of young stars. Ongoing accretion onto the star and the presence of a circumstellar disk can cause a near-infrared (NIR) excess of their spectral energy distributions making most of the NIR photometry unreliable for derivation of stellar parameters. At the same time, young stars are heavily extincted which affects short-wavelength photometry. These effects cannot be fully avoided but can be minimised by using the $J$-band filter \citep[e.g.][]{Kenyon&Hartmann1995, Luhman1999}.

The bolometric magnitude is calculated following the equation (\citetalias{Itrich+2024}),
\begin{equation} 
\label{eq:mag_bol}
    M_{\rm{bol}} = J - A_{\rm{J}} - \rm{DM} + (BC_{\rm{V}} + (\mathit{V}- \mathit{K}) - (\mathit{H} - \mathit{K}) - (\mathit{J} - \mathit{H})),
\end{equation}
where colours and correction values are taken from \cite{Kenyon&Hartmann1995}. We calculate the $J$-band extinction ($A_{\rm{J}}$) using the measured visual extinction (\av) and the extinction law from \cite{Cardelli+1989} and apply the distance modulus (DM) of 11.86 mag for Tr14, equivalent to a distance of 2.35~kpc (\citealt{Goppel&Preibisch2022}, \citetalias{Itrich+2024}). The final bolometric luminosity is calculated by subtracting the solar bolometric luminosity of 4.74~\citep{Cox2000} from the bolometric magnitude:
\begin{equation} 
\label{eq:lum_bol}
    \text{log}\: (L_{\rm{bol}}/L_{\odot}) = -0.4 \:(M_{\rm{bol}} - M_{\rm{bol},\odot}).
\end{equation}
The difference in the bolometric luminosity between this work and \citetalias{Itrich+2024} arises only from the difference in \av\ and \teff\ because all the other values, such as the $J$-band photometry, are the same.

In the upper panels of Fig.~\ref{fig:hr}, we present 2-dimensional histograms of all samples on the Hertzsprung–Russell (HR) diagram, where the left panel is based on the stellar parameters from \citetalias{Itrich+2024} and the right panel is based on the stellar parameters from this work. In the lower panels, we present the distribution of each sample group as red (\Clean) and blue (\Unc) contours. PARSEC theoretical isochrones (solid lines) and isomasses~\citep[dashed lines;][]{Bressan+2012} are plotted on each HR diagram.
The HR diagram from this work appears smoother than that from TF parameters due to the non-discretised temperature measurements. The luminosity for each given spectral type exhibits a considerable spread of about 1.5~dex in both studies. The scatter of the TF parameter-based luminosities for K1 -- G8 type stars spans 2~dex, covering isochrones from 0.2 to 20~Myr. 
Since our cINN assigns different \teff\ to these stars, they shift on the HR diagram, forming two distinct groups: one group of stars lying close to the 1~Myr isochrone and the other group of a few stars around the 20~Myr isochrone.  
The luminosity spread is broader for \Unc\ samples and M-type stars than for \Clean\ samples or K/G-type stars for both the MAP-based luminosity and TF parameter-based luminosity. This spread observed in the HR diagrams can be attributed to several factors such as stellar parameter uncertainties, intrinsic age spread, stellar variability, or unresolved multiplicity~\citep{Hartmann2001, Peterson+2008, Baraffe+2009, Costigan+2014, Venuti+2014, Claes+2022}.

One of the distinctive features in the HR diagram based on the MAP values is the high number density around the K5 type. As shown in Fig.~\ref{fig:group_1to1}, many stars classified as K4 -- K0 type by \citetalias{Itrich+2024} are classified as around K5 type by our cINN. This results in the vertical distribution along a temperature of about 4250~K.  
Based on the isomass lines, most stars have stellar mass lower than 1~\msol, while the HR diagram based on the TF parameters shows more massive stars up to 3~\msol. We will discuss the distribution of the mass more in detail in the following section.

\subsection{Stellar age and mass}
\label{sec:age_mass}

\begin{figure*} 
    \centering
    \includegraphics[width=1.9\columnwidth]{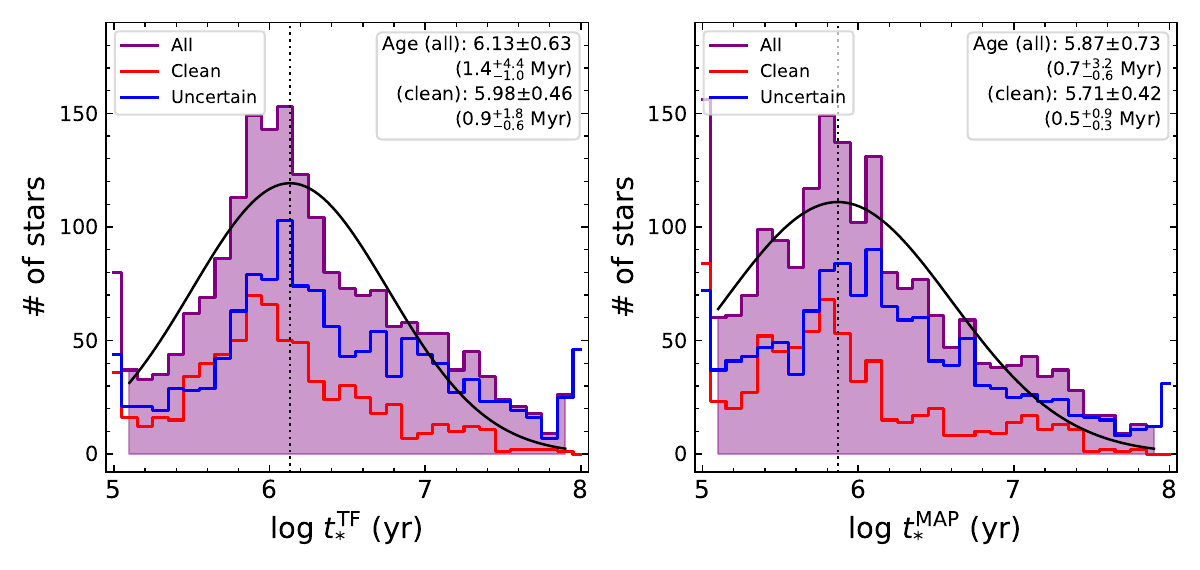}
    \caption{Histograms of stellar age calculated from the TF parameters (left) and stellar age derived from the MAP values (right). The red and blue lines show the distribution of the \Clean\ and \Unc\ sample groups, respectively. We fit the purple shaded area with a log-normal function and plot the best-fit curve as the black solid line. The peak of the best-fit curve, denoted by a vertical black dotted line, is used as an average stellar age.
     } \label{fig:age} 
\end{figure*}

\begin{figure*} 
    \centering
    \includegraphics[width=1.9\columnwidth]{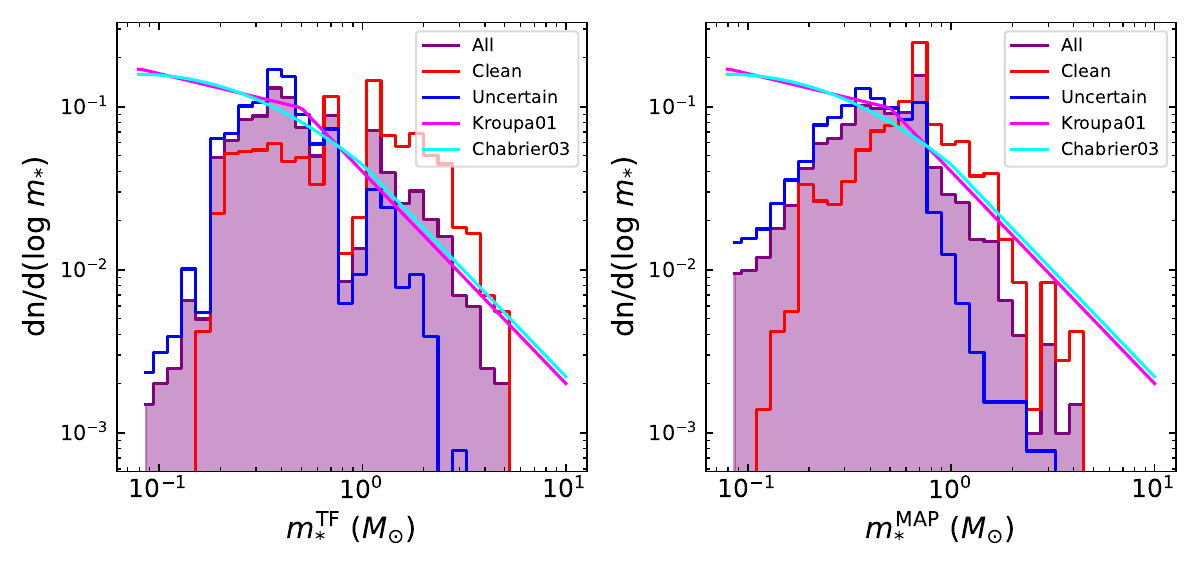}
    \caption{The number density of stars within a given mass bin as a function of stellar mass from the TF parameters (left) and from the MAP values (right). The red and blue lines show the distribution of the \Clean\ and \Unc\ sample groups, while the purple lines and purple-shaded areas indicate the distribution of all samples. We present the initial mass functions from \cite{Kroupa01} and \cite{Chabrier03} with pink and cyan lines, respectively.
     } \label{fig:mass} 
\end{figure*}

By interpolating the PARSEC theoretical evolutionary tracks~\citep{Bressan+2012}, we calculate the ages and masses of individual stars from their effective temperatures and bolometric luminosities. We note that we only use evolutionary tracks from \cite{Bressan+2012} in this paper and that the stellar age and mass can vary depending on the choice of the evolutionary track.
In Fig.~\ref{fig:age}, we compare the distribution of stellar ages from \citetalias{Itrich+2024} with that from our work. The purple lines are based on all samples, while the red and blue lines show the distribution for the \Clean\ and \Unc\ groups, respectively.
The first and the last bins of the histogram are overdense because of the lower and upper limits of the PARSEC theoretical models. We exclude these two bins when calculating the average age of the samples. The purple shading in Fig.~\ref{fig:age} indicates the area we used to calculate the average stellar age for all samples. Following the same methodology as \citetalias{Itrich+2024}, we fit the histogram with a log-normal function and use the peak and standard deviation of the best-fit function (black curves) as the average age and its uncertainty.

Based on the stellar parameters estimated by the cINN, the average stellar age of all samples is 0.7$^{+3.2}_{-0.6}$~Myr. It is 0.26~dex lower than the average age calculated based on the TF parameters (1.4$^{+4.4}_{-1.0}$~Myr). 
However, their average ages are still in good agreement within the 1-$\sigma$ error range because both age values are widely distributed. The individual stellar ages based on the TF parameters and MAP values are different because of different stellar parameters, but the overall distribution and mean value are similar.
The \Clean\ samples tend to be at a younger age in comparison to the \Unc\ samples in both MAP-based ages and TF parameter-based ages.

In the MAP-based HR diagram (Fig.~\ref{fig:hr}), the distribution of stars hotter than $\sim$3600~K (i.e. log \teff\ > 3.56) is divided roughly around the 5~Myr isochrone. When we examine the age distribution for these 874 stars, a break occurs in the age distribution at log $t_{*}$ of around 6.7 ($\sim$5~Myr). About 16\%\ of the 874 stars have log $t_{*}$ > 6.7, showing an age distribution with a peak at log $t_{*}$ $\sim$ 7.2 and an average age of 14.7$^{+11.7}_{-6.5}$~Myr. On the other hand, the average age of the other stars (log $t_{*}$ < 6.7) is 0.5$^{+0.7}_{-0.3}$~Myr, which is similar to the average age of the entire sample, but the standard deviation is smaller. The age break is not noticeable in the age distribution of lower-temperature stars. 
It is uncertain whether this break reflects two stellar populations with an age difference of $\sim$10~Myr or whether the observed separation is a spurious feature because stellar ages are subject to significant uncertainties due to several factors (e.g. parameter uncertainty, evolutionary tracks).

In Fig.~\ref{fig:mass}, we present the distributions of stellar masses obtained from the TF parameters and MAP estimates. The mass of the Tr14 sources ranges from 0.1~\msol\ to 5~\msol. To compare the mass distribution with well-known initial mass functions~\citep[IMFs][]{Kroupa01, Chabrier03}, we calculate the number density of the star at the given mass bin. 
In the TF parameter-based mass distribution, there is a valley around 0.8--1~\msol, which is also exhibited on the HR diagram (Fig.~\ref{fig:hr}). This feature is attributed to the lack of templates between K4 and K6 types. On the other hand, there is no such limitation in the MAP-based masses, but the MAP-based stellar masses show an overdense region around 0.7--0.8~\msol.

Both TF parameter-based and MAP-based stellar mass distributions follow the initial mass functions relatively well for $m_{*} > 0.3$~\msol. On the other hand, the number density decreases in the lower mass range $m_{*} < 0.3$~\msol\ unlike the IMFs. This is because of the selection bias rather than physical reasons. As it is difficult to detect cold and faint stars with enough S/N ratio, we suspect that most of these stars are not observed or are excluded from the catalogue of \citetalias{Itrich+2024}. Even if they are observed with enough S/N ratio, our classification methods cannot classify these stars properly because the lower limit of the template stars used in \citetalias{Itrich+2024} is M9 type (2400~K) and the lower training limit of our cINN is 2600~K. Therefore, very low-mass stars below 0.2-0.3~\msol\ are out of our scope and we lack the samples in this mass range. We do not focus much on the comparison of the mass distributions with the two IMFs because the stellar mass range of Tr14 samples in this study is too narrow to cover the overall mass range of the IMF.

\subsection{Which method fits better}
\label{sec:fit}

\begin{figure}
    \centering
    \includegraphics[width=1\columnwidth]{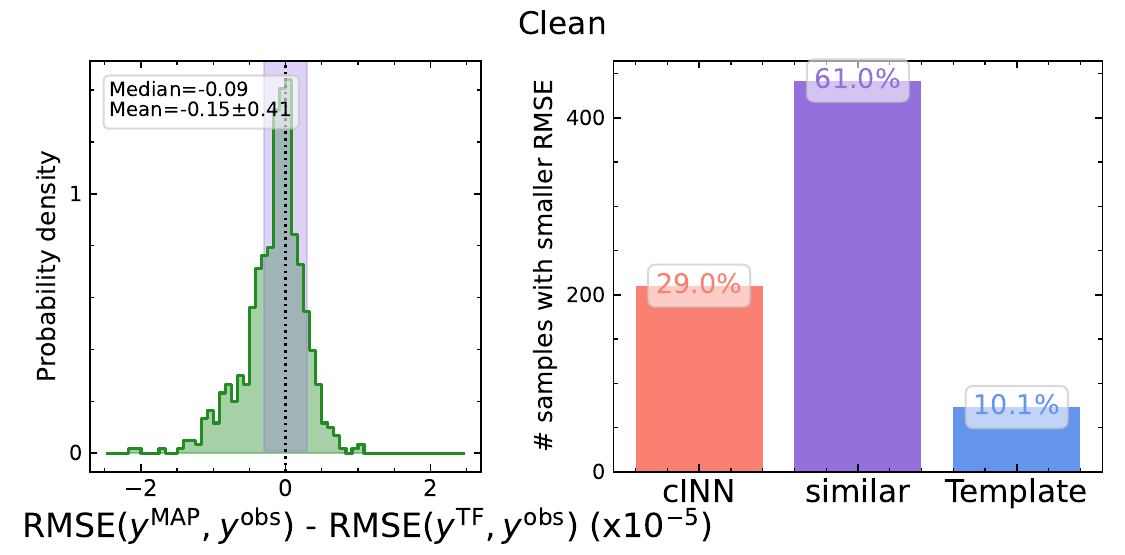}
    \includegraphics[width=1\columnwidth]{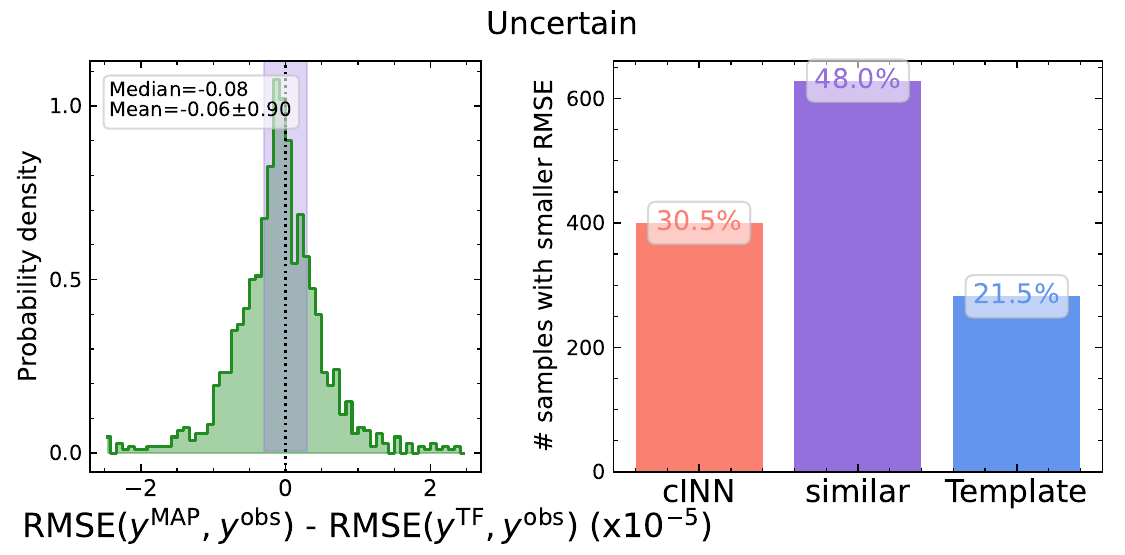}
    \caption{The left panel shows the comparison of the goodness-of-fit on the observational space: the difference between the RMSEs of the resimulated Phoenix spectra and the RMSEs of the modified template spectra. The right panel presents the number of samples where the corresponding method has a better fit (i.e. a smaller RMSE) to the original input spectrum. If the difference between the RMSEs is within $0.3 \times 10^{-5}$, they are classified as "similar". These cases are indicated by the purple shaded area in the left panel and a purple bar in the right panel. We use the \Clean\ group in the upper panels and the \Unc\ group in the lower panels.
     } \label{fig:bfit}
\end{figure}

\begin{figure*}
    \centering
    \includegraphics[width=1.8\columnwidth]{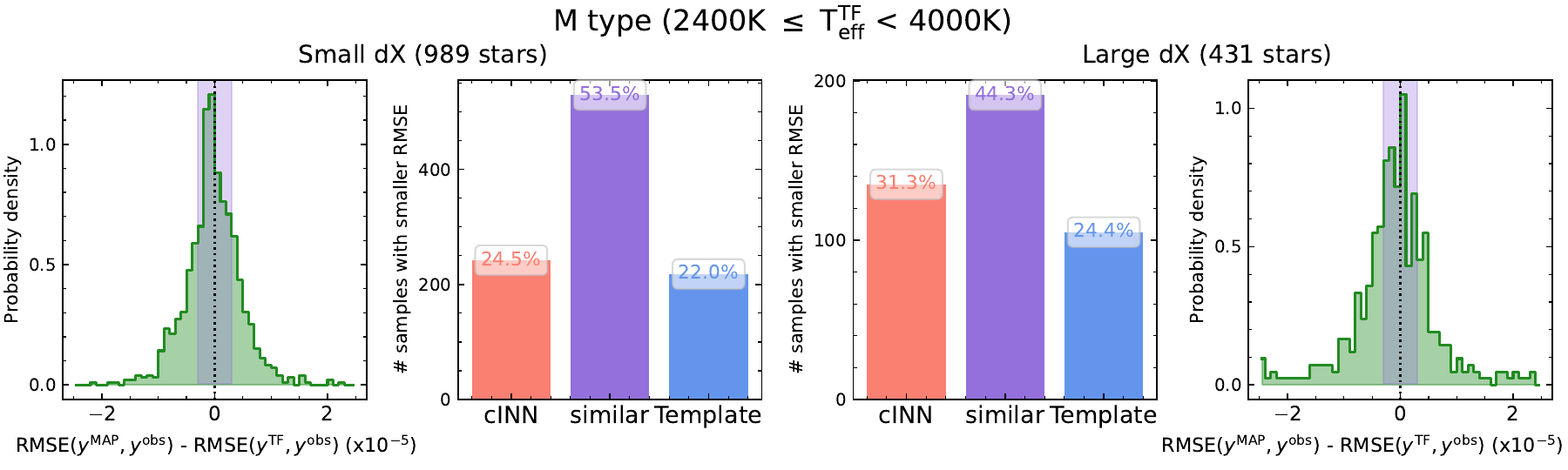}
    \includegraphics[width=1.8\columnwidth]{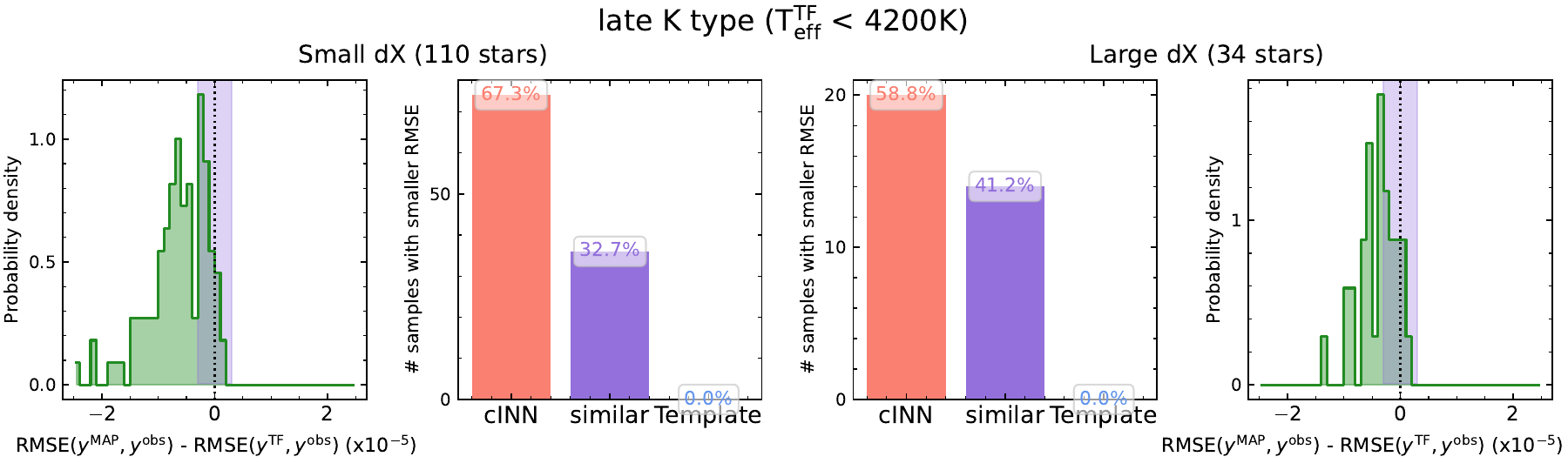}
    \includegraphics[width=1.8\columnwidth]{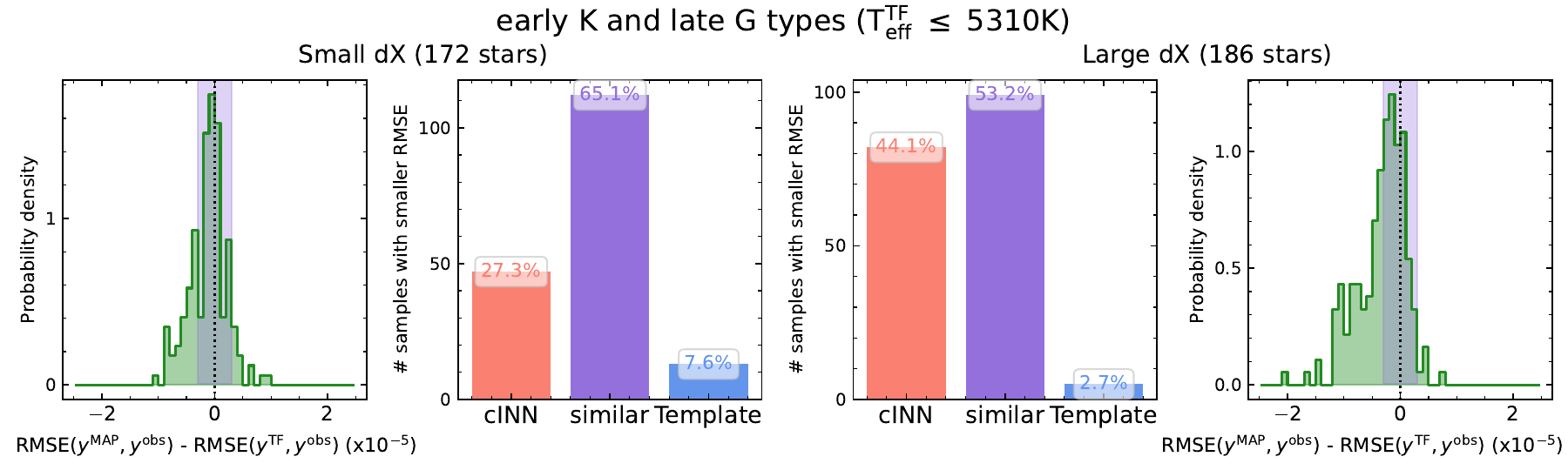}
    \caption{Comparison of RMSE on the observational space for small dX and large dX groups for different $T^{\rm{TF}}_{\rm{eff}}$ ranges. We use both the \Clean\ and \Unc\ groups in this figure. K- or G-type samples are divided into late K types (up to K6 type) and early K and late G types (K4 -- G9 types). G8-type stars are excluded in this figure. The first and the last columns present the difference between the RMSEs (goodness-of-fit) obtained by the two different methods and the second and third columns show the number of samples where the corresponding method has a better fit to the original input spectrum. Purple bars are the cases where RMSEs of the two methods are similar to within $0.3 \times 10^{-5}$.} \label{fig:bfit_types}
\end{figure*}

\begin{figure*}
    \centering
    \includegraphics[width=2\columnwidth]{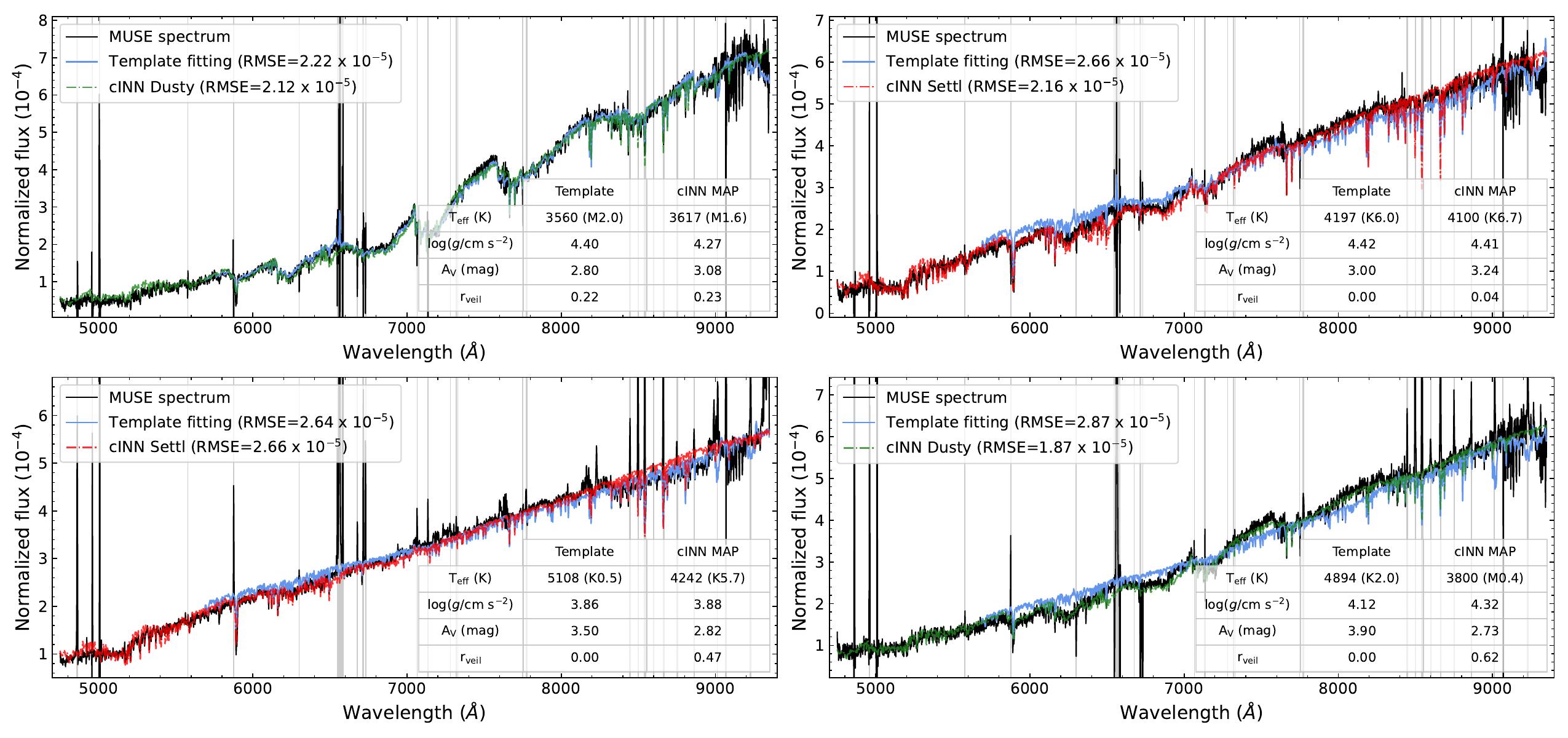} 
    \caption{Four examples of MUSE spectra (solid black line) together with the best-matching modified template spectra (solid blue line) chosen in \citetalias{Itrich+2024} and the resimulated Pheonix model spectra based on the cINN predictions (MAP estimates). The resimulated spectrum is described by either a red (Settl) or green (Dusty) dash-dot line depending on the MAP estimate of the library flag. For both the template fitting and cINN-based result, we present RMSEs between the fitted spectra and the original MUSE spectrum in the legend and list the measured stellar parameters in the table. All the spectra are normalised by the sum of the fluxes across all spectral bins within the common wavelength range between the template spectrum and MUSE spectrum (i.e. 5600 -- 9350~\AA) excluding the bins masked due to emission lines (grey shades). The surface gravity value for the template is the \logg\ value of the corresponding template star, not measured from the Tr14 spectrum.} 
    \label{fig:fit_ex4}
\end{figure*}

In this section, we verify whether the estimated parameters from each method reproduce the original input observation (i.e. stellar spectrum) well and investigate which method better explains the given observation.
We reproduce the modified template spectra used in \citetalias{Itrich+2024} and the synthetic Phoenix spectra based on the MAP estimates. We first select the appropriate Phoenix library by rounding off the MAP estimate of the library flag to either 0 (Settl) or 1 (Dusty). If the library flag estimate is between 0.2$\sim$0.8, we define this as an ambiguous case and produce two spectra using both Settl and Dusty, respectively. We feed the MAP estimates of effective temperature and surface gravity to our Phoenix interpolation pipeline and add the veiling effect and extinction. Here, we call this synthetic spectrum generated based on MAP estimates a resimulation or a resimulated spectrum. 

About 8\%\ of the Tr14 samples had MAP estimates of either \teff\ or \logg\ outside of the training range, where most of them were outliers in surface gravity. However, our interpolation pipeline is not able to extrapolate spectra for these cases. Considering that surface gravity has a comparatively small influence on the overall shape of the spectrum in contrast to \teff, \av, and \veil\, and that the degree of \logg\ extrapolation is small enough (see Sect.~\ref{sec:parameter_estimation}), we reproduce synthetic spectra for these cases by clipping the extrapolated \logg\ values to the upper or lower limit. In the case of Dusty model-based resimulated spectra, we computed resimulated spectra for \logg\ predictions of up to 5.5, even though that exceeds the formal upper limit of 5 because Dusty models with a \logg\ of up to 5.5 exist in our recent version of the interpolation pipeline.
We exclude 15 stars with extrapolated \teff\ estimates that we could not generate synthetic spectra.

On the other hand, to compare the cINN performance with the template fitting, we reproduce the modified template spectrum following the same process as implemented in \citetalias{Itrich+2024}, by choosing the same template star, adding the veiling effect, and reddening the spectrum using the \veil\ and \av\ values of TF parameters.
We normalise the spectra by the sum of the fluxes across the spectral bins within the common wavelength range between the template spectra and the MUSE spectra (i.e. 5600 -- 9350~\AA) excluding the bins masked due to the presence of emission lines (grey areas in Figs.~\ref{fig:flux_err_distr} and ~\ref{fig:fit_ex4}). This is different to the normalisation used in \citetalias{Itrich+2024} for the template fitting (i.e. the normalisation by the flux at 7500$\AA$).

To evaluate the goodness-of-fit, we compute the RMS of the residual between the input spectra and resimulated spectra or the input spectra and modified templates. When calculating this evaluation index, we only use the overlapping wavelength range between the template spectra and the MUSE spectra (5600 -- 9350\AA). We exclude spectral bins with emission lines, listed in Sect.~\ref{sec:training_data}, because the Phoenix model does not produce the emission lines and so those ranges were excluded in the input of the cINN as well. The calculated RMSEs are weighted by the S/N ratio at each spectral bin to consider the flux error.

The average RMSE of the resimulated spectra is $(2.5\pm1.1) \times 10^{-5}$ for the \Clean\ samples and $(6.2\pm4.9) \times 10^{-5}$ for the \Unc\ samples. The average RMSE of the modified templates is $(2.7\pm1.2) \times 10^{-5}$ and $(6.3\pm4.9) \times 10^{-5}$ for the \Clean\ and \Unc\ samples, respectively. The RMSEs are sufficiently small in general because both methods found their best-fit results. In both methods, there is a common trend where the lower the temperature of the stars, the larger the overall RMSE values are. For instance, in the \Clean\ samples, the average RMSE for M-type stars is approximately $3 \times 10^{-5}$, for K-types it is about $2 \times 10^{-5}$, and for G-types, it is about $1.7 \times 10^{-5}$. Regardless of methodological differences, stars with lower temperatures inherently exhibit larger RMSEs, but also their spectra are the most noisy. This trend has also been reported in \citetalias{Kang+23b}.

Considering the smaller RMSE in the observational space as indicative of a better fit, we investigate whether a resimulated spectrum or a modified template spectrum is chosen as a better fit for each star. 
The histogram in Fig.~\ref{fig:bfit} shows the distribution of the differences between the RMSE of the resimulated spectra and the RMSE of the modified templates. The median for the \Clean\ group is $-0.09 \times 10^{-5}$, while it is $-0.08 \times 10^{-5}$ for the \Unc\ group. As the RMSEs of both methods are small, their difference is also minor. The RMSEs of the resimulated spectra are slightly smaller than those of the modified template spectra.
The bar plot in the right panel of Fig.~\ref{fig:bfit} examines the frequency of which method is chosen as the better fit. If the absolute RMSE difference is smaller than $0.3\times 10^{-5}$, we consider it difficult to determine which of the two is superior and mark them as similar (purple bar). For 61\%\ of the \Clean\ samples, the two methods show comparable RMSE, while for the other 29\%\ of the samples, the resimulated spectrum has a smaller RMSE than the modified template, meaning that the resimulated spectrum is closer to the original input spectrum. Only in 10\%\ of the samples, the modified template is better than the resimulated spectrum.
In the case of the \Unc\ group, the modified template shows better fit for 21.5\%\ of the samples but the resimulation spectrum is still better for more samples (30.5\%).
These results show that statistically both methods find the best fit at a similar level for more than half of the samples, but for the rest, the resimulated spectra are slightly more likely to match the original spectra better than modified templates.

For the detailed analysis, we split the samples into two groups based on the stellar parameter differences: small dX and large dX. We expect one method to have a much smaller RMSE than the other if the parameter difference is large. If $|\Delta T_{\rm{eff}}| > 500$~K or $|\Delta A_{\rm{V}}| > 1$~mag or $|\Delta r_{\rm{veil}}| > 0.5$, it is classified as large dX, meaning that the parameter difference is large and otherwise as small dX. Here we exclude 75 stars classified as G8-type in \citetalias{Itrich+2024} because their \teff\ are just a lower limit. 
Then we further divide the sample groups according to the spectral type. In Fig.~\ref{fig:bfit_types}, we present the comparison of the goodness-of-fit as shown in Fig.~\ref{fig:bfit} by dividing the samples into small dX and large dX groups and separating them into three categories based on $T_{\rm{eff}}^{\rm{TF}}$: M types, late K types (up to K6 type), early K and late G types (K6--G9).
Additionally, we present four examples of MUSE spectra selected from the \Clean\ samples, together with the modified template spectra and the resimulated spectra in Fig.~\ref{fig:fit_ex4}.

Fig.~\ref{fig:bfit_types} shows that while the results in the large dX and small dX groups are similar, the results vary significantly with spectral types.
In the case of M-type stars, the RMSE of the resimulated spectrum and that of the modified template are similar for about half of the samples. For the remaining samples, neither method is dominant in the bar plot and has a similar amount of around 25\%.
The first example in Fig.~\ref{fig:fit_ex4} (upper left panel) shows the spectrum of an M2 star. In this case, the stellar parameters derived from template fitting and those from cINN are in good agreement and both the modified template spectrum and the resimulated spectrum satisfy the observed spectrum well with a similar RMSE.

For late K-type stars (up to K6 type), there is almost no difference between the small dX and large dX groups. Interestingly, for 65\%\ of the samples, the resimulated spectrum fits better than the modified template, which is a significantly larger fraction compared to M-types.
The second example in Fig.~\ref{fig:fit_ex4} (upper right panel) is one of the K6-type stars classified as a small dX group. The resimulated spectrum of this star fits better with the observed spectrum than the modified template spectrum.

Fig. \ref{fig:group_1to1} showed that stars earlier than K6 type often have large differences in \teff\ and \veil, where the cINN predicts smaller \teff\ and larger \veil\ values compared to TF parameters. When parameter differences are small for early K or late G type stars, the resimulated spectra and modified template spectra show similar RMSE for 65\%\ of the samples, while resimulated spectra have smaller RMSE for the majority of the other samples. On the other hand, when the parameter difference is large, it is more likely that the resimulated spectrum has a much smaller RMSE than the modified template spectrum.

The third and fourth panels (lower panels) in Fig.~\ref{fig:fit_ex4} show examples of early K-type stars, where differences between MAP estimates and TF parameters are large. In the former case (lower left panel), the parameter differences are considerably large whereas the RMSEs of the modified template spectrum and the resimulated spectra are similar. Both spectra satisfy the observation well, but the template spectrum fits better for $\lambda > 8000$~\AA, whereas the resimulated spectrum fits better for the rest of the wavelength range. The latter case (bottom right panel in Fig.~\ref{fig:fit_ex4}) is one where the parameter difference is large and one method clearly shows a better fit than the other. This star classified as K2 by template fitting is classified as M0.4 by the cINN and the corresponding resimulated spectrum satisfies the observation better than the template spectrum.

Summarising these results, for 53\%\ of the total samples, both methods satisfy the observed spectrum to a similar level. But for the other cases, it is more common that the resimulated spectrum fits the input spectrum better than the modified template. If the stellar parameter differences between the two methods are large, the fraction of samples whose resimulation has a smaller RMSE slightly increases, especially for the case of stars earlier than K6 type.


\section{Discussion}
\label{sec:discussion}

\subsection{Sources of uncertainty: cINN}
\label{sec:disussion_cinn}
We discuss the methodological limitations and weaknesses of the two different parameter estimation methods, the cINN and template fitting method used in \citetalias{Itrich+2024}, and outline the sources of uncertainty and inaccuracies in the estimation.
The main factors that influence the accuracy and reliability of the estimation by our cINN include the basic predictive performance, targets outside the training range, flux error, and the simulation gap.

The basic predictive performance refers to how well the cINN captures the underlying physical rules in the Phoenix models. This performance is validated immediately after the training by evaluating the performance of the cINN on the synthetic test models which follow the same physical rules as the training models but are not used in the training.
The network may fail to fully learn from the training data if the underlying physics of the training data is too complex relative to the capacity of the network (e.g. depth of the network). Training difficulty is influenced by several factors such as the complexity of the physical laws and whether the observational data contains sufficient information to constrain the target parameters. 
In our studies, we have demonstrated the excellent predictive performance of our cINNs. In \citetalias{Kang+23b}, the cINNs exhibited remarkable accuracy on the test models, demonstrating their capability to capture the physics of the Phoenix models. In this paper, we also demonstrated the good predictive performance of our cINN on the test set as a function of the relative flux error in Sect.~\ref{sec:synthetic}. Therefore, in \citetalias{Kang+23b} and this paper, the predictive performance of our cINN is highly reliable, with minimal impact on the estimation uncertainty. In addition, we demonstrated that the synthetic spectra resimulated based on the estimated parameters for Tr14 stars fit the input observations well, further confirming that the cINN effectively interprets the input observations based on the knowledge acquired during training.

Secondly, if the cINN encounters an unforeseen target, i.e. a star with properties outside the training range, its parameter estimates may be uncertain. Our cINN is trained on \teff\ between 2600 and 7000~K (M8--F3 types), \logg\ of 2.5--5, \av\ of 0--10~mag, and \veil\ between 0 and 2. The accuracy of the cINN for stars outside these ranges is currently untested and therefore uncertain. Plus, since the metallicity of our training data is fixed to solar metallicity, stars whose metallicity deviates significantly may also result in uncertain estimations.
However, we suggest that our training ranges sufficiently cover the characteristics of the Tr14 stars in this study and the young, low-mass stars we are interested in. 
The \av\ and \veil\ ranges span a sufficiently wide interval, encompassing the values reported in \citetalias{Itrich+2024} and the \logg\ range also includes most of the probable \logg\ values of young stellar objects found in the literature~\citep[e.g.][]{Frasca+2017, Manara2017, Olney2020}. Furthermore, we obtain parameter predictions falling well within the training ranges for most Tr14 stars in this study.

Still, about 10\%\ of Tr14 stars have at least one parameter outside the training range (Sect.~\ref{sec:parameter_estimation}). This may be due to the properties of those stars falling outside the training range, although the exact cause cannot be determined. We do not guarantee that the cINN can extrapolate accurately because we did not test on Phoenix models beyond the current \teff\--\logg\ space.
Nevertheless, we cautiously claim that the extrapolated values presented in this study are reliable as most of the extrapolations in this work are \logg\ values only slightly larger than the upper training limit of 5, falling mostly below 5.2. Given the \logg\ values measured between 5 and 5.5, it is worth extending the training range of \logg\ up to 5.5 in future studies.

The third factor is the relative flux error. Our cINN, the Noise-Net, accounts for the uncertainty caused by flux errors in parameter estimation. In Sect.~\ref{sec:synthetic}, we demonstrated that although the parameter estimation error increases with relative flux error ($\sigma_{\mathrm{med}}$), our cINN remains accurate at typical $\sigma_{\mathrm{med}}$ value of our Tr14 samples (1 -- 10\%, see Table~\ref{table:rmse_map} for detailed values).
Since the training is limited to log $\sigma_{\rm{med}}$ between -5 and -0.5, performance beyond this range is uncertain. We set this training range based on the relative flux errors in the Tr14 data so that for most Tr14 samples, the relative flux error falls within it (Fig.~\ref{fig:flux_err_distr}). In the \Clean\ samples, only one star exceeds the 
log $\sigma_{\rm{med}}$ limit, while in the \Unc\ samples, there are 164 cases (12.4\%), though most of their log $\sigma_{\rm{med}}$ are below 0 and are not excessively large.
\cite{Kang+23a}, which first proposed the setup of the Noise-Net, tested the performance of the Noise-Net on errors beyond its training limit and found that parameters obtained by feeding errors exceeding the upper limit are similar to those obtained by clipping the error to the upper limit. The posterior distribution for the former case was slightly wider than that for the latter, though the difference was negligible overall. \cite{Kang+23a} concluded that the Noise-Net does not produce erratic predictions, even for unexpectedly large observational errors but instead estimates parameters by self-clipping the observational error close to the upper limit. Therefore, even for log $\sigma_{\rm{med}}$ > -0.5, we suggest that the measured MAP value is reliable, although the width of the posterior distribution (i.e. estimation error) may be underestimated.

Additionally, we used one $\sigma$ value per spectrum, i.e. $\sigma_{\mathrm{med}}$, at all spectral bins because of the difficulties in training, which may affect the parameter estimation by underestimating or overestimating $\sigma_{\lambda}$ especially at outer wavelength ranges. In Sect.~\ref{sec:observation_samples}, we confirmed that the level of underestimation at bluer wavelength ranges was relatively small for the earlier types which should be more sensitive to this issue. However, concerning the underestimation of relative flux errors, we further investigated how cINN prediction changes if we increase $\sigma_{\mathrm{med}}$ of Tr14 samples and confirmed that the changes in MAP estimates were not significant enough to affect the overall results in this study. The RMSE between the original MAP estimates and the MAP estimates obtained at five times larger $\sigma_{\mathrm{med}}$ values was 53~K, 0.082~dex, 0.044~mag, and 0.071, for \teff, \logg, \av, and \veil, respectively, when we use the entire 2051 stars. We suggest that our methodological limitation may influence parameter estimation, but the impact is unlikely to be significant in the present study. We will adopt a better approach to handle wavelength-dependent relative flux errors in future studies.

The last factor, the simulation gap, has the most significant impact on the reliability of the parameters obtained by cINN. We proved through several experiments that the cINN effectively learned from the training data and correctly interpreted the observed data. However, the cINN cannot overcome the gap between the training data (i.e. Phoenix models) and actual observations. In most astronomical problems, it is common to use synthetic observations for training machine learning models because it is very difficult to collect a sufficient number of real data with highly accurate parameters and high-quality observations at the same time. In \citetalias{Kang+23b}, we roughly quantified the simulation gap between the Phoenix models and the class III template spectra and found that the gap is larger in M-type stars than in hotter stars. 

The simulation gap is an inherent limitation not only for our cINN but also for all methodologies that use theoretical models to analyse real data. For example, estimating stellar masses and ages using evolutionary tracks or spectral energy distribution (SED) fitting using synthetic spectra are both subject to the limitations of the simulation gap. However, interpreting real observations using theoretical models is a widely used and accepted methodology in many studies. Moreover, the Phoenix models used in this paper have also been used in numerous other studies to obtain stellar parameters from observations. For example, \cite{Frasca+2017} and \cite{Manara2017} measured the stellar parameters of young stars by fitting the observed spectra with the Settl model and validated that the measured parameters agreed well with the spectral type measured using spectral indices. \cite{Bayo+2017} also used Settl models for SED fitting and \cite{Rajpurohit+2018} classified M dwarfs using Settl models and demonstrated that the Settl model accurately reproduces the characteristic features of the stellar spectrum. Since we demonstrated that our cINN effectively captures the underlying rules in the Phoenix models, the uncertainty resulting from the simulation gap in our cINN is comparable to that in other studies using the same Phoenix models.

\subsection{Sources of uncertainty: template fitting in \citetalias{Itrich+2024}}
\label{sec:discussion_tf}

\begin{figure}
    \centering
    \includegraphics[width=1\columnwidth]{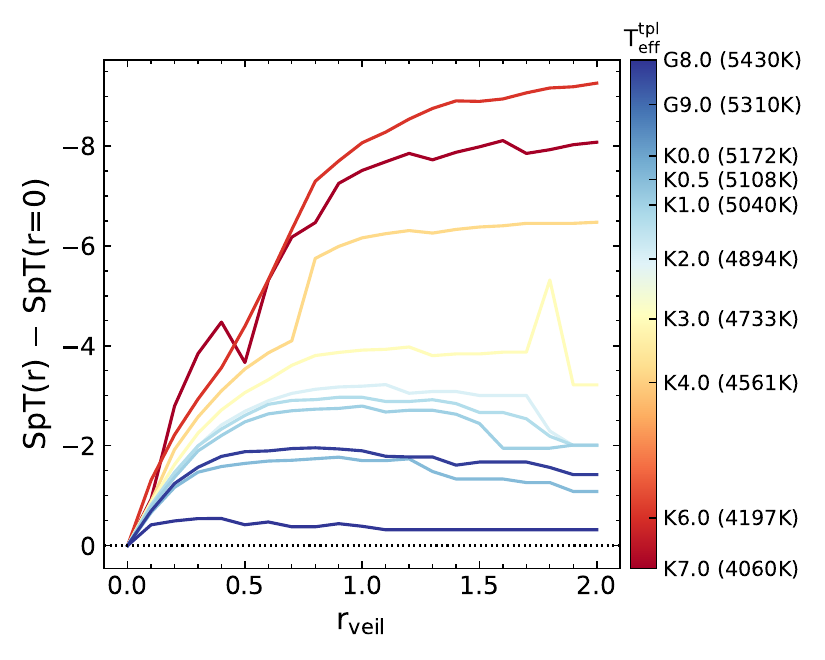}
    \caption{The difference of spectral type measured for veiled templates to the spectral type measured for pure templates without veiling as a function of the veiling factor. We measure the spectral types using the same methodology used in \citetalias{Itrich+2024}, which is based on the equivalent width of several absorption lines. The colour denotes the \teff\ of the template.
     } \label{fig:spt_veiltest}
\end{figure}

\begin{figure}
    \centering
    \includegraphics[width=1\columnwidth]{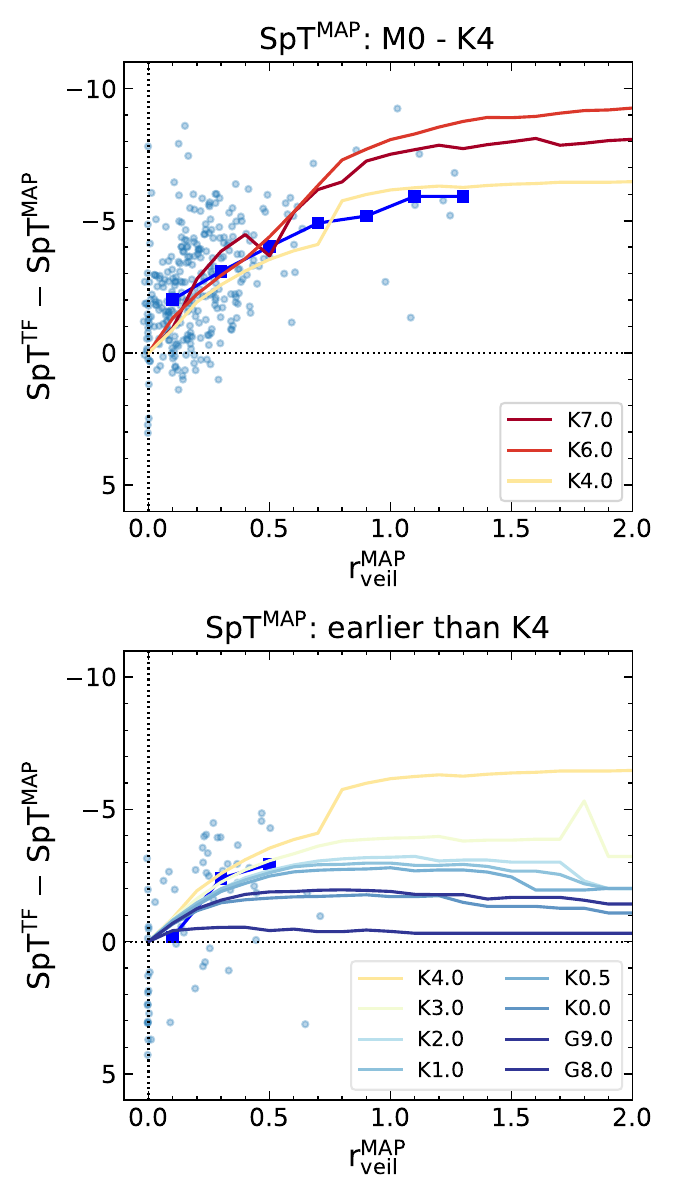}
    \caption{Comparison of spectral type differences between TF parameters and MAP values to the MAP veiling values for the \Clean\ samples of Tr14 (blue dots). The samples are divided into two groups depending on the spectral type based on the MAP values. Stars classified as G8 by template fitting are excluded. The blue squares represent the average values calculated within the veiling interval of 0.2. When calculating the average values, we exclude stars whose MAP-based temperature exceeds the maximum template fitting temperature of 5430~K (G8 type). The coloured lines correspond to the curves presented in Fig.~\ref{fig:spt_veiltest}, which show the change of the spectral type as a function of the veiling tested on the template stars.
     } \label{fig:dspt_r}
\end{figure}

In this section, we discuss the limitations and weaknesses of the classification method used in \citetalias{Itrich+2024} that can affect the reliability of the estimated parameters. In particular, we focus on the classification method used for K- or G-type stars where three parameters (\teff, \av, and \veil) were not considered simultaneously. The details of the method are described in Sect.~\ref{sec:template fitting}.

The main sources of uncertainty in \citetalias{Itrich+2024} include the low density of the template grid, the uncertainty of the template parameters, and the failure to consider all three parameters simultaneously for K/G-type stars. The first two factors represent the general limitations of the template fitting method. 
The grid of templates based on real data is less dense than the training data of cINN unless interpolation-like techniques~\citep[e.g.][]{Fang+2021, Claes+2024} are employed. \citetalias{Itrich+2024} used 37 template stars ranging from M9.5 to G8, most of which have 0.5 subclass intervals for M types and 1 subclass intervals for K/G types. 
However, \citetalias{Itrich+2024} lacked a template between K6 and K4, corresponding to a temperature interval of approximately 350~K, and could not analyse the stars within this temperature range (see Figs. \ref{fig:hr} and \ref{fig:mass}). Additionally, the maximum temperature of the template grid was G8 (5430~K), which is too low for some hotter stars. Several stars were classified as between K4--K6 type or as hotter than G8 by the cINN, highlighting the importance of the diversity of templates in improving the accuracy of parameter estimation in the template fitting method. Furthermore, as the template grid had only one or two stars per spectral type, the influence of surface gravity was neglected in \citetalias{Itrich+2024}.

The most critical factor we focus on is that for K- and G-type stars, the temperature was determined independently, without considering other parameters, particularly veiling. The equivalent widths (EWs) of absorption lines decrease in the presence of veiling~\citep{Herczeg2014}. According to the constant veiling model used in both \citetalias{Itrich+2024} and this paper (Eq.~\ref{eq:veiling}), the EWs decrease by about 50\%\ when \veil\ is 1. Since EWs are sensitive to veiling, measuring the spectral type without accounting for veiling may result in incorrect spectral classification. \cite{Rugel+2018} highlighted the importance of considering the degeneracy between spectral type, extinction, and accretion of pre-main-sequence stars, and \cite{Fang+2021} showed that the temperature is overestimated if the veiling is not considered in the classification process. If veiling is determined after fixing the spectral type, we are likely to measure very small or zero veiling because the spectral type is already determined based on the veiled spectrum. This trend is revealed in the fact that about 84\%\ of the hot stars have \veil\ of 0 when estimated by template fitting.

To investigate the influence of veiling on the spectral type determination based on EWs, we examine the change of spectral type by increasing the veiling on the class III template stars used in \citetalias{Itrich+2024} within the K7--G8 range. We first calculate the spectral type on a pure template spectrum without any extinction and veiling, i.e. SpT($r=0$). Next, we add veiling to the template spectrum following Eq.~\ref{eq:veiling} and measure the spectral type again, where \veil\ increases from 0 to 2 in 0.1 intervals. Fig.~\ref{fig:spt_veiltest} exhibits how dramatically the spectral type changes under the influence of veiling. A difference of 1 in the y-axis corresponds to a one-subclass difference. The impact of veiling varies depending on the original spectral type. Late K types are more sensitive to veiling than earlier types. According to this experiment, at a \veil\ of 0.5, the spectral type of K6 and K7 types is overestimated by about four subclasses (about 700~K higher) and for early K types, it is overestimated by 2--3 subclasses (200--300~K). In the case of G types, the spectral type increases by about 1--2 subclasses (100--200~K).

We suspect that the temperature difference between the TF parameters and the MAP values in K/G-types is highly affected by this veiling problem because the MAP values are mostly smaller than the TF parameters by $\sim$600~K whereas the difference in \teff\ for M-type stars is about 200~K. Therefore, we convert each temperature value into spectral type (SpT$^{\rm{TF}}$ and SpT$^{\rm{MAP}}$) and compare the SpT difference with MAP-based veiling ($r_{\rm{veil}}^{\rm{MAP}}$). In Fig.~\ref{fig:dspt_r}, we split the \Clean\ samples into two groups depending on the MAP-based SpT and present the correlations between SpT difference and veiling in Tr14 samples. Here, we exclude stars classified as G8 type by template fitting. The reason for using MAP-based veiling is that if SpT$^{\rm{TF}}$ was influenced by veiling, it is highly probable that $r_{\rm{veil}}^{\rm{TF}}$ was underestimated. We also compared SpT difference with veiling difference ($r_{\rm{veil}}^{\rm{MAP}}$ $-$ $r_{\rm{veil}}^{\rm{TF}}$) and found no significant change from Fig.~\ref{fig:dspt_r}, as most of $r_{\rm{veil}}^{\rm{TF}}$ is 0.
Generally, the higher the veiling, the larger the SpT difference, as we showed in Sect.~\ref{sec:param_comparison} (see Fig.~\ref{fig:clean_3d}). In each panel, we plot the SpT difference in comparison to the curves of the template stars presented in Fig.~\ref{fig:spt_veiltest}, for corresponding SpT$^{\rm{MAP}}$ ranges. Interestingly, the overall distribution of the Tr14 stars follows the SpT difference curves of the templates well. The change of SpT is more dramatic in late K types than earlier types in both Tr14 stars and templates. Despite the scatter in individual stars (blue circles), the average trends marked with the blue squares are almost identical to the curves obtained from the templates. The average values are calculated within a veiling interval of 0.2. We exclude stars in the average calculations whose MAP-based SpT are earlier than G8 (5430~K), the maximum temperature of the template fitting, because their temperature differences may be due to the lack of hotter templates.

The overall results in Fig.~\ref{fig:dspt_r} support our hypothesis that the main cause of the temperature differences in K/G-type stars is because the veiling is not considered in the SpT determination in \citetalias{Itrich+2024}. This may not be the only cause of the temperature difference considering the scatter shown in Fig.~\ref{fig:dspt_r}. However, the distributions varying on the spectral type match well with the results of templates at the same temperature range, implying that the veiling is a major cause of the temperature difference.
Moreover, in the case of the M-type stars, MAP-based temperature and template fitting-based temperature agree well even though non-neglectable veiling exists (Fig.~\ref{fig:perform_bar_clean}). We claim that this is because \citetalias{Itrich+2024} considered temperature and veiling simultaneously for M-type stars.

In the case of K/G-type stars, extinction is not considered simultaneously with temperature either, but we expect the impact of neglecting extinction to be smaller than veiling because reddening the spectrum does not significantly change the EWs. We also examine the SpT change as a function of the extinction using the template stars, in the same way we tested the influence of veiling. We redden the template spectrum by increasing the \av\ from 0 to 5~mag in 0.2~mag intervals. Fig.~\ref{fig:spt_avtest} shows that the influence of extinction on SpT is very small compared to the veiling. The trend varies a lot depending on the original SpT but the SpT changes within 0.3 subclass in \av\ of 1--4~mag, the typical \av\ range of the Tr14 samples. As the maximum SpT difference is about 0.4 subclass, below the precision level of SpT measured in \citetalias{Itrich+2024}, we expect that extinction does not significantly affect the measured SpT in \citetalias{Itrich+2024}.

\subsection{Veiling and NIR excess}
\label{sec:discussion_veil}

Factors contributing to the observed veiling include ultraviolet and optical emission from the mass accretion process of young stars, as well as infrared (IR) emission from the protoplanetary disk. The continuum emission by accreting young stars is more easily detected as Balmer continuum emission and characteristic Balmer jump~\citep{Gullbring+1998, Herczeg&Hillenbrand2008, Rigliaco+2012, Alcala+2014}, but neither are observable within the MUSE wavelength range. Given the MUSE spectral range (4750--9350~\AA), the veiling observed in our sample primarily results from Paschen continuum emission (3646--8203~\AA), which is dimmer than the Balmer emission~\citep{Alcala+2014}. Although mass accretion generates various emission lines (e.g. helium and calcium lines and Paschen series), neither \citetalias{Itrich+2024} nor our study utilised these lines for \veil\ measurements. IR emission from the disk can partially affect the longest wavelengths of the spectrum. Therefore, the veiling measured in \citetalias{Itrich+2024} and this study primarily reflects mass accretion rather than contributions from the disk.

One of the key limitations of the veiling values measured in both \citetalias{Itrich+2024} and this study is the assumption of a wavelength-independent veiling model (Eq.~\ref{eq:veiling}). While this approximation can be applicable within the MUSE spectral range (4750--9350~\AA), which does not include the Balmer jump, it has clear limitations that become particularly significant for stars with high veiling.
Both \citetalias{Itrich+2024} and this study limited the veiling factor, \veil, to 0--2. As a result, heavily veiled stars requiring \veil\ > 2 may not have been properly classified and were likely excluded from the final catalogue in \citetalias{Itrich+2024} although such stars could exist given the age of Tr14. However, we note that there may be some heavily veiled stars remaining in our samples with underestimated veiling values, although clearly identifying such cases is challenging.
As the measured high veiling values may be uncertain due to limitations in the current simple veiling model, we consider that the reliability of veiling measurements, both from template fitting and cINN, is relatively low for high veiling values (\veil\ > 1).
However, in Tr14 samples, only 110 stars (5\%) exhibited MAP-based veiling values larger than 1, with just 27 stars among the \Clean\ sample. Similarly, veiling values larger than 1 obtained in \citetalias{Itrich+2024} also constituted about 4\%\ of the total sample, indicating that high veiling values are relatively rare in this study.  

Additionally, the measured veiling values from both the cINN and in \citetalias{Itrich+2024} are less reliable for early K and G-type stars because veiling becomes less influential in changing the overall shape of the spectra for these stars compared to that of M-type stars. The average error of \veil\ from \citetalias{Itrich+2024} is 0.67 for K/G-type stars while it is 0.27 for M-type stars. This implies that veiling has minimal impact on chi-square fitting for K/G types, making accurate veiling measurements challenging. 
Fig.~\ref{fig:rmse_curve_spt} shows that the average intrinsic error of \veil\ from cINN is also higher for K/G-type models than M-type models at the same level of relative flux error, indicating that measuring \veil\ becomes challenging for hotter stars.
To improve the veiling measurement, we need a more complex, wavelength-dependent model, such as the hydrogen slab model which has been widely used in many studies~\citep[e.g.][]{Valenti+1993, Herczeg&Hillenbrand2008, Rigliaco+2012, Manara+2013b, Alcala+2014}. Although improving the model might not entirely overcome the difficulties in measuring veiling within the MUSE spectral range due to the absence of distinct features like the Balmer jump, we will adopt a more sophisticated veiling model in future studies.

\citetalias{Itrich+2024} determined whether stars exhibited NIR excess based on NIR photometry data ($J$-, $H$-, $K$-bands) by matching with the HAWK-I~\citep{Preibisch+2011a} and VISTA~\citep{Preibisch+2014} catalogues and following the definition from \cite{Zeidler+16}. We use this NIR excess flag and check the validity of the measured veiling values in this study and \citetalias{Itrich+2024}. However, it is important to note that this flag is only based on NIR photometry data and we need MIR data to accurately detect IR excess by protoplanetary disks.
We also investigate the fraction of stars with a non-negligible amount of veiling ($f_{\mathrm{veil}}$) by setting a threshold of 0.2 (i.e. \veil\ > 0.2). The average estimation uncertainty for \veil\ of around 0.1 from the cINN is very small, but it does not quantitatively reflect other uncertainty factors such as the simulation gap. Therefore, when calculating $f_{\mathrm{veil}}$, we adopted the \veilmap\ error of 0.15 for all Tr14 stars based on our test results on class III template stars (Sect.~\ref{sec:templates}).

In the entire sample, only 25\%\ of stars exhibit NIR excess ($f_{\mathrm{NIR}}$), while $f_{\mathrm{veil}}$ is 0.45$^{+0.34}_{-0.19}$ and 0.26$^{+0.56}_{-0.14}$ based on MAP estimates and TF parameters, respectively. The low $f_{\mathrm{NIR}}$ is surprising given the age of Tr14 and \cite{Preibisch+2011b} interpreted it as an impact of the harsh environment accelerating disk dispersal.
We found that both template fitting and cINN results show a decreasing trend in $f_{\mathrm{NIR}}$ and $f_{\mathrm{veil}}$ for M type stars. 
Based on TF parameters, $f_{\mathrm{veil}}$ decreases from 0.64 to 0.25 and $f_{\mathrm{NIR}}$ decreases from 0.68 to 0.16 when $T^{\mathrm{TF}}_{\mathrm{eff}}$ increases from 2700 to 4000~K. When using MAP estimates, $f_{\mathrm{veil}}$ is overall higher than when derived from TF parameters, but it similarly decreases from 0.84 to 0.31 with increasing $T^{\mathrm{MAP}}_{\mathrm{eff}}$, while $f_{\mathrm{NIR}}$ decreases from 0.49 to 0.24.
However, for \teff\ > 4000~K, $f_{\mathrm{veil}}$ and $f_{\mathrm{NIR}}$ exhibit different trends depending on the methods used. Based on MAP estimates, $f_{\mathrm{NIR}}$ gradually decreases from 0.24 to 0.1 with increasing $T^{\mathrm{MAP}}_{\mathrm{eff}}$, whereas $f_{\mathrm{veil}}$ increases from 0.31 to 0.65. However, high values of $f_{\mathrm{veil}}$ at around 0.6 for stars with \teff\ > 4800~K are less reliable due to the small sample size. On the other hand, when using TF parameters, $f_{\mathrm{NIR}}$ gradually increases from 0.16 to 0.35 within the K7--G9 type interval. Within this temperature range, $f_{\mathrm{veil}}$ based on \veiltf\ is small with a value of 0.08$^{+0.92}_{-0.08}$ and without significant change but the uncertainty is too large to for a meaningful interpretation.

We examined how the distribution of stars on the colour-colour diagram varies with the spectral type (M or K/G types) and \veil\ using the NIR colours from the HAWK-I catalogue~\citep{Preibisch+2011a}. For this analysis, we only use the \Clean\ sample, where stellar parameters are expected to be more reliable.
In both \citetalias{Itrich+2024} and this study, most M-type stars with \veil\ < 0.1 do not exhibit NIR excess ($f_{\mathrm{NIR}}=0.06$), suggesting that \veil\ < 0.1 is a reliable indicator of minimal extra emission from either accretion or a disk. In Fig.~\ref{fig:ccd_tf}, 20\%\ of K/G-type stars with \veiltf\ < 0.1 still exhibit NIR excess with large $H-K$ colour. As discussed in Sect.~\ref{sec:discussion_tf}, this supports our hypothesis that \citetalias{Itrich+2024} likely underestimated the veiling of these stars.

In Figs.~\ref{fig:ccd_map} and \ref{fig:ccd_tf}, stars with higher veiling tend to have a higher $f_{\mathrm{NIR}}$. This trend appears consistently across spectral types in the MAP-based results (Fig.~\ref{fig:ccd_map}). However, even among M-type stars with \veil\ > 0.3, nearly half do not show NIR excess. Given that the stellar parameters for M-type stars are reliable, as shown by the consistency between template fitting and cINN results, it is likely that a significant fraction of these stars have correctly measured veiling but undetected IR excess. In the case of K/G-type stars, some stars with high veiling but without NIR excess may have inaccurate veiling measurements because of the limitations of the veiling model and difficulty in veiling measurement for K/G types. However, some K/G-type stars do show clear excess and $f_{\mathrm{NIR}}$ is also higher for stars with high veiling. Like the case of M-type stars, it is still likely that veiling is correctly measured for a significant number of stars without NIR excess.

We examined the validity of the measured \veil\ using the NIR photometry but could not filter out misclassified veiling values. We suspect that the NIR excess flag did not fully capture the presence of actual IR excess because it is based on NIR photometry. The fact that the average \veil\ is higher for stars with NIR excess than those without suggests that the NIR excess does indicate a higher probability of a disk and accretion. However, more than 30\%\ of stars without NIR excess still have veiling values larger than 0.2 in this study and \citetalias{Itrich+2024}.
Although the NIR excess flag offers some insight into the validity of veiling, it has the following limitations: it is based solely on NIR photometry without MIR data, and \veil\ likely traces mass accretion rather than IR excess. A more detailed comparison, incorporating MIR data for accurate IR excess measurement and other accretion tracers like emission lines, is necessary but beyond the scope of this study so we leave this for future work.

\section{Summary}
\label{sec:summary}
In this paper, we present an updated version of the cINN for young low-mass stars and apply it to 2051 young stars in Tr14. In the previous study (\citetalias{Kang+23b}), we introduced a cINN approach to estimate stellar parameters (\teff, \logg, \av) from optical stellar spectra. cINNs were trained on Phoenix stellar atmosphere models and pretested on the class III template stars well-analysed in the literature. After confirming the applicability of the cINN in the study of low-mass stars in \citetalias{Kang+23b}, we upgraded the cINN in this study for the application to the spectra of young stars observed with VLT/MUSE.

The major updates in the new cINN include a Noise-Net setup, the addition of veiling, and the combination of two Phoenix libraries for training. Firstly, the cINN in this paper adopts a Noise-Net architecture which reflects the influence of random noise of observation on the parameter prediction. This update is motivated by the fact that the S/N ratios of Tr14 spectra in this study are on average lower than the template stars used in \citetalias{Kang+23b}. We feed the stellar spectrum and a median relative flux error ($\sigma_{\rm{med}}$), calculated across the wavelength range, together to the cINN as an input. 
Secondly, we added the veiling factor as a fourth target parameter to measure the influence of veiling on the observed spectra. The third update constitutes the construction of a new training dataset by combining two Phoenix libraries, Settl and Dusty. As the new cINN is trained on two libraries, the last target parameter of the new cINN is a library origin label which distinguishes Settl and Dusty models. 

We tested the performance of the new cINN on synthetic test models, confirming that although relative flux errors widen the posterior distributions, the network still provides very accurate predictions within the typical error range of our Tr14 data (Sect.~\ref{sec:synthetic}). Then, we tested the network on 36 class III template stars well-interpreted by \cite{Manara2013, Manara2017} and \cite{Stelzer2013}, and confirmed that our new cINN estimates \teff\ and \logg\ well within 1-$\sigma$ estimation uncertainty, similar to previous cINNs presented in \citetalias{Kang+23b}. The average error of the new parameter, veiling, was about 0.13--0.15.
 
Using our new cINN, we measured the stellar parameters (\teff, \logg, \av, and \veil) of young stars in Tr14 and compared them with those measured by \citetalias{Itrich+2024} using the template fitting method on the same data. We only compare \teff, \av, and \veil\ because \citetalias{Itrich+2024} could not measure \logg. Our main results of the comparison between cINN estimates (MAP values) and values obtained by \citetalias{Itrich+2024} (TF parameters) are the following.
\begin{enumerate}

\item The MAP values and TF parameters are in good agreement in general, but there is a non-negligible difference for earlier type stars (early K types and G types), especially in the case of \teff\ and \veil. All three parameters show good agreements for stars within M4-- K6 types (3270 -- 4197~K).

\item We found correlations between parameter differences. For example, when the cINN-based temperature is larger than the TF parameter, the cINN-based extinction is larger whereas the cINN-based veiling is smaller than the TF parameters. This trend is exhibited in K/G types, while the trend is the opposite in M-type stars.

\item The temperature values from template fitting are discrete because of the limited number of templates, whereas the cINN-based temperature estimates are continuous. As \citetalias{Itrich+2024} lacked templates between K6 --K4 types and templates hotter than G8, the template fitting-based temperature is missing in these ranges, whereas many stars have temperatures between K6 --K4 types and temperatures hotter than G8 when measured by the cINN.

\item Using the HR diagram and evolutionary tracks, we determine the average stellar age of the Tr14 stars to be 0.7$^{+3.2}_{-0.6}$~Myr, which is slightly smaller than the template fitting-based value, but still in good agreement within a 1-$\sigma$ error.

\item Both the resimulated Phoenix spectra based on the MAP values and the modified template spectra based on the TF parameters match the original input spectra well. When we evaluate the goodness-of-fit with the RMSE, the two methods had similar RMSE for half of the samples. However, for most of the other half, the resimulated spectra fit better to the input spectra than the modified templates.

\item The simulation gap, the gap between the synthetic models and reality, is the most influential factor affecting the reliability of the cINN outputs. This is because the cINN cannot overcome the intrinsic discrepancy between the Phoenix model and reality, even though the cINN can perfectly capture the underlying rules in the training models. However, this is a common limitation of all methodologies that apply theoretical models to real observations. 

\item We found that the parameter differences for K/G-type stars are affected by the limitation of the spectral type classification method used in \citetalias{Itrich+2024} based on equivalent widths of absorption lines. Using the template stars, we demonstrated that neglecting the veiling in the spectral type classification leads to an overestimation of the temperature. For example, the spectral type of a K6-type star is measured as four subclasses earlier if \veil\ is about 0.5. The temperature difference between the MAP values and the TF parameters agrees well with the amount of overestimation incurred by ignoring the amount of veiling measured by the cINN.

\end{enumerate}

Our cINN performs comparably to the multi-dimensional template fitting method used in \citetalias{Itrich+2024} and is more advantageous in some aspects: the cINN is overwhelmingly faster than the multi-dimensional fitting, it analyses stars over a wide temperature range (2600 -- 7000~K) in a consistent manner, and it provides a posterior distribution. We suggest that our cINN is a useful enough tool for analysing large amounts of data observed with VLT/MUSE.

\section*{Data availability}
A full version of Table~\ref{table:catalog} and spectra images like Fig.~\ref{fig:fit_ex4} for all sources are only available in electronic form at the CDS via anonymous ftp to \href{url}{cdsarc.u-strasbg.fr} (130.79.128.5) or via \href{url}{http://cdsweb.u-strasbg.fr/cgi-bin/qcat?J/A+A/}.


\begin{acknowledgements}
We thank the anonymous referee for insightful comments that helped to improve the flow of argument.
This work was partly supported by European Union’s Horizon 2020 Research and Innovation Program and the European Research Council via the ERC Synergy Grant ``ECOGAL'' (project ID 855130), by the Marie Sklodowska-Curie grant DUSTBUSTERS (project No 823823), by the Heidelberg Cluster of Excellence (EXC 2181 - 390900948) ``STRUCTURES'', funded by the German Excellence Strategy, and by the German Ministry for Economic Affairs and Climate Action in project ``MAINN'' (funding ID 50OO2206). We also thank for computing resources provided by the Ministry of Science, Research and the Arts (MWK) of the State of Baden-W\"{u}rttemberg through bwHPC and DFG through grants INST 35/1134-1 FUGG and  INST 35/1597-1 FUGG, as well as for data storage at SDS@hd through grants INST 35/1314-1 FUGG and INST 35/1503-1 FUGG. RSK is greatful to the 2024/25 Class of Radcliffe Fellows for highly interesting and stimulating discussions. DI acknowledges support from collaborations and/or information exchange within NASA’s Nexus for Exoplanet System Science (NExSS) research coordination network sponsored by NASA’s Science Mission Directorate under Agreement No. 80NSSC21K0593 for the program “Alien Earths”.

\end{acknowledgements}

\bibliographystyle{aa}

\begin{appendix}
\section{Supplemental materials}

In this appendix, we present supplementary figures mentioned in the main contents.

\subsection{Validation of cINN on synthetic test models and class III stars}

\begin{figure*}
    \includegraphics[width=2\columnwidth]{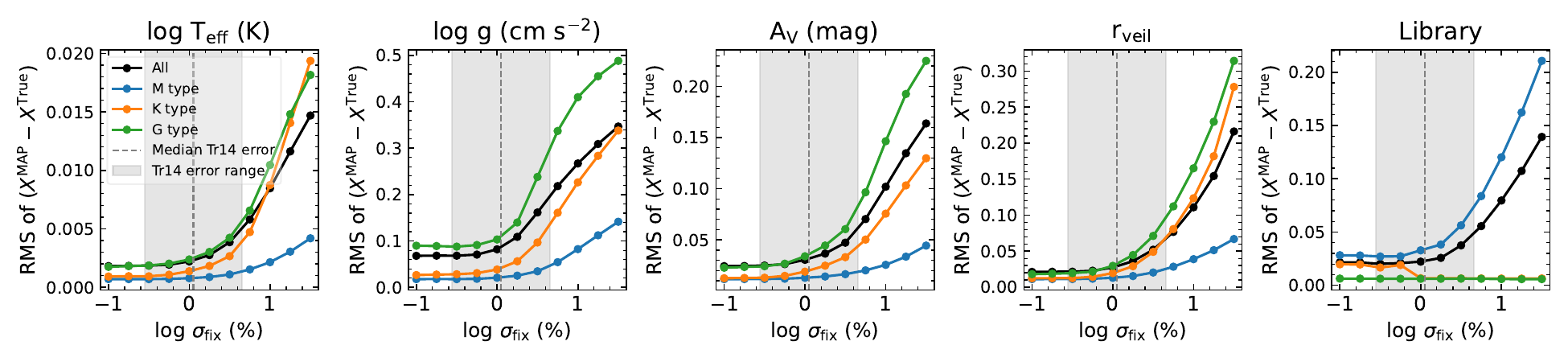}
    \caption{The RMSE of the MAP estimates for all synthetic test models (black), M-type test models (blue, 2600~K $\leq$ \teff\ $\leq$ 4000~K), K-type test models (orange, 4000~K < \teff\ $\leq$ 5250~K), and G-type test models (green, 5250~K < \teff\ $\leq$ 6000~K) at 11 different relative flux errors. 
    The grey horizontal dashed line and shaded area represent the median (1.12\%) and standard deviation (0.77~dex) of the relative flux error ($\sigma_{\rm{med}}$) of Tr14 samples in the \Clean\ group.
     } \label{fig:rmse_curve_spt} 
\end{figure*}

We evaluated the performance of our network on 26,214 synthetic test models and 36 class III template stars observed with VLT/X-shooter in Sect.~\ref{sec:validation}. 
By splitting the test models into three groups, M type (2600 -- 4000~K), K type (4000 -- 5250~K), and G type (5250 -- 6000~K), we investigated how the prediction performance varies with \teff\ of the models. Figure~\ref{fig:rmse_curve_spt} shows the cINN performance as a function of relative flux error ($\sigma_{\mathrm{med}}$) for the three groups along with the one for the entire test models (the same as the blue curve in Fig.~\ref{fig:rmse_curve}).

\begin{figure*}
    \centering
    \includegraphics[width=1.99\columnwidth]{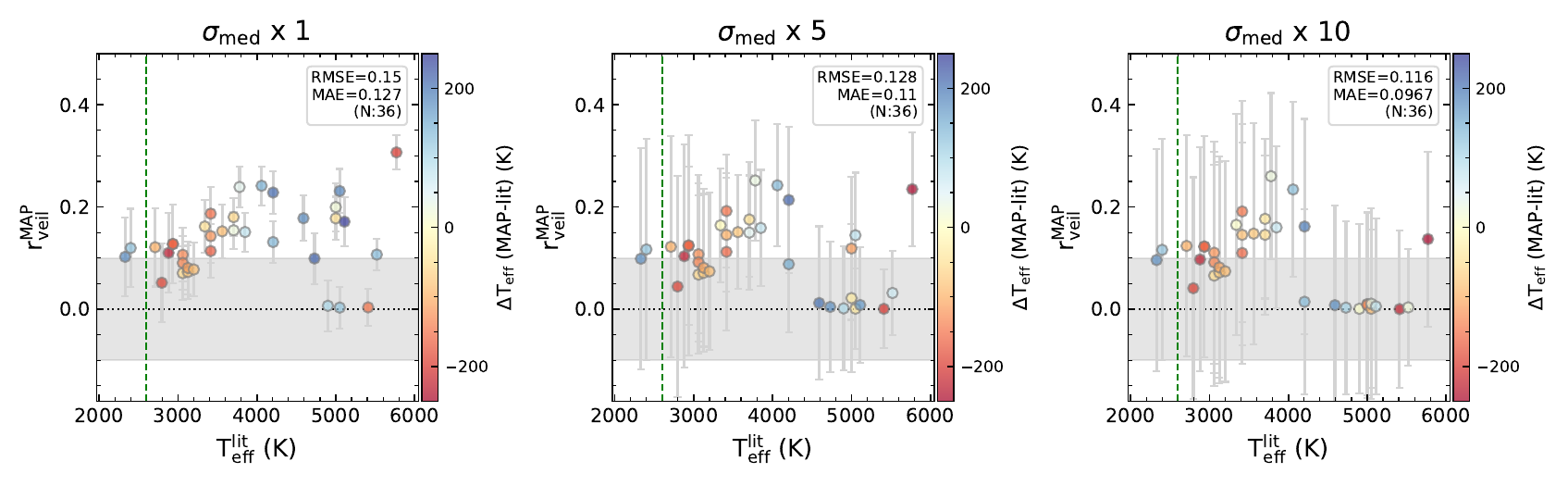}
    \caption{Comparison of MAP estimates of \veil\ measured by cINN with the literature \teff\ values for 36 template stars. For the second and third panels, we increase $\sigma_{\rm{med}}$ values by a factor of 5 and 10, respectively.
     } \label{fig:tpl_veil} 
\end{figure*}

We examined the performance of our network on class III template stars in Sect.~\ref{sec:templates}. In Table~\ref{table:tpl_param}, we listed the stellar parameters of the template stars obtained in literature~\citep{Manara2013, Manara2017, Stelzer2013} and stellar parameters measured in this work by cINN. We tested the network performance at different levels of flux error by increasing the $\sigma_{\mathrm{med}}$ of each star, which we measured from the spectrum within the 6500--7500~\AA\ range. Fig.~\ref{fig:tpl_veil} shows the change of the veiling measurements with increasing $\sigma_{\mathrm{med}}$, which is the most noticeable between the four stellar parameters.

\begin{table*}
    \centering
    \caption{Stellar parameters of class III template stars obtained in literature and in this work by cINN}
    \resizebox{\textwidth}{!}{
    \begin{tabular}{l cccc cccc c}
    \toprule
    Object Name  &  Spectral Type &  $T^{\mathrm{lit}}_\mathrm{eff}$ (K) 
    &  log $(g^{\mathrm{lit}}/\mathrm{cm\,s}^{-2})$  & $\sigma_{\mathrm{med}}$ (\%) 
    & $T^{\mathrm{MAP}}_\mathrm{eff}$ (K) &  log $(g^{\mathrm{MAP}}/\mathrm{cm\,s}^{-2})$ 
    & $A^{\mathrm{MAP}}_{\mathrm{V}}$ (mag) & $r^{\mathrm{MAP}}_{\mathrm{veil}}$ 
    & Reference $\log(g^{\mathrm{lit}})$  \\
    \midrule
    RXJ0445.8+1556 & G5.0 & 5770 & 3.93$\pm$0.11 & 1.0 & 5571$\pm$45 & 4.35$\pm$0.11 & 0.06$\pm$0.04 & 0.31$\pm$0.03 & (1) \\ 
    RXJ1508.6-4423 & G8.0 & 5520 & 4.06$\pm$0.11 & 1.0 & 5656$\pm$44 & 4.24$\pm$0.11 & 0.02$\pm$0.04 & 0.11$\pm$0.03 & (1) \\ 
    RXJ1526.0-4501 & G9.0 & 5410 & 4.38$\pm$0.18 & 1.4 & 5264$\pm$48 & 4.01$\pm$0.13 & 0.08$\pm$0.05 & 0.00$\pm$0.04 & (1) \\ 
    HBC407 & K0.0 & 5110 & 4.33$\pm$0.26 & 1.4 & 5409$\pm$62 & 4.77$\pm$0.13 & 0.22$\pm$0.07 & 0.17$\pm$0.05 & (1) \\ 
    PZ99J160843.4-260216 & K0.5 & 5050 & 3.48$\pm$0.24 & 1.5 & 5236$\pm$54 & 4.44$\pm$0.13 & 0.13$\pm$0.05 & 0.23$\pm$0.04 & (1) \\ 
    RXJ1515.8-3331 & K0.5 & 5050 & 3.86$\pm$0.25 & 1.5 & 5171$\pm$71 & 4.24$\pm$0.15 & 0.30$\pm$0.09 & 0.00$\pm$0.04 & (1) \\ 
    PZ99J160550.5-253313 & K1.0 & 5000 & 3.81$\pm$0.24 & 1.5 & 4952$\pm$52 & 3.96$\pm$0.13 & 0.16$\pm$0.06 & 0.18$\pm$0.04 & (1) \\ 
    RXJ0457.5+2014 & K1.0 & 5000 & 4.51$\pm$0.23 & 1.6 & 5020$\pm$55 & 4.53$\pm$0.13 & 0.13$\pm$0.06 & 0.20$\pm$0.05 & (1) \\ 
    RXJ0438.6+1546 & K2.0 & 4900 & 4.12$\pm$0.30 & 1.7 & 5000$\pm$64 & 4.18$\pm$0.14 & 0.55$\pm$0.09 & 0.01$\pm$0.05 & (1) \\ 
    RXJ1547.7-4018 & K3.0 & 4730 & 4.22$\pm$0.16 & 1.7 & 4946$\pm$63 & 4.49$\pm$0.13 & 0.32$\pm$0.08 & 0.10$\pm$0.05 & (1) \\ 
    RXJ1538.6-3916 & K4.0 & 4590 & 4.21$\pm$0.13 & 1.8 & 4771$\pm$51 & 4.58$\pm$0.13 & 0.37$\pm$0.06 & 0.18$\pm$0.04 & (1) \\ 
    RXJ1540.7-3756 & K6.0 & 4205 & 4.42$\pm$0.22 & 2.0 & 4345$\pm$41 & 4.17$\pm$0.12 & 0.38$\pm$0.05 & 0.13$\pm$0.04 & (1) \\ 
    RXJ1543.1-3920 & K6.0 & 4205 & 4.12$\pm$0.23 & 1.8 & 4416$\pm$42 & 4.48$\pm$0.12 & 0.45$\pm$0.05 & 0.23$\pm$0.04 & (1) \\ 
    SO879 & K7.0 & 4060 & 3.90$\pm$0.50 & 2.1 & 4205$\pm$39 & 4.41$\pm$0.11 & 0.89$\pm$0.04 & 0.24$\pm$0.04 & (2) \\ 
    Tyc7760283\_1 & M0.0 & 3850 & 4.70$\pm$0.50 & 2.1 & 3931$\pm$40 & 5.17$\pm$0.11 & 0.41$\pm$0.04 & 0.15$\pm$0.04 & (2) \\ 
    TWA14 & M0.5 & 3780 & 4.70$\pm$0.60 & 1.8 & 3822$\pm$40 & 5.28$\pm$0.11 & 0.72$\pm$0.05 & 0.24$\pm$0.04 & (2) \\ 
    RXJ1121.3-3447\_app2 & M1.0 & 3705 & 4.60$\pm$0.60 & 2.3 & 3732$\pm$47 & 5.13$\pm$0.11 & 0.79$\pm$0.07 & 0.15$\pm$0.04 & (2) \\ 
    RXJ1121.3-3447\_app1 & M1.0 & 3705 & 4.80$\pm$0.50 & 2.2 & 3645$\pm$46 & 5.13$\pm$0.11 & 0.71$\pm$0.07 & 0.18$\pm$0.04 & (2) \\ 
    CD\_29\_8887A & M2.0 & 3560 & 4.40$\pm$0.40 & 2.4 & 3473$\pm$55 & 5.01$\pm$0.16 & 0.60$\pm$0.05 & 0.15$\pm$0.05 & (2) \\ 
    CD\_36\_7429B & M3.0 & 3415 & 4.50$\pm$0.60 & 3.0 & 3247$\pm$54 & 4.52$\pm$0.16 & 0.59$\pm$0.05 & 0.11$\pm$0.05 & (2) \\ 
    TWA7 & M3.0 & 3415 & 4.40$\pm$0.40 & 2.4 & 3262$\pm$54 & 4.70$\pm$0.16 & 0.78$\pm$0.05 & 0.19$\pm$0.05 & (2) \\ 
    TWA15\_app2 & M3.0 & 3415 & 4.60$\pm$0.60 & 2.6 & 3293$\pm$55 & 4.77$\pm$0.16 & 0.68$\pm$0.05 & 0.14$\pm$0.05 & (2) \\ 
    TWA15\_app1 & M3.5 & 3340 & 4.50$\pm$0.30 & 2.4 & 3293$\pm$55 & 4.82$\pm$0.16 & 0.70$\pm$0.05 & 0.16$\pm$0.05 & (2) \\ 
    SO797 & M4.5 & 3200 & 3.90$\pm$0.40 & 3.4 & 3116$\pm$54 & 3.79$\pm$0.16 & 1.10$\pm$0.05 & 0.08$\pm$0.05 & (2) \\ 
    SO641 & M5.0 & 3125 & 3.80$\pm$0.40 & 3.5 & 3015$\pm$55 & 3.76$\pm$0.16 & 0.88$\pm$0.05 & 0.08$\pm$0.05 & (2) \\ 
    Par\_Lup3\_2 & M5.0 & 3125 & 3.70$\pm$0.40 & 3.3 & 3051$\pm$54 & 3.87$\pm$0.16 & 0.73$\pm$0.05 & 0.07$\pm$0.05 & (2) \\ 
    SO925 & M5.5 & 3060 & 3.80$\pm$0.40 & 4.1 & 2928$\pm$55 & 3.75$\pm$0.16 & 1.09$\pm$0.06 & 0.11$\pm$0.05 & (2) \\ 
    SO999 & M5.5 & 3060 & 3.80$\pm$0.40 & 3.9 & 3005$\pm$55 & 3.61$\pm$0.16 & 0.93$\pm$0.06 & 0.07$\pm$0.05 & (2) \\ 
    Sz107 & M5.5 & 3060 & 3.70$\pm$0.30 & 3.5 & 2929$\pm$55 & 3.64$\pm$0.16 & 0.86$\pm$0.06 & 0.09$\pm$0.05 & (2) \\ 
    LM717 & M6.5 & 2935 & 3.50$\pm$0.21 & 4.8 & 2777$\pm$80 & 3.53$\pm$0.22 & 1.93$\pm$0.07 & 0.13$\pm$0.08 & (2) \\ 
    Par\_Lup3\_1 & M6.5 & 2935 & 3.60$\pm$0.40 & 4.6 & 2759$\pm$80 & 3.49$\pm$0.22 & 2.56$\pm$0.08 & 0.13$\pm$0.08 & (2) \\ 
    J11195652-7504529 & M7.0 & 2880 & 3.09$\pm$0.20 & 6.9 & 2647$\pm$83 & 3.41$\pm$0.23 & 1.94$\pm$0.14 & 0.11$\pm$0.08 & (1) \\ 
    LM601 & M7.5 & 2795 &      & 6.5 & 2606$\pm$90 & 3.42$\pm$0.26 & 1.04$\pm$0.21 & 0.05$\pm$0.08 &      \\ 
    CHSM17173 & M8.0 & 2710 &      & 5.2 & 2611$\pm$80 & 3.45$\pm$0.22 & 2.31$\pm$0.08 & 0.12$\pm$0.08 &      \\ 
    TWA26 & M9.0 & 2400 & 3.60$\pm$0.50 & 5.7 & 2527$\pm$80 & 3.42$\pm$0.22 & 2.78$\pm$0.07 & 0.12$\pm$0.08 & (2) \\ 
    DENIS1245 & M9.5 & 2330 & 3.60$\pm$0.30 & 7.3 & 2513$\pm$80 & 3.34$\pm$0.22 & 2.68$\pm$0.08 & 0.10$\pm$0.08 & (2) \\
    \bottomrule
    \end{tabular}} 
    \label{table:tpl_param} 
    \tablefoot{The fifth column indicates the relative flux error measured within the 6500--7500~\AA\ range that is fed into the cINN. The last column indicates the literature source of the $\log(g)$ values. No measurement was available in the literature for two stars.
    }
    \tablebib{(1)~\citet{Manara2017}; (2) \citet{Stelzer2013}.}
\end{table*}

\subsection{Influence of extinction on spectral type classification}

\begin{figure}
    \centering
    \includegraphics[width=1\columnwidth]{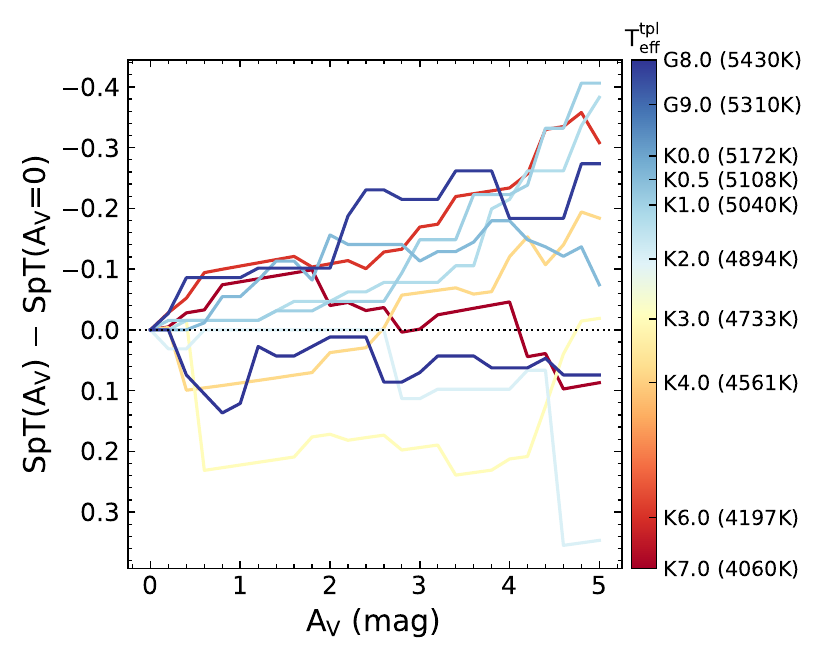}
    \caption{The difference of spectral type measured for reddened templates to the spectral type measured for the corresponding, pure template without extinction as a function of extinction. Spectral types are measured using the same methodology used in \citetalias{Itrich+2024}, which is based on the equivalent width of several absorption lines. The colour denotes the \teff\ of the template.
     } \label{fig:spt_avtest}
\end{figure}

In Sect.~\ref{sec:discussion_tf}, to examine the influence of extinction (\av) on the spectral type classification method used in \citetalias{Itrich+2024}, we redden the spectra of template stars between K7 type to G8 type artificially applying \av\ from 0 to 5~mag in 0.2~mag intervals. We measure the spectral type from each reddened spectrum and compare it with the spectral type measured from the unreddened, pure template spectrum where \av\ is 0~mag (SpT(\av\ = 0)). In Fig.~\ref{fig:spt_avtest}, we plot the difference between the spectral type of the reddened spectrum and the pure spectrum as a function of \av. Unlike the case of veiling (Fig.~\ref{fig:spt_veiltest}), the reddening changes the spectral type measurement mostly within 0.3 subclass and up to 0.4 subclass. We conclude that the influence of \av\ on the spectral type classification is negligible in our \av\ range.

\subsection{NIR colour-colour diagram}
\begin{figure*}
    \centering
    \includegraphics[width=1.99\columnwidth]{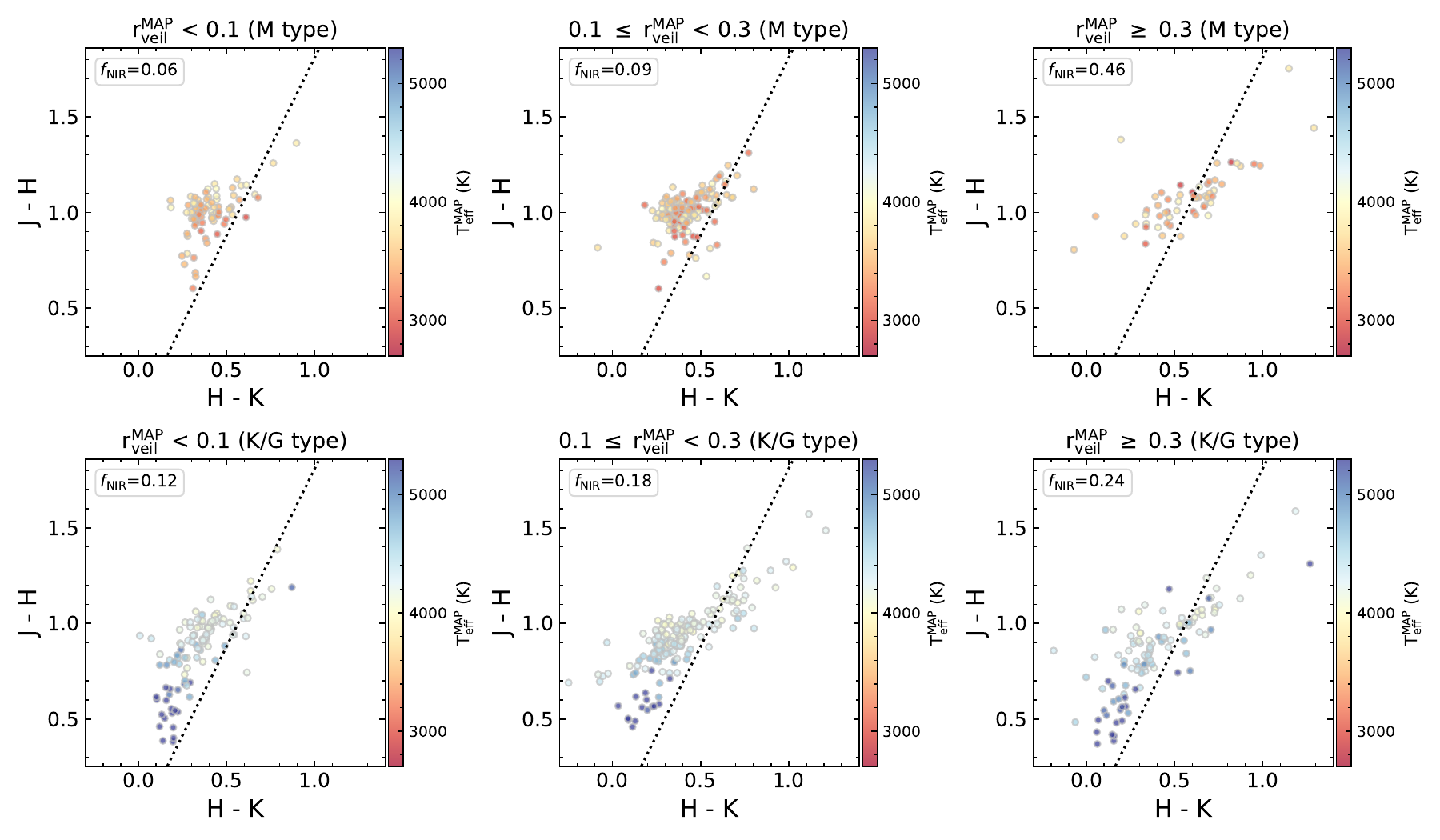}
    \caption{Colour–colour diagrams for the \Clean\ samples using the NIR photometry data from HAWK-I catalouge~\citep{Preibisch+2011a}. We divide samples into six groups using MAP estimates: M-type stars (\teff\ < 4000~K), K/G-type stars (\teff\ $\geq$ 4000~K), \veil\ < 0.1, 0.1 $\leq$ \veil\ < 0.3, and \veil\ > 0.3. The colour of each star indicates the corresponding \teff\ value. The black dashed lines with a slope of 1.86 indicate the NIR excess selection threshold from \cite{Zeidler+16}, where stars located to the right of the dashed line are classified as having NIR excess in \citetalias{Itrich+2024}. We present the fraction of stars with NIR excess at the upper left corner of each panel ($f_{\mathrm{NIR}}$).
    } \label{fig:ccd_map} 
\end{figure*}

\begin{figure*}
    \centering
    \includegraphics[width=1.99\columnwidth]{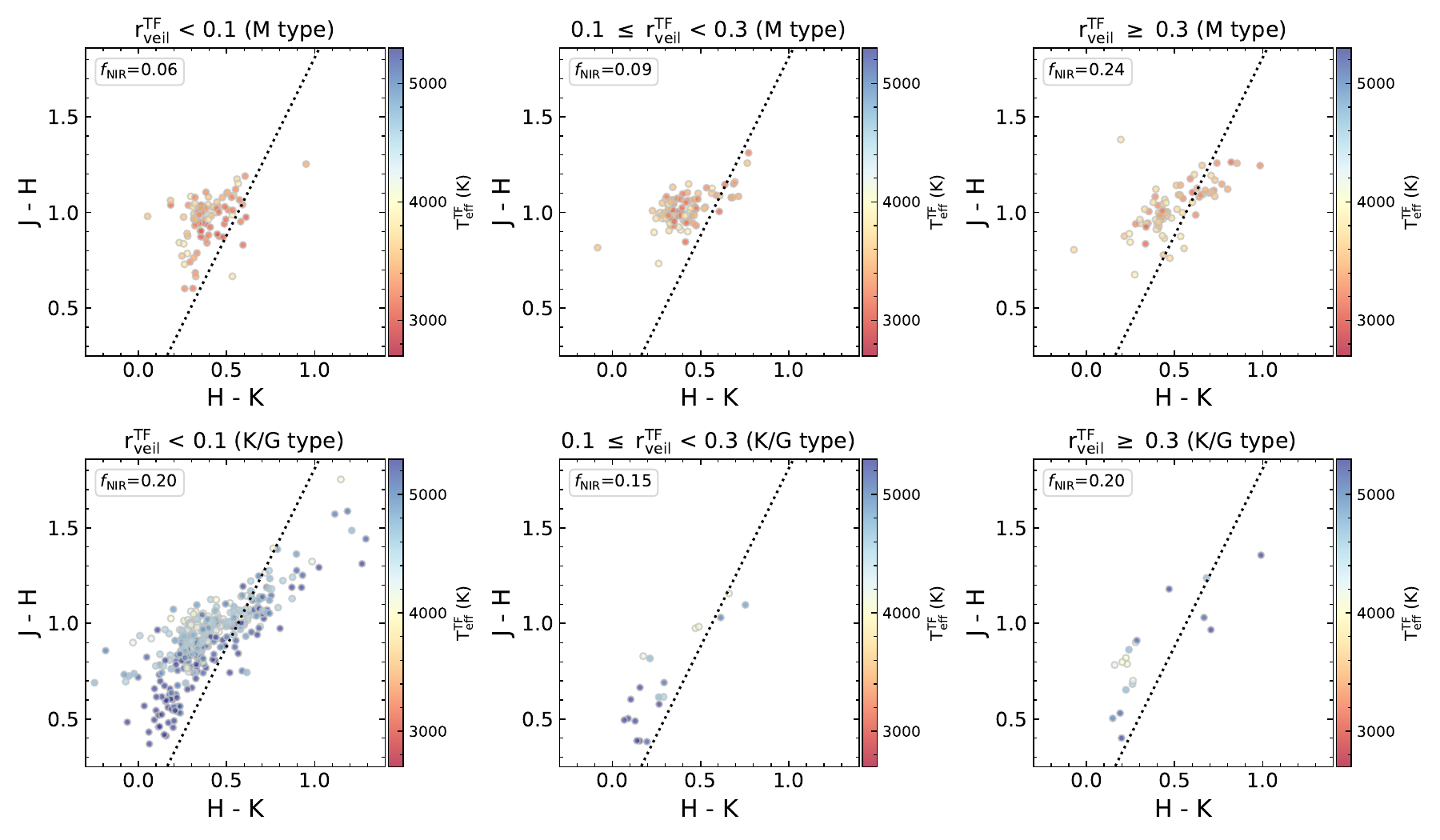}
    \caption{The colour–colour diagrams of the \Clean\ samples, similar to Fig.~\ref{fig:ccd_map}, but with sample groups divided based on TF parameters from \citetalias{Itrich+2024}. The colour of each star indicates the corresponding \teff\ value. The black lines and the numbers on the upper left corner indicate the NIR excess selection threshold from \cite{Zeidler+16} and the fraction of stars with NIR excess, respectively.
    } \label{fig:ccd_tf} 
\end{figure*}

In Sect.~\ref{sec:discussion_veil}, we investigate the distribution of the \Clean\ samples on the colour-colour diagrams using their NIR photometry data ($J$-, $H$-, and $K$-bands) obtained from HAWK-I catalouge~\citep{Preibisch+2011a}. Among 727 \Clean\ samples, we only use the 694 stars with matched HAWK-I counterparts. 
In Fig.~\ref{fig:ccd_map}, we divide the samples into 6 groups depending on their spectral type classification (i.e. M-type or K/G types) and \veil\ ranges based on MAP estimates. The colour code indicates the corresponding \teff. In Fig.~\ref{fig:ccd_tf}, we present similar diagrams but classify the samples according to stellar parameters from \citetalias{Itrich+2024}.


\clearpage

\section{Surface gravity}
\label{sec:gravity}

\begin{figure}
    \centering
    \includegraphics[width=0.95\columnwidth]{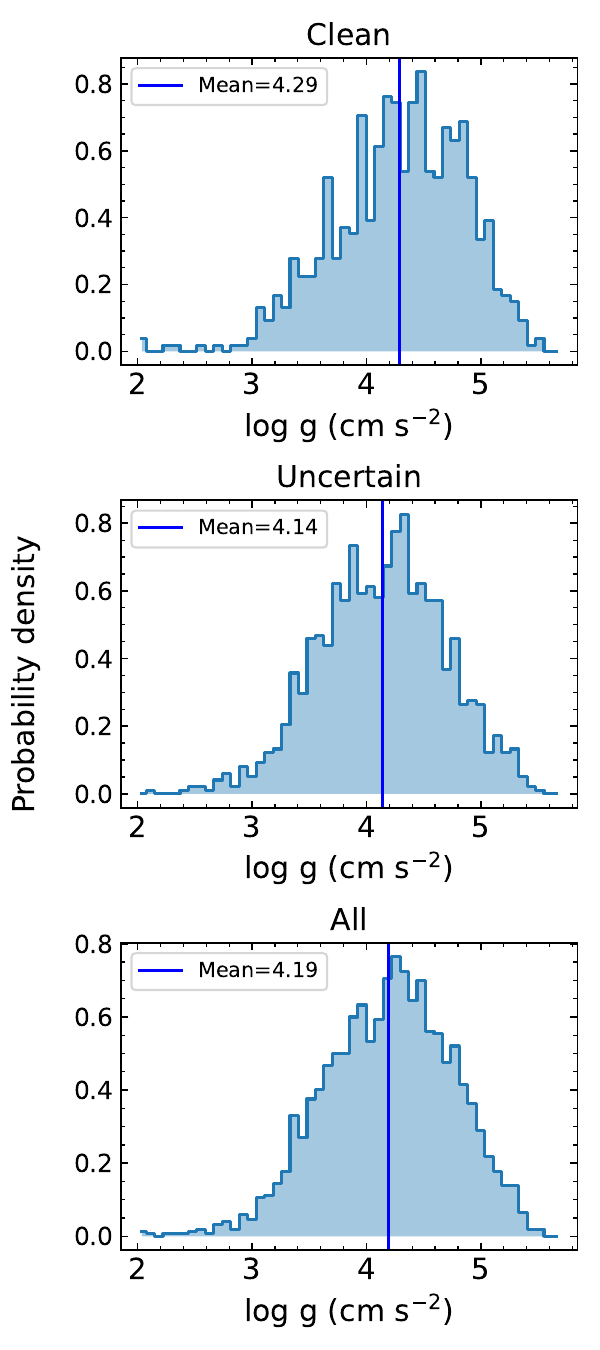}
    \caption{Distribution of \logg\ measured by the cINN for each sample group: \Clean, \Unc, and All. The blue vertical line indicates the mean value of the samples.}
    \label{fig:gravity}
\end{figure}

In this section, we analyse the surface gravity (\logg) of Tr14 stars measured by our cINN, which were excluded from the main sections. Fig.~\ref{fig:gravity} presents the distributions of surface gravities for different sample groups (\Clean, \Unc, and All). The \logg\ of Tr14 stars mostly ranges from 3 to 5.5, matching well with the typical range of \logg\ of pre-main sequence stars~\citep{Herczeg2014, Stelzer2013, Frasca+2017, Manara2017, Olney2020}. The average \logg\ for all samples is about 4.2. The overall distributions and mean values are almost the same for \Clean\ and \Unc\ groups.

For 39 stars (about 2\%\ of the entire sample), we measure a relatively small \logg\ values of 2--3 but considering the 1-$\sigma$ estimation uncertainty, only 15 of them have \logg\ values less than 3. Among these, \logg\ values of 6 stars are below 2.5, the lower limit of the training range. These low surface gravities may be wrong estimates, especially for cases outside the training range. One possible explanation for the low surface gravity is that they are background giant stars whose luminosity is underestimated. The background stars may have higher extinction than other stars and we found that 6 of 15 stars have high extinction (\av\ > 4), much higher than the average of 2.6~mag for the entire sample and \citetalias{Itrich+2024} also measured high \av\ values for these stars. On the other hand, for about 7\%\ of the sample, \logg\ is larger than 5, the upper training limit. Nevertheless, as shown in Fig.~\ref{fig:gravity}, \logg\ values exceeding the upper limit of 5 are not significantly extrapolated and are all below 5.5. These values are relatively uncertain in that they are extrapolated and a bit high for the surface gravity of young stars. We discuss these stars later in Appendix~\ref{sec:high_logg}.

\begin{figure*} 
    \centering
    \includegraphics[width=1.9\columnwidth]{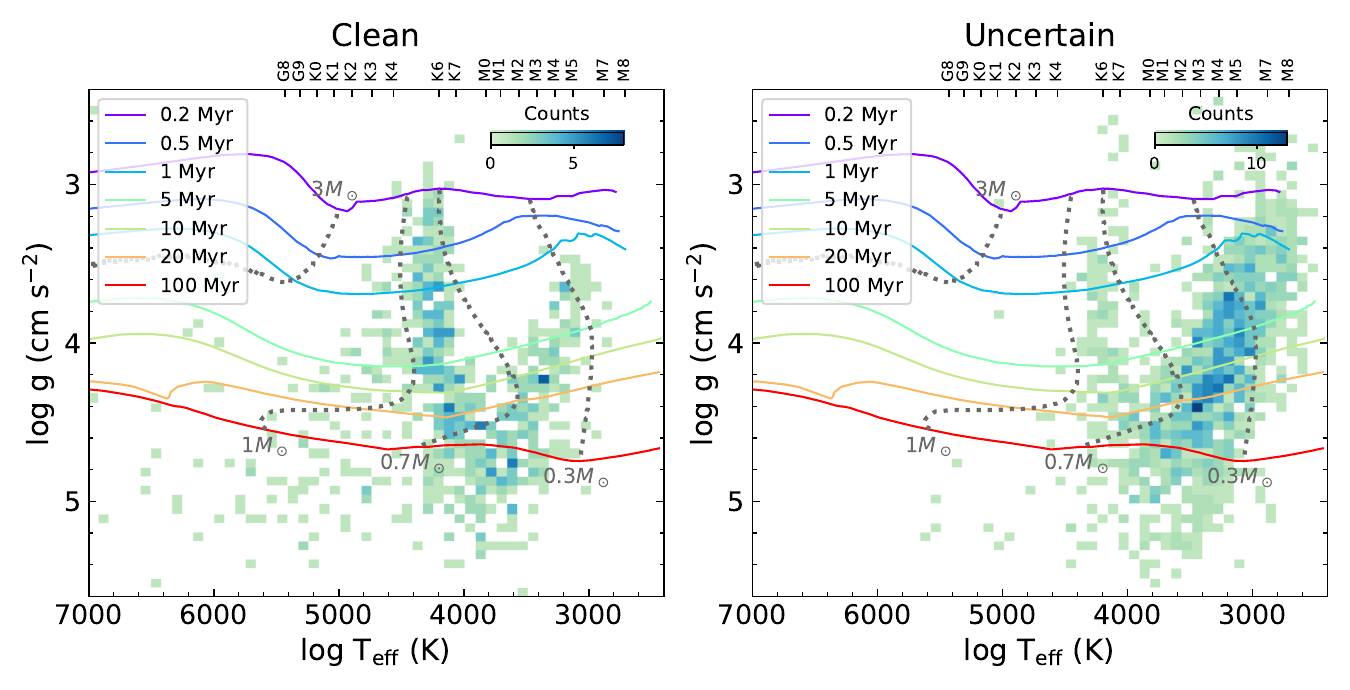}
    \caption{\logg\ -- \teff\ diagrams of the \Clean\ samples (left) and the \Unc\ samples (right) based on the stellar parameters obtained by the cINN in this work. The colour indicates the number density of the stars. We overplot PARSEC theoretical evolutionary tracks~\citep{Bressan+2012} where solid lines indicate isochrones from 0.2 to 100~Myr and dashed lines indicate isomasses of 0.3, 0.7, 1, and 3~\msol.
     } \label{fig:tg} 
\end{figure*}

Fig.~\ref{fig:tg} shows the distribution of \teff\ and \logg\ as a 2D histogram for the \Clean\ and \Unc\ samples, respectively. Interestingly, we find a V-shaped distribution in both sample groups. Centred at 3800$\sim$4000~K, the cooler stars tend to have higher gravity with lower temperatures, whereas the hotter stars have higher gravity with higher temperatures. This feature reveals that there are fewer or no stars around 3800$\sim$4000~K with \logg\ below 4.0. A similar V-shaped distribution in the T-g plot has been reported in \cite{Frasca+2017} and \cite{Manara2017} as well. 
We overplot the isochrone and isomass lines from the PARSEC evolutionary tracks~\citep{Bressan+2012}. Below one solar mass, isomass lines are nearly vertical, meaning that the stellar mass varies with \teff\ rather than \logg, while isochrones are mostly horizontal, varying with \logg. Because \logg\ increases with stellar age due to the photospheric contraction, it is often used as a tracer of stellar age instead of luminosity~\citep{Takagi+2010}. However, in Fig.~\ref{fig:tg}, most stars are distributed near the isochrones much older than the $\sim$1Myr ages derived from HR diagrams (Fig.\ref{fig:hr}). To obtain a similar range of stellar age around 1~Myr, \logg\ should be less than 3.5, but \logg\ are mostly higher than this. Moreover, between \logg\ of 4 and 4.5, the stellar age changes rapidly from 5 to 100~Myr depending on gravity.

Similar to Sect.~\ref{sec:age_mass}, we calculate the stellar masses and ages by interpolating the same evolutionary tracks along \teff\ and \logg\ and compare these values with those inferred from \teff\--$L_{\rm{bol}}$ relation. The difference between the two masses ($m_{*,\rm{Tg}}$ and $m_{*,\rm{TL}}$), is the smallest for K-type stars (0.12 dex). For M-type stars, $m_{*,\rm{Tg}}$ is consistently larger than $m_{*,\rm{TL}}$ by about 0.23 dex, whereas for G-type stars, the scatter is large.
Meanwhile, the scatter is too large to find a correlation between the two ages obtained from the different relations ($t_{*,\rm{Tg}}$ and $t_{*,\rm{TL}}$) and the RMSE between them is large (1.3 dex). As shown in Fig.~\ref{fig:tg}, $t_{*,\rm{Tg}}$ is in general older than $t_{*,\rm{TL}}$. The stellar ages of individual stars, both $t_{*,\rm{Tg}}$ and $t_{*,\rm{TL}}$, may have large uncertainties because of various factors but we also find a large difference in the mean stellar age obtained through the distribution of individual ages. In Fig.~\ref{fig:age}, the mean age of Tr14 stars obtained through \teff\ and $L_{\rm{bol}}$ is 0.7$^{+3.2}_{-0.6}$~Myr (5.87$\pm$0.73 in log scale), whereas the mean age measured through \teff\ and \logg\ using the same methodology is 8.1$^{+43}_{-6.8}$~Myr (6.91$\pm$0.80 in log scale), where the difference between the two is more than 1 sigma. \cite{Frasca+2017} and \cite{Manara2017} also reported the ages inferred from \teff\--\logg\ higher than those inferred from \teff\--$L_{\rm{bol}}$.

\subsection{High surface gravity}
\label{sec:high_logg}
As shown in Fig.~\ref{fig:gravity}, some stars exhibit surface gravity higher than the typical values for pre-main sequence stars. Specifically, 20\%\ of the sample (410 stars) have \logg\ values higher than 4.5 when considering the 1-$\sigma$ estimation uncertainty, and this fraction decreases to 8.6\%\ (177 stars) when considering the 3-$\sigma$ uncertainty. We further examine the 177 stars classified as having higher \logg\ than typical pre-main sequence stars. 

\begin{figure*} 
    \centering
    \includegraphics[width=2.0\columnwidth]{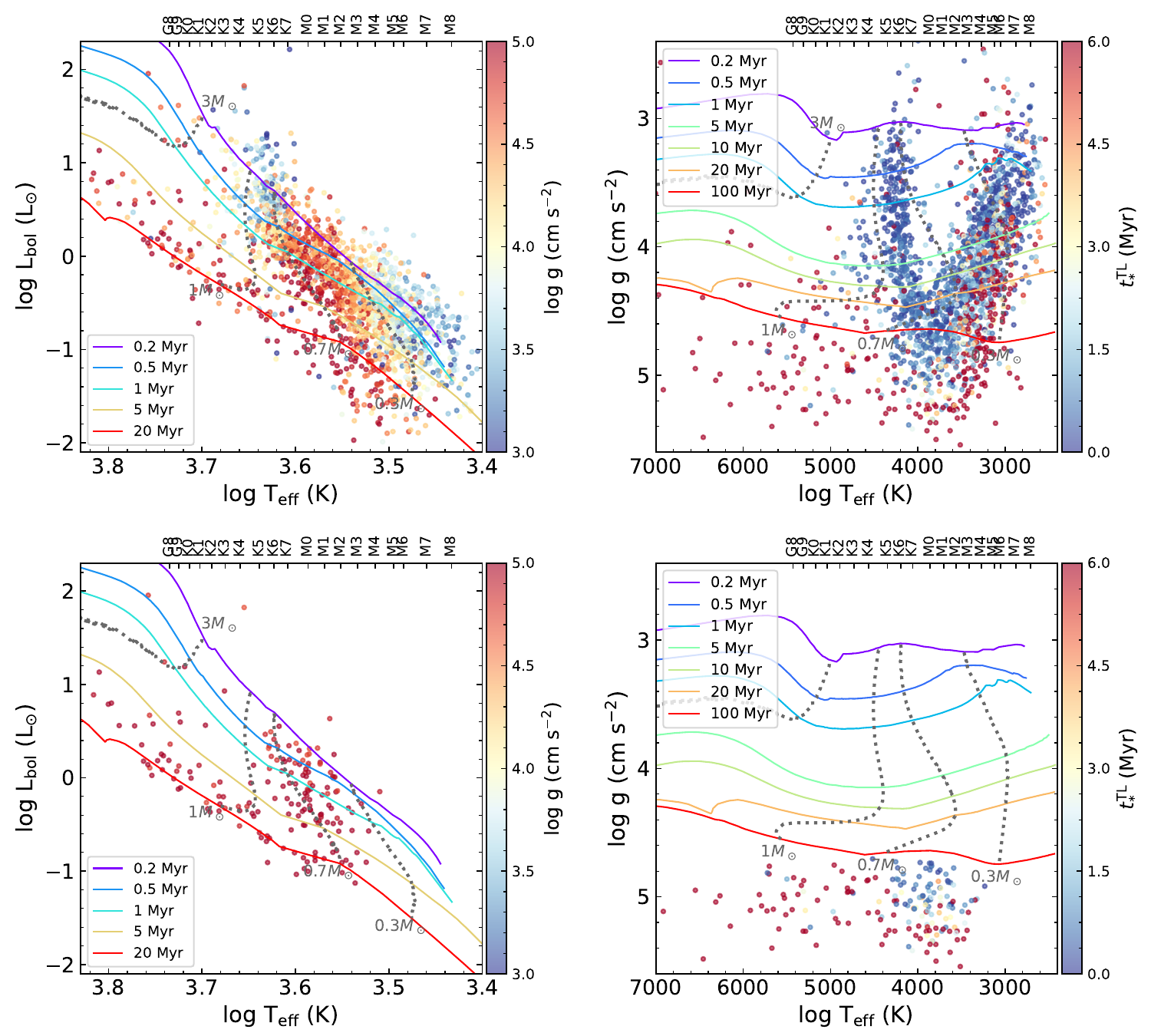}
    \caption{HR diagrams (left) and \logg\ -- \teff\ diagrams (right) based on parameters obtained by cINN in this work using all 2051 Tr14 samples (upper panels) and using only 177 stars with surface gravities higher than 4.5 considering 3-$\sigma$ estimation uncertainty. The colour in the HR diagrams indicates \logg\ estimates while the colour in the \logg\ -- \teff\ diagrams represents stellar age derived from bolometric luminosity and \teff\ using PARSEC evolutionary tracks~\citep{Bressan+2012}.
    We overplot PARSEC theoretical evolutionary tracks~\citep{Bressan+2012} where solid lines indicate isochrones from 0.2 to 100~Myr and dashed lines indicate isomasses of 0.3, 0.7, 1, and 3~\msol.
     } \label{fig:hr_tg} 
\end{figure*}

In Fig.~\ref{fig:hr_tg}, we present HR diagrams with \logg\ values represented by colour and \logg\ -- \teff\ diagrams where age estimates derived from the HR diagram are colour-coded. The first row shows the entire sample, while the second row shows only the 177 stars with \logg\ values above 4.5.
These high-surface gravity stars split into two groups on the HR diagram: stars near the 20~Myr isochrone and stars appearing younger with estimated ages around 1~Myr. Of the 177 stars, approximately 55\%\ (i.e. 98 stars) have $t_{*,\rm{TL}}$ below 5~Myr, while the remaining 45\%\ are older than 5~Myr. When we increase the \logg\ threshold from 4.5 to 5, the proportion of high-\logg\ stars with younger age estimates decreases to 30--40\%. Thus, for about half of the stars with high \logg\, the ages inferred from the HR diagram are also relatively higher than the average stellar age of Tr14 ($\sim$1~Myr).
Stars with high \logg\ but young age estimates generally lie near the vertex of the V-shaped distribution on the \logg\ -- \teff\ diagram, corresponding to \teff\ between 3400 and 4500 K (M3--K5).

For these 98 stars with high \logg\ but low age estimates ($t_{*,\rm{TL}}$ < 5~Myr), we can consider three cases: 1) \logg\ is overestimated 2) $t_{*,\rm{TL}}$ is underestimated or 3) both measurements are accurate.
First, if the surface gravity is overestimated, potential causes are the \logg\ accuracy of cINN, the simulation gap of the Phoenix model, and spectral features that mimic higher surface gravity.
Surface gravity is typically measured based on line broadening in the spectrum with higher spectral resolution than that of MUSE. Our network, however, uses spectra with R$\sim$4000, which is generally considered insufficient for precise gravity measurements. Nevertheless, as demonstrated in Sect.~\ref{sec:validation}, our network accurately measured \logg\ from both synthetic and real spectra. Despite the relatively low spectral resolution, the \logg\ estimates for template stars were consistent with literature values measured from X-shooter spectra with higher spectral resolution within 1-$\sigma$ uncertainty.

Additionally, in \citetalias{Kang+23b}, we showed which spectral features the network relies on to estimate \logg\ (see Sect. 6 of \citetalias{Kang+23b} for more detail) and found that important spectral features varied depending on the spectral type of the target star and that the network prioritised several spectral lines known as typical stellar parameter tracers. 
For example, the \nai\ doublet 8183, 8195~\AA\ lines and the \ki\ doublet 7665, 7699~\AA\ lines were important for M-type stars, while for hotter stars, the importance of the \ki\ doublet lines decreased and the \nai\ doublet 5890, 5896~\AA\ lines became more important. 
Since the measured surface gravity for Tr14 stars also falls mostly within the expected range for pre-main sequence stars, we suggest that the overall surface gravity accuracy of our cINN is sufficient and reliable.

The second potential cause is related to the simulation gap in the Phoenix model. As shown in Table~\ref{table:tpl_param}, cINN measured \logg\ over 4.7 for some class III template stars whose spectral types (M3.5–M0) are similar to those of 98 high-\logg\ stars with low age estimates. Our estimates still agree well with the literature values since the literature~\citep{Stelzer2013} also measured high \logg\ values above 4.5 for these template stars. We note that \logg\ of these template stars are also measured by using Phoenix models (BT-Settl). Notably, the V-shaped \logg\ -- \teff\ distribution with high \logg\ near 4000~K has also been reported by \cite{Manara2017} and \cite{Frasca+2017}, where both used Phoneix models to measure \logg. It is possible that the effects of \logg\ were not fully reflected in the Phoenix synthetic spectra at these temperatures, potentially leading to the overestimation of \logg\ when applying Phoenix models to real observations.
The third possibility is that spectral features sensitive to surface gravity could be influenced by factors other than the surface gravity itself, mimicking signatures of high surface gravity. In such cases, even alternative methods other than cINN might yield similarly high \logg\ values.

On the other hand, the age estimates based on luminosity may also be underestimated. The luminosity spread at the same temperature on the HR diagram is commonly observed and is attributed to several factors such as uncertainties in stellar parameters, intrinsic age spread, stellar variability, etc. However, the observed positions of these young high-\logg\ stars on the HR diagram deviate non-negligibly from their expected positions based on \logg\ estimates. We propose that the main sequence foreground stars may have been incorrectly measured as young stars because of the overestimated luminosity. As mentioned in \citetalias{Itrich+2024}, significant contamination by foreground stars is expected because of the large distance of 2.35~kpc to Tr14 and because our samples with good Gaia astrometric measurements were limited.

Lastly, some stars may have higher surface gravity than expected for their age. If a star has undergone a slightly different evolutionary process than other stars in the cluster, its photospheric parameters may differ from those of stars with the same age and mass. If the stellar mass significantly increased due to strong accretion before the radius expanded, this could result in high \logg. Such cases would likely exhibit characteristics of strong accretors, such as veiling or prominent emission lines. However, the average \veil\ for the 98 stars is 0.31, which is neither small nor extremely high. Among these stars, only 12 have \veil\ larger than 0.5, and only 8 exceed 0.7, suggesting that strong accretors are relatively rare among these 98 stars.

\clearpage

\section{Degeneracy in posterior distributions}
\label{sec:posterior_peak}

In this paper, we sample 4096 posterior estimates per observation to get a posterior distribution and use a MAP estimate as a representative value from a marginalised, one-dimensional posterior distribution of each stellar parameter. To use the most probable estimate (i.e. peak value) as a representative, we need to check if degeneracy remains in the posterior distribution which is usually exhibited as a multimodal distribution. We check the number of peaks (i.e. local maxima) in each posterior distribution for Tr14 samples to investigate how often the degeneracy remains.

Using the same methodology as in \cite{Kang+22}, we fit the marginalised 1D posterior distributions with multiple Gaussians and describe the posterior distribution with a maximum of 6 Gaussian components. Then we count the number of local maxima using the first derivative of the fitted function.   
For the entire Tr14 samples, 93.7\%\ of them have a clear single peak in their posterior distributions for all parameters. Only 6 samples in the \Clean\ group have multiple peaks in at least one parameter while 124 samples in the \Unc\ group (i.e. 9.4\%\ of the group) have multiple peaks. 
We found that multi-peaks arise from the estimation of the Phoenix library origin flag and that the double modes in the posterior distribution of the other stellar parameters are clearly separated based on the posterior estimates of the Phoenix library origin. In most of the multimodal posterior distributions, one mode is clearly more dominant than the other mode so we can use the MAP estimate as a representative which is determined from the dominant mode. 

There are 13 samples (three from the \Clean\ group and ten from the \Unc\ group) which have two distinct modes with a comparable probability in the posterior distribution of stellar parameters (i.e. \teff, \logg, \av, and \veil). For 11 of them, we confirm that the MAP estimates are determined from the same mode in all parameters. However, for the other two samples, which all belong to the \Unc\ group, the MAP estimates are determined in different modes for each parameter. It is better not to use a MAP estimate as a representative in these cases. Still, we will include these samples in our statistical analysis of the \Unc\ group, considering that the difference between the two modes is almost negligible because of the very narrow width of the posterior distribution even in the multimodal cases. From the experiment, we conclude that we can use the MAP estimate as a representative in this paper. 

\begin{figure*}
    \centering
    \includegraphics[width=1.65\columnwidth]{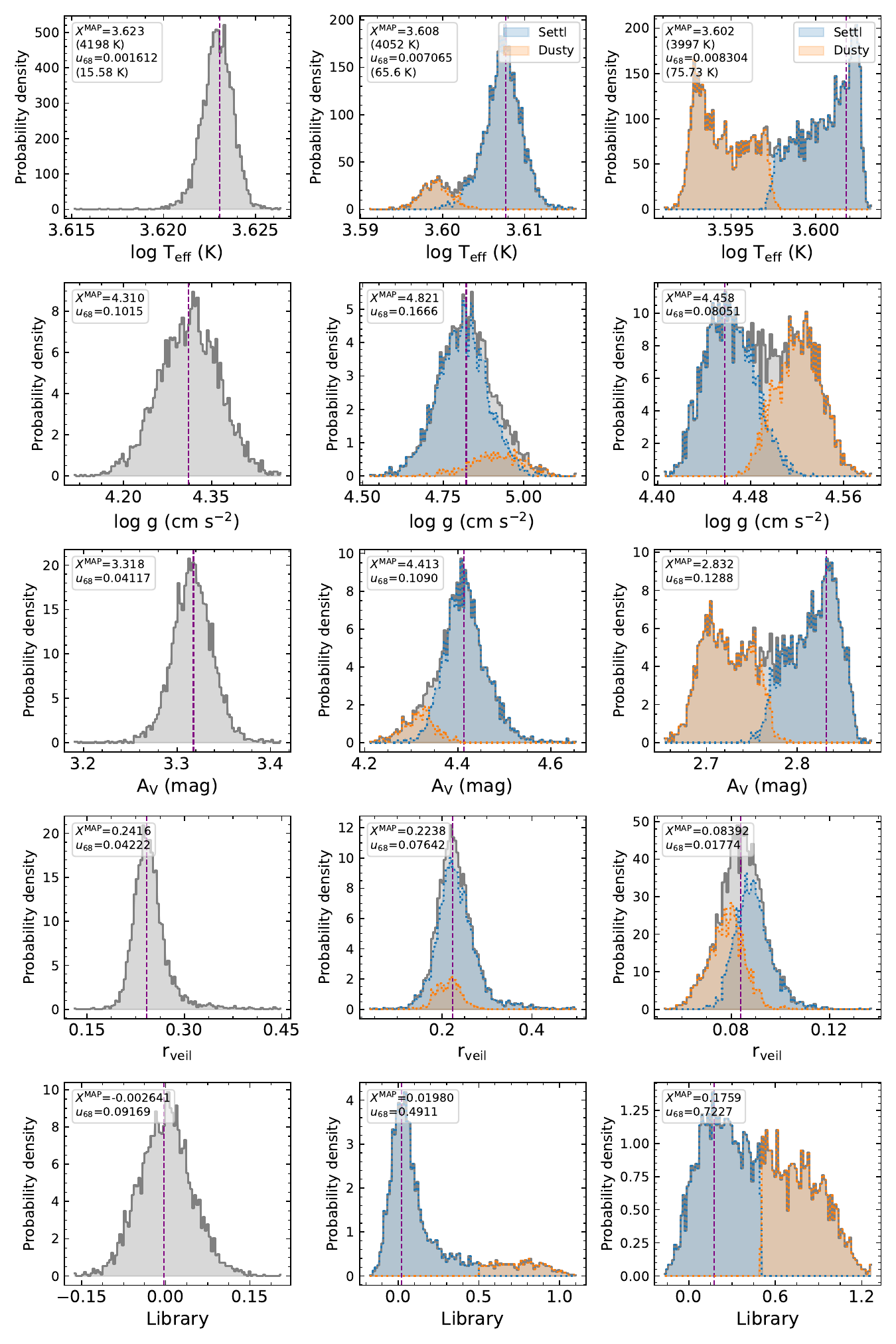}
    \caption{Posterior probability distributions (grey) of the five target parameters. Each column corresponds to a different sample drawn from the \Clean\ group.
    The first column shows an example of an unimodal posterior distribution with a clear single peak (i.e. MAP estimate), which is the most common case in our results. The other columns represent degenerate cases where the posterior distribution of some parameters has a multimodal shape. In these cases, we divide the posterior distribution into two groups depending on the posterior estimate of the  Phoenix library origin: a library value of 0 denotes Settl (blue shading) and a value of 1 denotes Dusty (orange shading). In each panel, we present the MAP estimate and the uncertainty at 68\%\ confidence interval ($u_{68}$) in the text box.
     } \label{fig:ex_post} 
\end{figure*}

In Fig.~\ref{fig:ex_post}, we present the posterior distributions of the five target parameters for three examples selected from the \Clean\ group. The first example (the first column of Fig.~\ref{fig:ex_post}) shows the most common case, an unimodal posterior distribution with a clear single peak, where we determine a MAP estimate. The second and third examples represent the cases of a multimodal posterior distribution, where the different modes can be separated depending on the posterior estimate of the Phoenix library origin flag. In the second example, the mode of one library is more dominant than the other one so MAP estimates are determined by the dominant one. The third one is an example of multimodal posterior distributions where two modes have a comparable probability. By dividing the posterior distribution into two groups depending on the posterior estimate of the Phoenix library origin (Settl and Dusty), we confirm that the two peaks in the posterior distribution are due to the degeneracy in the library origin estimate.

\end{appendix}

\end{document}